\documentclass[secnumarabic, reprint, superscriptaddress, tightenlines, amsmath,amssymb, aps, prl, nobibnotes,
]{revtex4-1}

\usepackage{hyphenat}
\usepackage{float}
\usepackage{natbib}
\usepackage{graphicx,caption,subcaption,xcolor}
\usepackage[colorlinks=true,citecolor=blue,
linkcolor=blue,urlcolor=blue]{hyperref}
\usepackage{blindtext}
\usepackage[utf8]{inputenc} 
\usepackage[english]{babel}
\usepackage[OT2, T1]{fontenc}
\usepackage{upquote}
\usepackage{mathtools}
\usepackage [autostyle, english = american]{csquotes}
\usepackage{bigints}
\usepackage{subcaption}

\usepackage[format=plain]{caption}
\usepackage{stackengine}

\usepackage{amsmath,accents}

\usepackage[format=plain,labelfont=bf]{caption}

\begin{document}

\title{
Tailoring Superconductivity with Two-Level Systems
}

\author{Joshuah T. Heath}
\email{joshuah.t.heath@su.se}
\author{Alexander C. Tyner}

\affiliation{Nordita, Stockholm University and KTH Royal Institute of Technology, Hannes Alfvéns väg 12, SE-106 91 Stockholm, Sweden}
\affiliation{Department of Physics, University of Connecticut, Storrs, Connecticut 06269, USA}


\author{\\ S. Pamir Alpay}
\affiliation{Institute of Material Science, University of Connecticut, Storrs, Connecticut 06269, USA}
\affiliation{Department of Materials Science and Engineering, University of Connecticut, Storrs, Connecticut 06269, USA}

\author{Peter Krogstrup}

\affiliation{NNF Quantum Computing Programme, Niels Bohr Institute, University of Copenhagen, Denmark}


\author{Alexander V. Balatsky}
\email{balatsky@kth.se}
\affiliation{Nordita, Stockholm University and KTH Royal Institute of Technology, Hannes Alfvéns väg 12, SE-106 91 Stockholm, Sweden}
\affiliation{Department of Physics, University of Connecticut, Storrs, Connecticut 06269, USA}

\date{\today}

\begin{abstract}
\noindent 
We investigate the impact of two-level systems (TLSs) on superconductivity, treating them as soft modes localised in real space. We show that these defects can either enhance or suppress the superconducting critical temperature, depending on their surface density and average frequency. Using thin-film aluminium as a case study, we quantitatively describe how TLSs modify both the critical temperature and the zero-temperature superconducting gap. Our results thus highlight new opportunities for tailoring material properties through TLS engineering.
\end{abstract}

\pacs{1}

\maketitle

\indent {\it Introduction--} The critical temperature $T_c$ of conventional (phonon-exchange) superconductors is 
principally governed by the electronic band structure and phonon spectrum~\cite{Bardeen1957,McMillan1968Mar,RickayzenBook,Allen1975Aug,RevModPhys.62.1027}. In materials like aluminium (Al), $T_c$ may be enhanced by introducing structural disorder~\cite{Garland1968Oct,Mayoh2014Oct,Pracht2016Mar,Leavens1981Nov,Tamura1993Mar,Croitoru2016Mar,Grankin2024Aug,Lee2025Nov} or by approaching the monolayer limit~\cite{,Douglass1964Jul,Strongin1965Jun,Abeles1966Sep,Cohen1968Apr,Strongin1968Oct,chubov1969dependence,Meservey1971Jan,Townsend1972Jan,Allen1974May,Pettit1976Apr,Leavens1981Nov,Feibelman1984Jun,Tamura1993Mar,Bose2009Apr,Walmsley2011Feb,Cherney2011Feb,Lozano2019Feb,Houben2020Mar,Guo2004Dec,Shanenko2006Sep,Adams2017Mar,Nguyen2019May,Yu2022Jun,vanWeerdenburg2023Mar,Allen2024Jun,Deshpande2025Jan}. While the precise origin of this enhancement is debated~\cite{ginzburg1964superconductivity,Ginzburg1964Nov,Little1964Jun,Little1987,Little1996,Ginzburg1989Jan,Baggioli2020Jun,Travaglino2023Jan,Zaccone2024Apr,Zaccone2025May}, it is generally attributed to one of two mechanisms: i) modification of the electronic density of states (DoS) or ii) some change in the electron-phonon spectral function $\alpha^2F(\nu)$. In this Letter, we introduce a third mechanism for enhancing conventional superconductivity: iii) engineering the density and frequency of two-level systems (TLSs) on the surface of the superconductor.

Two-level systems are localised microscopic defects commonly found in 
the native oxides of substrate-metal and air-metal interfaces~\cite{Phillips1987Dec,W.Anderson1972Jan,Wisbey2010Nov,Wenner2011Sep,DuBois2013Feb,Earnest2018Nov,Muller2019Oct,Chayanun2024Jun}. They serve as a major source of decoherence in 
superconductor-based quantum technology~\cite{Simmonds2004Aug,Shnirman2005Apr,Ku2005Jul,Martinis2005Nov,Muller2019Oct}, and can persist into the epitaxial limit~\cite{Hung2023Feb,Zhang2024Oct,Oh2006Sep,Oliver2013Oct}. Reasonable agreement between first principles simulations of amorphous Al$_2$O$_3$~\cite{Tyner2025Jul} and experimental studies of polycrystalline $\gamma$-Al$_2$O$_3$ and amorphous $\alpha$-AlO$_x$~\cite{Hung2022Mar} suggest that localised soft modes (or "vibrons") are a likely candidate for dipole-activated TLSs. As a consequence, a large TLS density will function as an ensemble of dynamical impurities~\cite{Balatsky2006May,Heath2025Mar}, with each TLS contributing to the low-frequency spectral weight of the bulk phonon population and influencing the superconductor's critical temperature.

\begin{figure}
  \centering
\vspace{1mm}
  \begin{minipage}{0.51\textwidth}
    \begin{minipage}{0.05\textwidth}
      \raggedleft
      (a)
    \end{minipage}%
    \hspace{0.01\textwidth}%
    \vspace{-2mm}
    \begin{minipage}{0.9\textwidth}
      \hspace{-7mm}\includegraphics[width=0.95\linewidth]{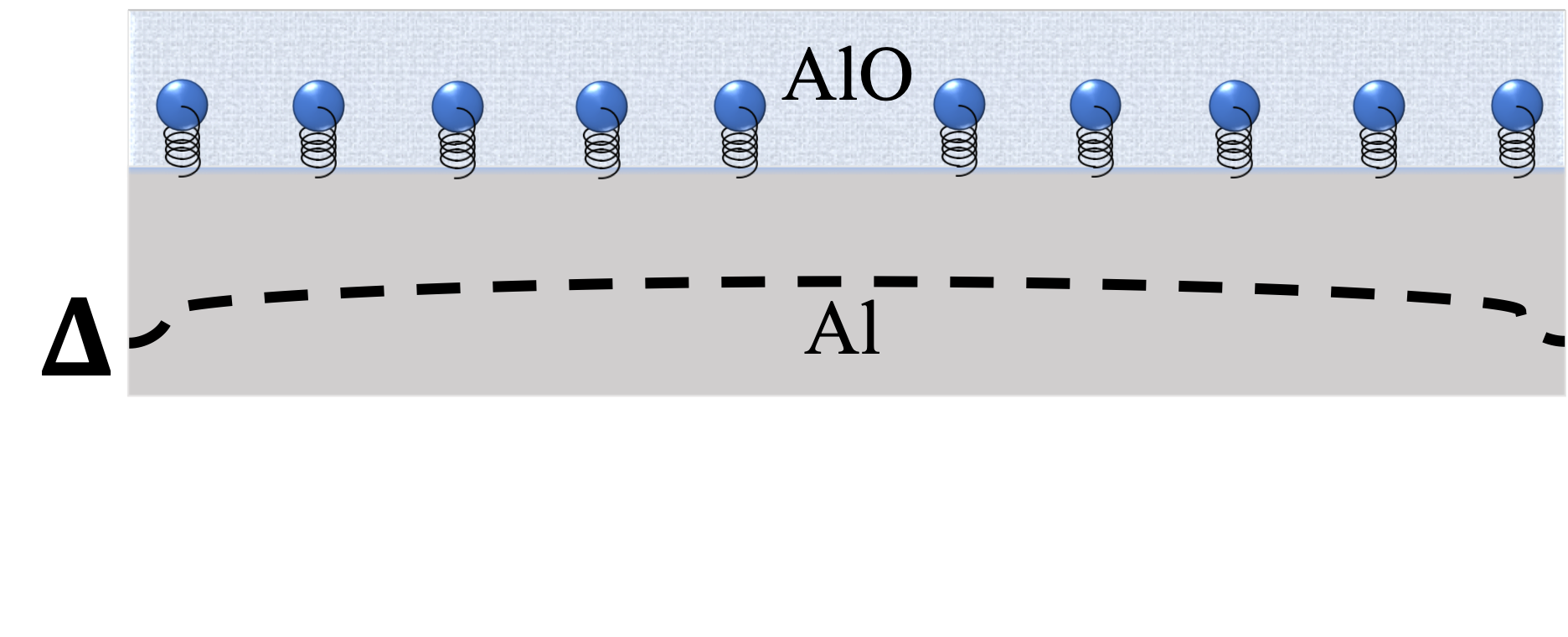}
    \end{minipage}
  \end{minipage}
\vspace{-6mm}
  \hfill

  \begin{minipage}{0.51\textwidth}
    \begin{minipage}{0.05\textwidth}
      \raggedleft
      (b)
    \end{minipage}%
    \hspace{0.01\textwidth}%
    \begin{minipage}{0.9\textwidth}
      \hspace{-7mm}\includegraphics[width=0.95\linewidth]{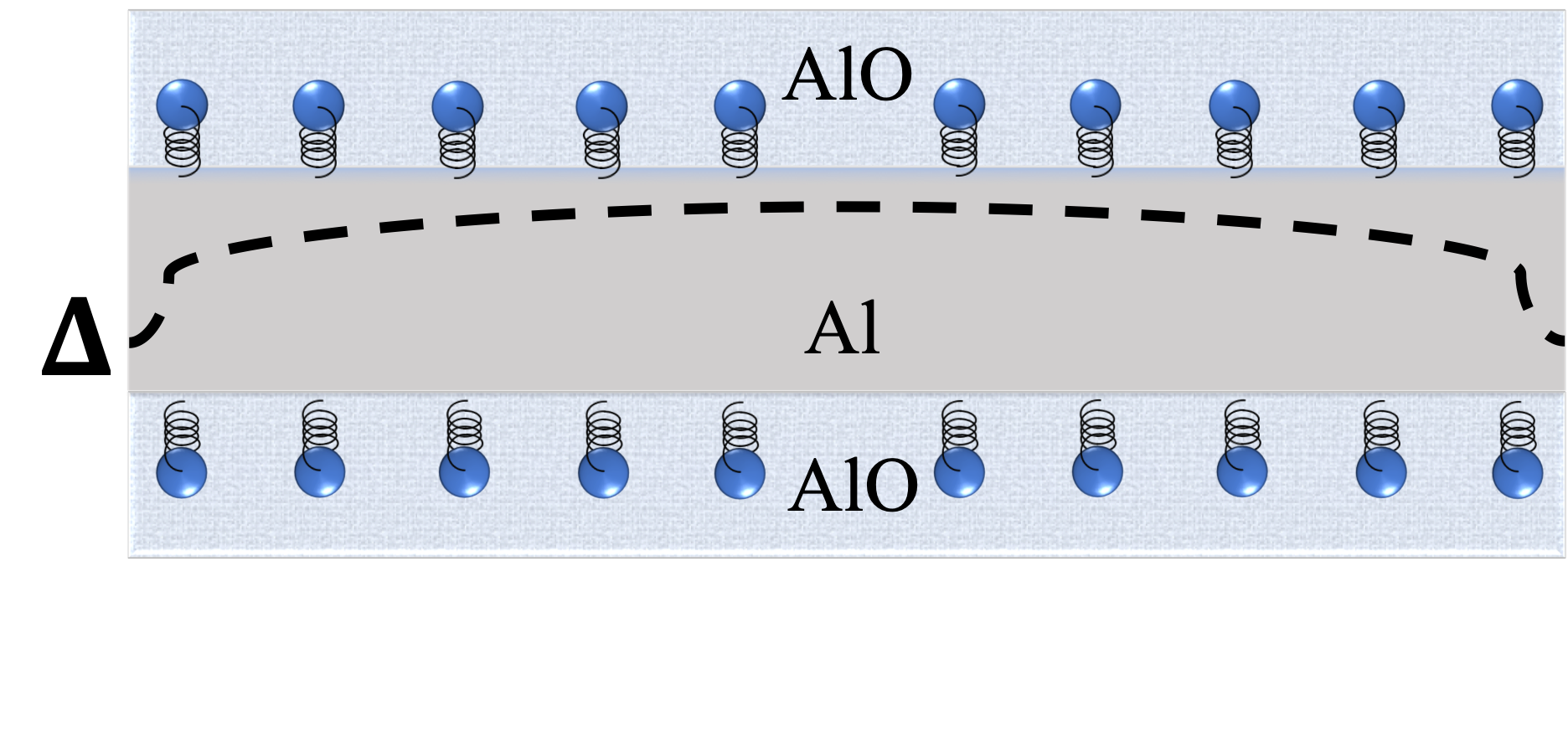}
    \end{minipage}
  \end{minipage}
  \phantom{.}
\vspace{-7 mm}
  \caption{\sloppy \small 
  Two-level systems (TLSs) on (a) the oxidised top surface and (b) the oxidised top and bottom surfaces of an Al sample, raising the local superconducting gap $\Delta$.  
  \vspace{-0mm}}
  \vspace{-4mm}
\end{figure}

\vspace{0mm}
We pinpoint under what conditions a finite TLS population will strengthen superconductivity in thin film Al. This is done by introducing a "two-boson" model where the vibron and bulk phonon contributions to $\alpha^2 F(\nu)$ are explicitly taken into account in the calculation of $T_c$. We focus on multilayer aluminium, with parametrization of the TLS contribution to $\alpha^2 F(\nu)$ motivated by I-V data taken on epitaxially-grown Al near the monolayer limit~\cite{vanWeerdenburg2023Mar} and the bulk phonon contribution to $\alpha^2 F(\nu)$ found via first principles calculations. The critical temperature $T_c$ and the gap edge $\Delta_1$~\cite{Fibich1965Apr} are then found numerically by solving the Eliashberg equations~\cite{Eliashberg1960,Eliashberg1961,Parks1,Scalapino1965Jan,
Marsiglio1991Mar,Chubukov2020Jun,Travaglino2023Jan,Ummarino2024Nov}. Our work subsequently proposes a new avenue to
enhance or suppress superconductivity 
by engineering the TLS frequency~\cite{Lisenfeld2023Jan,Pappas2024Aug,Chen2024Sep,Chen2025Mar} in such a way as to optimize its contribution to the bulk phonon spectral weight.
\\
\indent {\it Enhancement and suppression of superconducting $T_c$ from two-level systems}– Expressions for $T_c$ which take into account the underlying form of the phonon spectrum were put forth by McMillan~\cite{McMillan1968Mar}, Allen and Dynes~\cite{Allen1975Aug} and, most rigorously, Leavens and Carbotte~\cite{Leavens1974Jan,Mitrovic1984Jan,Leavens2011Feb,RevModPhys.62.1027}, with the latter deriving $T_c =1.13\omega_{\ln}\exp\left[-(1+\lambda)/(\lambda-\mu^*)\right]$ from the self-consistent Eliashberg equations~\cite{Eliashberg1960,Eliashberg1961,RevModPhys.62.1027,Marsiglio2020Jun}. Here, $\mu^*$ is the Coulomb pseudopotential, $\lambda$ is the electron-phonon coupling strength, and $\omega_{\ln}$ is the log average phonon frequency~\cite{RevModPhys.62.1027}. Note that, in general, there is an anti-correlation between $\lambda$ and $\omega_{\ln}$: namely, that a larger intensity of low-frequency phonons will result in an enhanced $\lambda$ and a suppressed $\omega_{\ln}$, while a larger intensity of higher-frequency phonons results in the converse~\cite{Balatsky2006Sep}. This results in a non-trivial relationship between $T_c$ and the underlying phonon spectrum, which may be further complicated by the varying influence of different phonon frequencies~\cite{Bergmann1973Aug}.

We will model the influence of TLSs on $\lambda$, $\omega_{\ln}$, and, ultimately, $T_c$ by adding a new term to the electron-phonon spectral function; namely, ${ \alpha^2 F(\nu)}=\alpha^2F^{(0)}(\nu)+n_{\textrm{TLS}}\alpha^2F_{\textrm{TLS}}(\nu)$, where $n_{\textrm{TLS}}$ is the areal density of TLSs. The full electron-boson coupling $\lambda$ and log-average boson frequency $\omega_{\ln}$ can be written in terms of the total $\alpha^2 F(\nu)$ as 

\begin{subequations}
\begin{equation}
    \lambda \equiv 2\int_0^\infty d\nu\,\dfrac{\alpha^2F(\nu)}{\nu}= \lambda^{(0)}+n_{TLS}\lambda_{TLS}
    \label{Eqn1a}
\end{equation}
\begin{align}
    \omega_{\ln}&\equiv \omega^{(0)}\exp\bigg[\dfrac{2}{\lambda}\int_0^\infty d\nu\,\dfrac{\alpha^2F(\nu)}{\nu} \log(\nu/\omega^{(0)})\bigg]
\end{align}
\end{subequations}
where $\omega^{(0)}$ is the characteristic phonon frequency of the material in the absence of TLS defects. In our notation, the $(0)$ superscript always denotes a quantity in the absence of TLSs. The areal TLS density $n_{TLS}$ is roughly the number of TLSs per unit cell~\cite{Wang2025Feb}.

\begin{figure}[t!]
    \centering
    
        \includegraphics[width=1\columnwidth]{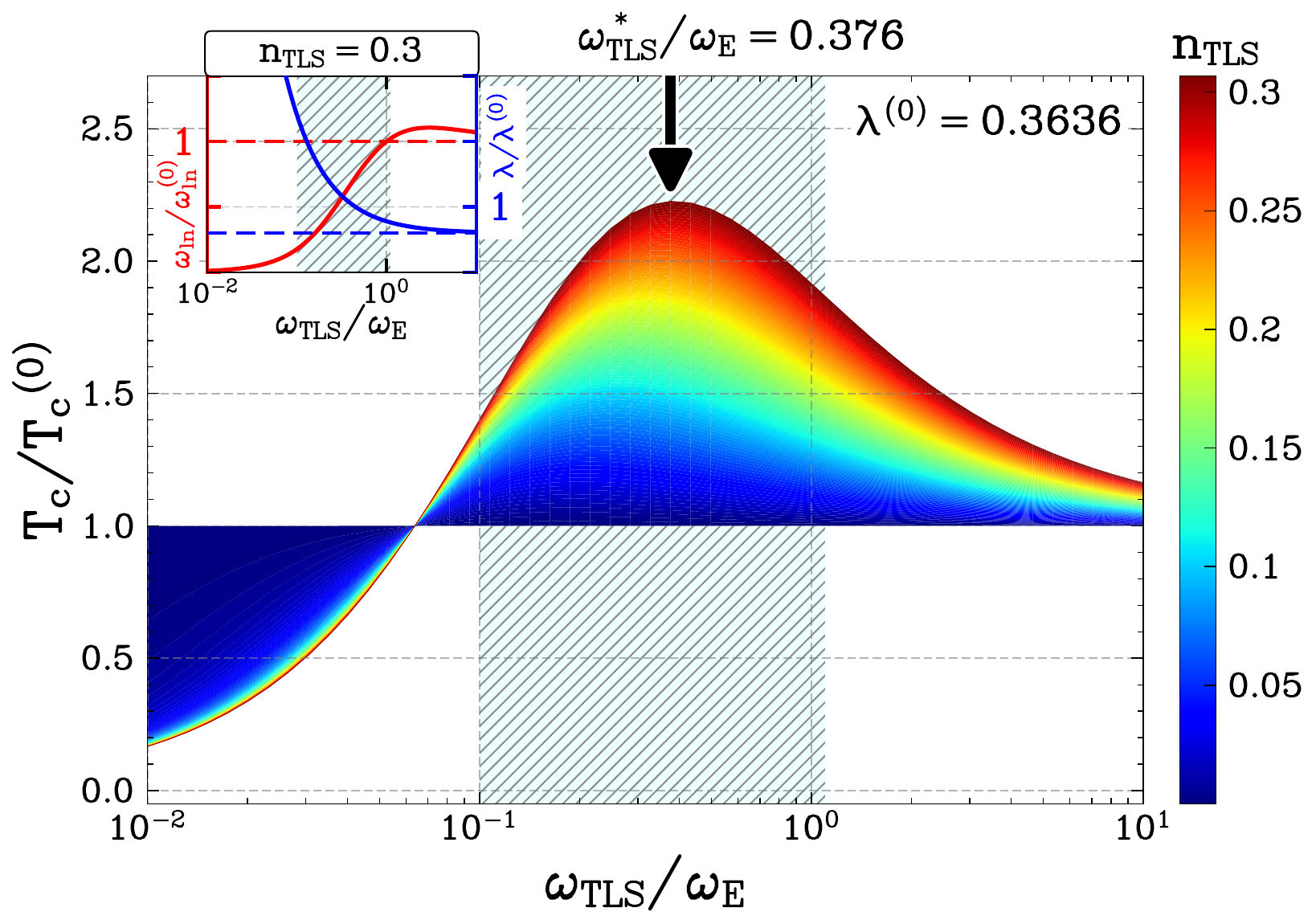}
        \label{fig:sub1}
    \caption{\small 
    Change in $T_c/T_c^{(0)}$ versus TLS (vibron) frequency $\omega_{TLS}$ normalized by Einstein frequency $\omega_E$ and with $\lambda=0.3636$ held fixed. An Einstein boson model is used for both the bulk phonon and vibron, and $T_c$ is found via the Leavens-Carbotte formula~\cite{Leavens2011Feb}. The hatched cyan region corresponds to the frequency regime of amplified $T_c$, which becomes more pronounced for weaker coupling. The black arrow marks the maximum $T_c$, located at $\omega_{TLS}^*/\omega_E\approx 0.38$. Inset: Comparison of $\omega_{\ln}/\omega_{\ln}^{(0)}$ (red) and $\lambda/\lambda^{(0)}$ (blue) as $\omega_{TLS}$ changes for $n_{TLS}=0.3$.}
    \label{Fig2}
\end{figure}

    

\begin{figure}[htbp]
    \centering
    \begin{subfigure}[b]{0.45\textwidth}
        \centering
        \includegraphics[width=\textwidth]{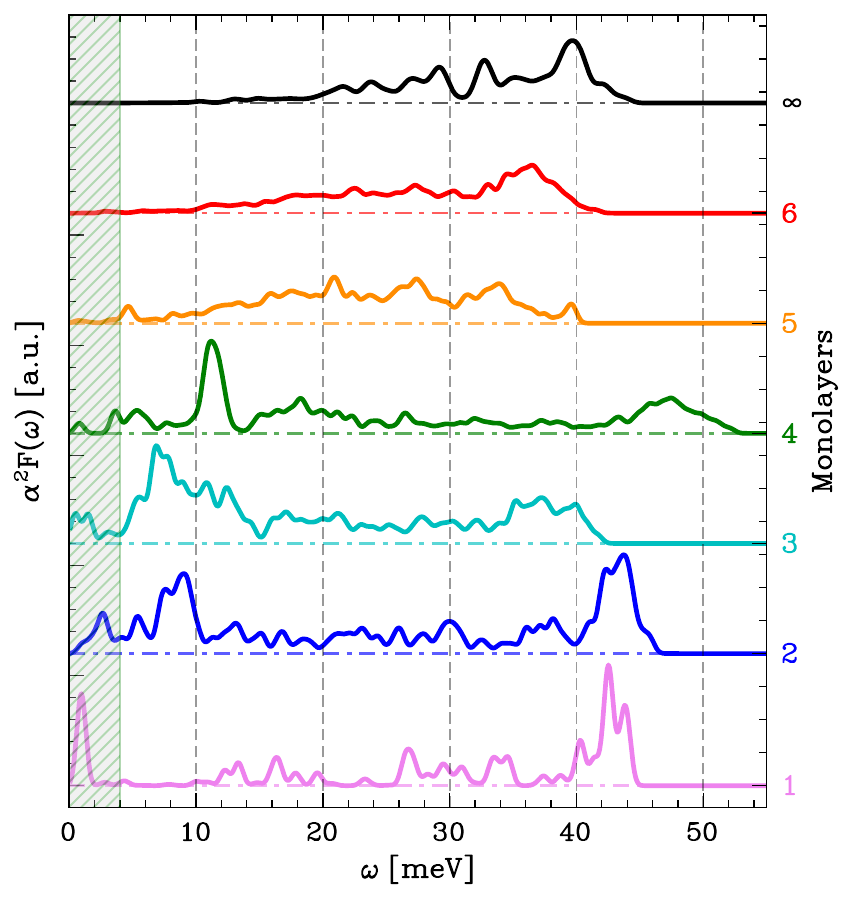}
        \caption{\small $\alpha^2F(\nu)$ for bulk and multilayered systems}
        \label{AbInitio}
    \end{subfigure}
    \hfill
    \begin{subfigure}[b]{0.45\textwidth}
        \centering
        \hspace{-7mm}\includegraphics[width=0.95\textwidth]{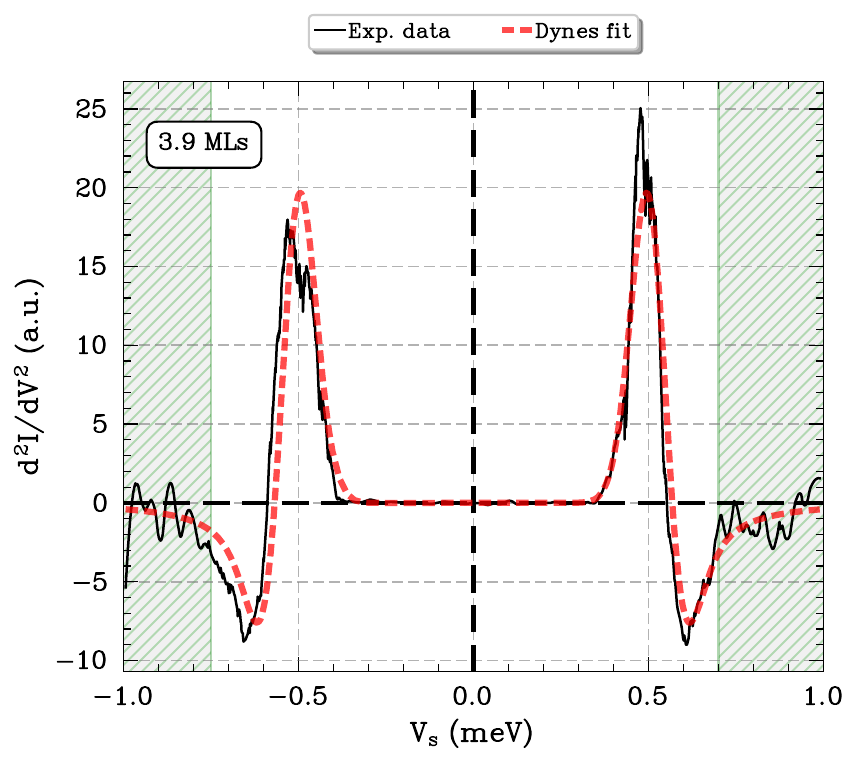}
        \caption{\small $d^2I/dV^2$ for 3.9 monolayer Al}
        \label{IETS}
    \end{subfigure}
    \caption{\small(a) Electron-phonon spectral function in bulk (top) and multi-layered Al found via first principles. In the monolayer limit, more spectral weight is pushed to lower frequencies. The green hatched region corresponds to the regime potentially populated by TLSs ($\omega<4$ meV). (b) Second derivative of the experimental I-V data taken on an epitaxially-grown thin film of Al $3.9$ monolayers thick~\cite{vanWeerdenburg2023Mar} (black) overlain with the Gaussian-broadened Dynes fit (red-dashed). The voltage regime characterized by localised dips in $d^2I/dV^2$ is highlighted in green hatch.  
    \vspace{-3mm}
    }
\end{figure}

While both $\alpha^2 F_{TLS}(\nu)$ and $\alpha^2 F^{(0)}(\nu)$ may have a more complicated structure, a useful first approximation is to assume that both the phonon and TLS contributions to the spectral weight are that of local (Einstein) oscillators; i.e., $\alpha^2F(\nu)=N(0)g^2\delta(\nu-\omega')$ where $N(0)$ is the electronic DoS at the Fermi level, $g$ is the electron-phonon coupling matrix element, and $\omega'=\omega_{\ln}=\omega_E$ ($\omega_{TLS}$) for phonons (TLSs). A plot of $T_c/T_c^{(0)}$ vs. $\omega_{\textrm{TLS}}/\omega_E$ for $n_{TLS}=0.3$ is shown in Fig.~\ref{Fig2}, with $\omega_{\ln}/\omega_{\ln}^{(0)}$ and $\lambda/\lambda^{(0)}$ plotted in the inset and $\lambda^{(0)}=0.3636$ (i.e., the bulk electron-phonon coupling we find from first principles). We take the Leavens-Carbotte (LC) analytical formula for $T_c$, as it avoids bulk-specific fitting parameters inherent to the Allen-Dynes theory while still capturing the salient features produced by modifying $\alpha^2 F(\omega)$. As $\mu^*$ is typically much smaller than $\lambda^{(0)}$, it is ignored for now~\cite{Lee1995Jul,Leavens2011Feb,Simonato2023Aug}.

From Fig.~\ref{Fig2}, we see that $T_c$ is suppressed in the limit of small $\omega_{TLS}/\omega_E$ due to the low-frequency vibrons driving down the log average phonon frequency. Note however that both $\omega_{\ln}/\omega_{\ln}^{(0)}$ and $\lambda/\lambda^{(0)}$ remains appreciably large for $\omega_{TLS}/\omega_E>0.1$. This results in a "sweet spot" within the window of $0.1<\omega_{\textrm{TLS}}/\omega_E<1$ (hatched cyan region of Fig.~\ref{Fig2}) where the influence of $\alpha^2 F_{TLS}(\nu)$ increases $\lambda$ but doesn't significantly reduce $\omega_{\ln}$, thereby resulting in an {\it increase in the superconducting critical temperature}. In the limit of $n_{TLS}\rightarrow 0$, we can see from Fig.~\ref{Fig2} that $T_c/T_c^{(0)}\rightarrow 1$. In the limit of $\omega_{TLS}\rightarrow 0$, vibron modes with $\omega_{TLS}\ll 2\pi T_c$ will have reduced influence on $T_c$, and the Leavens-Carbotte formula breaks down~\cite{Leavens2011Feb,Bergmann1973Aug}. For Al, this occurs when $\omega_{TLS}\ll 0.1$ meV (or equivalently $\ll24$ GHz), and thus we limit $\omega_{TLS}/\omega_E>10^{-2}$ in the Einstein boson model. As we will see below, both the suppression and enhancement of superconductivity is not restricted to an Einstein approximation of the spectral function.

\indent {\it Estimation of $T_c$ from the thin film electron-phonon spectral function--} In order to move beyond the Einstein model and make predictions for more realistic systems, we require more detailed structure for both $\alpha^2F^{(0)}(\nu)$ and $\alpha^2F_{TLS}(\nu)$. For the former, we perform first principles simulations on bulk and N-layer Al, with $N=1-6$. The lattice parameters and atomic positions of all structures are relaxed until all forces are below $1.0 \times 10^{-5} \textrm{Ry}/\textrm{Bohr}$. Electron-phonon properties are computed on a coarse grid using density functional perturbation theory (DFPT) in the Quantum Espresso software package~\cite{Perdew1996,Giannozzi2020Apr} and fully relativistic, norm-conserving pseudopotentials~\cite{kresse1999VASP,van2018pseudodojo}. We extract $\alpha^2F^{(0)}(\nu)$ using the EPW package~\cite{PhysRevB.13.5188,Noffsinger2010Dec,lee2023electron}, interpolating the electron-phonon matrix elements to dense grids in the Wannier basis~\cite{Pizzi2020}. Results are shown in Fig.~\ref{AbInitio}. Upon decreasing the number of monolayers, a shift of spectral weight towards both the low and high frequency regimes is seen, which may be attributed to the simultaneous softening~\cite{bozyigit2016soft,Yeh2023Nov} and hardening~\cite{Luo2013Aug} of surface phonons in the single monolayer limit.

Note that the first principles calculation does not take into account the presence of TLSs. If we assume the TLSs are clustered about a small frequency interval, their net effect on the electron-phonon spectral weight will be a narrow-width, low-frequency Lorentzian~\cite{Marsiglio2020Jun} addition to $\alpha^2 F^{(0)}(\nu)$. The free parameters that go into the determination of $\alpha^2 F_{TLS}(\nu)$ are thus the width $\epsilon_{TLS}$ of this Lorentzian, the average TLS frequency $\omega_{TLS}$, and the density of TLSs $n_{TLS}$. Based upon recent experiments~\cite{Martinis2005Nov,Schreier2008May,Queen2013Mar,Gunnarsson2013Jul,Muller2019Oct,Bejanin2021Sep}, the areal TLS density may be estimated to be up to $n_{TLS}\approx 0.3$ (the value used earlier for Fig.~\ref{Fig2}). A suitable range for TLS frequencies may be inferred by turning to recent STM measurements on ultra-thin epitaxial Al films on Si(111)~\cite{vanWeerdenburg2023Mar} (see Fig.~\ref{IETS}). As the bias voltage increases to $(\Delta_1+\hbar \omega_{TLS})/e$ (where $\Delta_1$ is the gap edge~\cite{Fibich1965Apr}), the resonant exchange of vibrons should enhance the electron-pair binding energy, leading to a local reduction in the effective tunnelling DoS and subsequent dips in $d^2I/dV^2$ about the TLS frequencies~\cite{Giaever1962May,Schrieffer1963Apr,Scalapino1966Aug,Jaklevic1966Nov,RevModPhys.62.1027,Schackert2014}. Such deviations of 
the measured $d^2I/dV^2$~\cite{vanWeerdenburg2023Mar}
from the Gaussian convolution of a Dynes-type fit~\cite{Dynes1978Nov,Oda1996Feb,Howald2001Aug} are observed over a broad frequency range extending upwards of $0.1$ THz above the superconducting gap. This frequency range is in agreement with numerical~\cite{Tyner2025Jul} and experimental~\cite{Hahnle2021Jul,Shan2024Sep,Buijtendorp2025Jan} estimates of $\omega_{TLS}$ in related materials, thus supporting our underlying hypothesis. We will therefore consider values of $\omega_{TLS}$ in the frequency range of $0.15-4\,\,\textrm{meV}$ ( around $0.05-1.0$ THz).

\begin{figure}[htbp]
    \centering
    \begin{subfigure}[b]{0.45\textwidth}
        \centering
        \includegraphics[width=\textwidth]{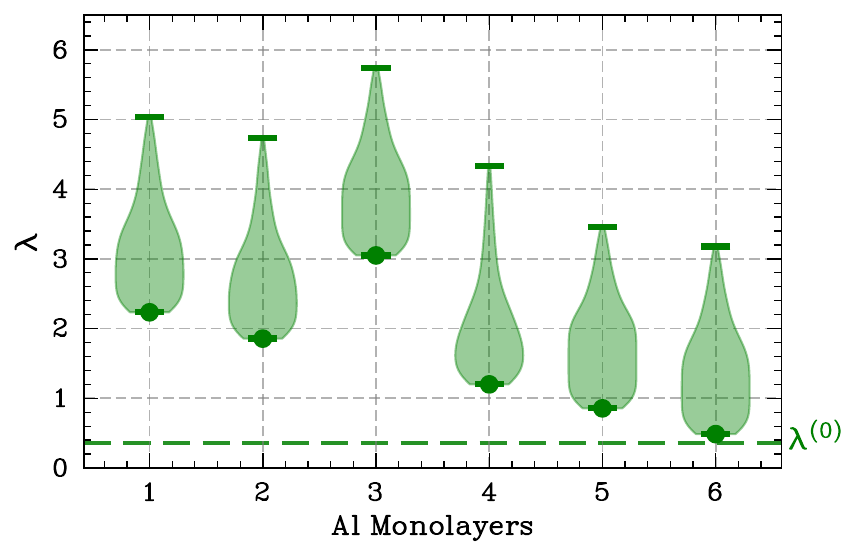}
        \caption{\small Electron-phonon coupling $\lambda$ versus monolayers}
        \label{lambVSmonolayers}
    \end{subfigure}
    \hfill
    \begin{subfigure}[b]{0.45\textwidth}
        \centering
\hspace{-1.8mm}\includegraphics[width=1.02\textwidth]{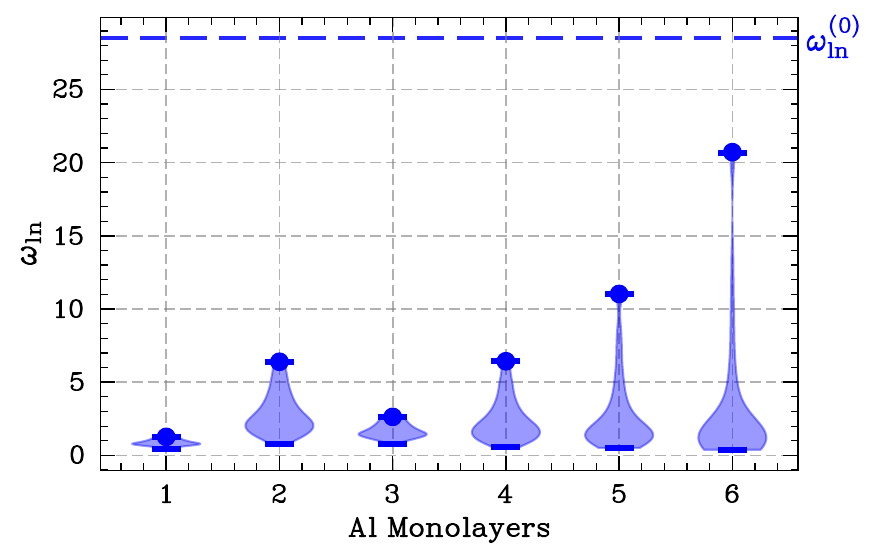}
        \caption{\small Log average frequency $\omega_{\ln}$ versus monolayers}
        \label{omegalnVSmonolayers}
    \end{subfigure}
    \hfill
    \begin{subfigure}[b]{0.45\textwidth}
        \centering
        \hspace{-10.5mm}\includegraphics[width=0.99\textwidth]{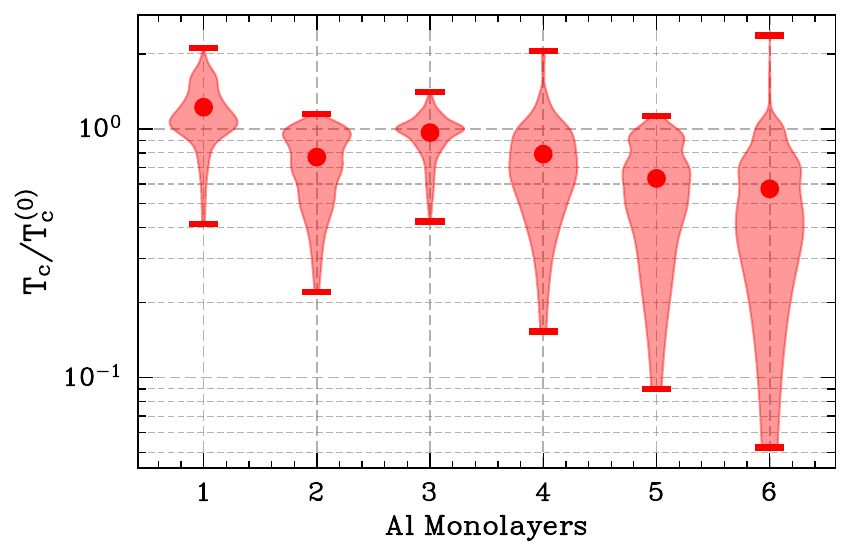}
        \caption{\small Critical temperature $T_c/T_c^{(0)}$ versus monolayers}
        \label{TcVSmonolayers}
    \end{subfigure}
    \label{Fig4}
    \caption{\small TLS-induced modification of (a) the electron-phonon coupling, (b) the log average phonon frequency, and (c) the ratio of $T_c$ (with TLSs) to $T_c^{(0)}$ (without TLSs) in layered Al. Values for $T_c$ are found via the Leavens-Carbotte eqn.~\cite{Leavens1974Jan} with $\alpha^2F(\nu)$ found via first principles. The coloured error bars correspond to variations of $T_c$ upon changing the TLS density $n_{TLS}$ and the TLS frequency $\omega_{TLS}$, with the coloured shadow representing the spread of data. We use flat top distribution of $\omega_{TLS}$ in the range of 0.15 to 4 mev and $n_{TLS}$ in the range 0.01 to 0.3 in generating these plots. 
    As we approach the single monolayer limit, $\lambda$ grows while $\omega_{\ln}$ shrinks. The net result is larger variation of $T_c$ in the single monolayer limit. As discussed in the main text, the suppression of $T_c/T_c^{(0)}$ in the analytic solution is an artifact of the constant gap approximation taken by Leavens and Carbotte~\cite{Bergmann1973Aug}. 
    }
\end{figure}

The values of $\lambda$, $\omega_{\ln}$, and $T_c$ for multi-layered Al is shown in Figs.~\ref{lambVSmonolayers}, \ref{omegalnVSmonolayers}, and \ref{TcVSmonolayers}, respectively. For each, we have utilized the Leavens-Carbotte equation with a value of the Coulomb pseudopotential $\mu^*\sim0.091$ given by first principles~\cite{Lee1995Jul} and the total $\alpha^2 F(\nu)$ including the TLS and bulk contribution found via our own first principles calculation.
%
%
While severe suppression of $T_c$ occurs for a larger number of monolayers, noticeable amplification of the critical temperature is seen in a single monolayer of Al (i.e., aluminene~\cite{Kamal2015Aug}). The latter can be understood as a consequence of phonon softening in the thin film, which pushes significant spectral weight to the GHz range. As suggested in the simple Einstein model, when the phonon frequency is reduced to around $10\times\omega_{TLS}$, $\lambda$ is amplified while $\omega_{\ln}$ remains appreciably large, leading to a boost in $T_c$. For a larger number of monolayers, TLSs typically have a net effect of lowering $T_c$ due to a higher phonon frequency relative to $\omega_{TLS}$ (as seen in Fig.~\ref{AbInitio}). This results in a severely suppressed $\omega_{\ln}$ and, consequently, a suppressed $T_c$, regardless of the "boost" in $\lambda$ induced by the TLSs. Finally, note that from the Einstein phonon prediction we may write an approximate expression for the average TLS frequency $\omega_{TLS}^*$ that maximises $T_c$:
\begin{align}
    \omega_{TLS}^{*}=\alpha \omega_{E}^{\textrm{eff}},\quad \alpha\lessapprox  0.38
    \label{omegastar}
\end{align}
The unitless number $\alpha$ is a function of $n_{TLS}$ and $\omega_{E}^{\textrm{eff}}$ is an effective Einstein frequency that roughly corresponds to the lowest salient peak(s) in $\alpha^2F(\omega)$. Increasing the TLS frequency $\omega_{TLS}<\omega_{TLS}^*$ will result in an increase in $T_c$, while increasing $\omega_{TLS}>\omega_{TLS}^*$ will result in a decrease in $T_c$. 
While $\omega_{E}^{\textrm{eff}}$ is highly dependent on the precise form of the electron-phonon spectral function, from our ab initio calculation (see Fig.~\ref{AbInitio}) and Eqn.~\ref{omegastar}, we can estimate $\omega_{TLS}^*$ around $0.33-1$ THz for $n_{TLS}=0.3$ and $1-4$ monolayers of Al.

\begin{figure}[t!]
    \centering
    \begin{subfigure}[b]{0.45\textwidth}
        \centering
        \hspace{-10.5mm}\includegraphics[width=1\textwidth]{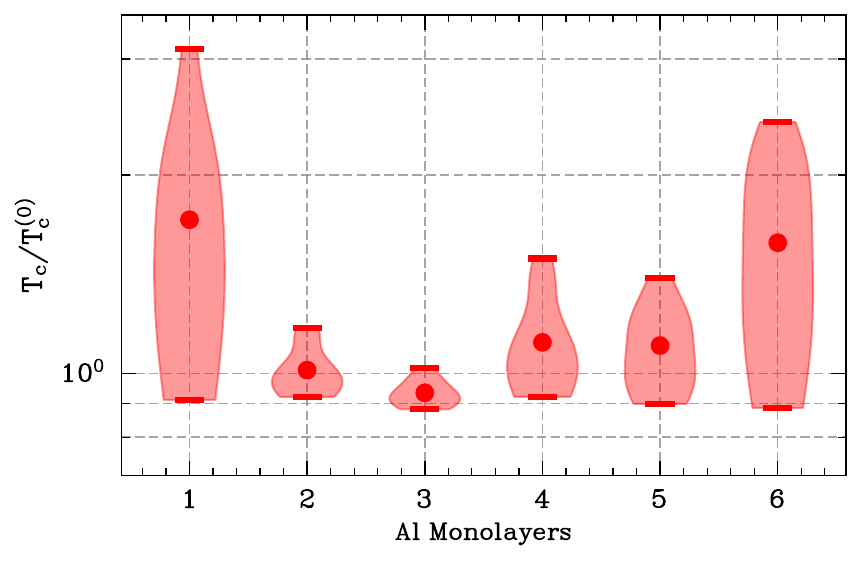}
        \caption{\small Critical temperature $T_c/T_c^{(0)}$ (found numerically) versus monolayers}
        \label{Fig5a}
    \end{subfigure}
     \begin{subfigure}[b]{0.45\textwidth}
        \centering
        \hspace{-10.5mm}\includegraphics[width=1\textwidth]{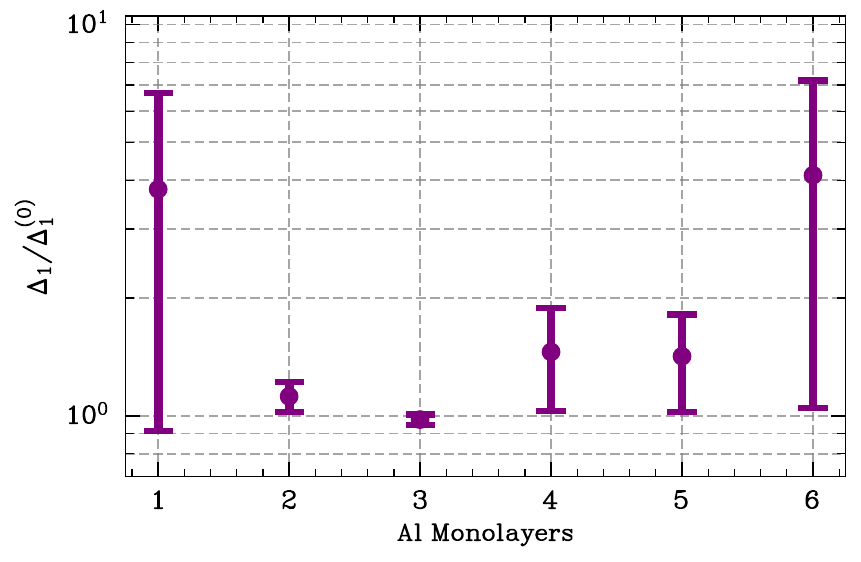}
        \caption{\small Gap edge $\Delta_1/\Delta_1^{(0)}$ (found numerically) versus monolayers}
        \label{Fig5b}
    \end{subfigure}
    \label{Fig5}
    \caption{\small TLS-induced modification of (a) the critical temperature $T_c$ and (b) the gap edge $\Delta_1$. Both $T_c$ and the gap edge are found by numerically solving the Eliashberg equations. Error bars correspond to changes in the respective quantities upon introducing TLSs, and the dots correspond to average values. TLS parameters are those found in the previous section that led to maximised enhancement and suppression of $T_c/T_c^{(0)}$ when calculated numerically. As compared to the analytical calculation, TLSs tend to enhance (as opposed to suppress) the critical temperature. The change in the local gap edge $\Delta_1/\Delta_1^{(0)}$ follows a similar trend as the critical temperature, except with a more pronounced enhancement. Larger relative enhancement of the gap edge is expected, as the gap-to-$T_c$ ratio is expected to be locally enhanced from the amplified $\lambda$ induced by the TLSs~\cite{Carbotte1987,Carbotte1986May}.
    }
\end{figure}

\indent {\it Numerical calculation of $T_c$ and the gap edge}– 
Up till this point, we have utilized the Leavens-Carbotte formula for extracting TLS-induced modification of $T_c$.
To more accurately find $T_c$ and the gap, we will now turn to numerically solving the Eliashberg equations.
The results of the numerical calculation are shown in Figs.~\ref{Fig5a} and \ref{Fig5b}. In Fig.~\ref{Fig5a}, $T_c/T_c^{(0)}$ is found by linearizing the Eliashberg equations on the Matsubara frequency axis and solving the resulting eigenvalue equation via Von Mises iteration~\cite{RevModPhys.62.1027,Owen1971Oct,Kresin1987Jun,Tajik2018,Marsiglio2020Jun}. TLS parameters are varied within $\omega_{TLS}\in [0.1,4.0]$ meV (slighly lower than experimentally observed), with $n_{TLS}=0.3$ held fixed to obtain the maximal effects of a finite TLS population on $T_c$. We see that the Leavens-Carbotte estimation is most accurate for predicting the maximal $T_c/T_c^{(0)}$ values, with good approximate agreement for $6-2$ Al monolayers. Slight deviations from the analytic prediction of the maximum $T_c/T_c^{(0)}$ may be attributed to the fine-structure of $\alpha^2 F(\omega)$ (see Supplemental Material for why this is especially important for $3$ Al monolayers). Most notably, however, we observe that the numerical values of $T_c/T_c^{(0)}$ do not exhibit the severe suppression predicted in the analytic solution. As mentioned before, the reduction of $T_c/T_c^{(0)}$ in the presence of low-frequency TLSs is an artifact of the Leavens-Carbotte (and, likewise, the Allen-Dynes) formulation of $T_c$. As low-frequency modes will have reduced influence on $T_c$~\cite{Bergmann1973Aug}, such deviation from the analytic solution in the presence of low-frequency vibrons is expected, and is in agreement with Anderson's theorem~\cite{Anderson1959Sep}.

In addition to the critical temperature, we also obtain the $T=0$ local gap function by numerically solving the Eliashberg equations on the Matsubara frequency axis to obtain $\Delta(i\omega_n)$, and then analytically continuing to real frequencies~\cite{Marsiglio1988Apr} to get the real axis gap function $\Delta(\omega)$. From the gap function, we can then extract the local gap edge $\Delta_1\equiv \Re(\Delta(\omega=\Delta_1))$~\cite{Fibich1965Apr}. The smallest TLS frequency is taken to be $\omega_{TLS}\approx 0.15$, while the maximum TLS frequency is taken to be the value that maximised the value of $T_c/T_c^{(0)}$. Much like the critical temperature, we find that the local gap edge is enhanced by virtue of including TLSs, however in the case of the latter we find a greater relative enhancement than seen in the former. This is expected, as the ratio of $\Delta_1$ over $T_c$ is expected to range from $1.76$ in the BCS limit~\cite{Bardeen1957Dec} to $\sim 6-6.5$ in the very strongly coupled limit~\cite{RevModPhys.62.1027,Carbotte1987}. As such, an increase of $\Delta_1/\Delta_1^{(0)}\sim 7$ is not unreasonable if we assume $T_c/T_c^{(0)}\sim 2.5-3$.

\indent {\it Conclusions}– 
By treating two-level systems as localised soft modes, we have shown that TLSs add to the spectral weight of bulk phonons, directly influencing the critical temperature and gap of the superconductor. As quantum confinement effects alone cannot explain the large boost in $T_c$ observed in epitaxial Al films grown on Si$(111)$~\cite{vanWeerdenburg2023Mar, Ummarino2024Nov}, extreme enhancement of the critical temperature as a function of film thickness may be a result of dimensional reduction in tandem with a finite TLS population. As a whole, we emphasise that our work opens new avenues for TLS-based material engineering and phase stabilisation, and consequently shifts the focus of TLSs from a ubiquitous source of decoherence to an abundant resource for enhancing superconductivity in the two-dimensional limit.
\\\\
\indent {\it Acknowledgements}– We thank Rufus Boyack, Jared Cole, Charlie Marcus, Vincent P. Michal, Dave Pappas,  Mark Kamper Svendsen, Thue Christian Thann and Xiqiao Wang for valuable discussions. AB and JH  are supported by the Novo Nordisk Foundation, Grant number NNF22SA0081175, NNF Quantum Computing Programme. A.T. acknowledges the hospitality of the Novo Nordisk Foundation Quantum Computing Programme.  AB is also  supported by the OSR program FA9550-25-1-0103.  SPA was supported by the University of Connecticut, RB was supported by Dartmouth College, 
and PK was supported by the NNF Quantum Computing Programme. A portion of the computations were enabled by resources provided by the National Academic Infrastructure for Supercomputing in Sweden (NAISS), partially funded by the Swedish Research Council through grant agreement no. 2022-06725.

\bibliography{main}{}

\begin{thebibliography}{49}%
\makeatletter
\providecommand \@ifxundefined [1]{%
 \@ifx{#1\undefined}
}%
\providecommand \@ifnum [1]{%
 \ifnum #1\expandafter \@firstoftwo
 \else \expandafter \@secondoftwo
 \fi
}%
\providecommand \@ifx [1]{%
 \ifx #1\expandafter \@firstoftwo
 \else \expandafter \@secondoftwo
 \fi
}%
\providecommand \natexlab [1]{#1}%
\providecommand \enquote  [1]{``#1''}%
\providecommand \bibnamefont  [1]{#1}%
\providecommand \bibfnamefont [1]{#1}%
\providecommand \citenamefont [1]{#1}%
\providecommand \href@noop [0]{\@secondoftwo}%
\providecommand \href [0]{\begingroup \@sanitize@url \@href}%
\providecommand \@href[1]{\@@startlink{#1}\@@href}%
\providecommand \@@href[1]{\endgroup#1\@@endlink}%
\providecommand \@sanitize@url [0]{\catcode `\\12\catcode `\$12\catcode
  `\&12\catcode `\#12\catcode `\^12\catcode `\_12\catcode `\%12\relax}%
\providecommand \@@startlink[1]{}%
\providecommand \@@endlink[0]{}%
\providecommand \url  [0]{\begingroup\@sanitize@url \@url }%
\providecommand \@url [1]{\endgroup\@href {#1}{\urlprefix }}%
\providecommand \urlprefix  [0]{URL }%
\providecommand \Eprint [0]{\href }%
\providecommand \doibase [0]{http://dx.doi.org/}%
\providecommand \selectlanguage [0]{\@gobble}%
\providecommand \bibinfo  [0]{\@secondoftwo}%
\providecommand \bibfield  [0]{\@secondoftwo}%
\providecommand \translation [1]{[#1]}%
\providecommand \BibitemOpen [0]{}%
\providecommand \bibitemStop [0]{}%
\providecommand \bibitemNoStop [0]{.\EOS\space}%
\providecommand \EOS [0]{\spacefactor3000\relax}%
\providecommand \BibitemShut  [1]{\csname bibitem#1\endcsname}%
\let\auto@bib@innerbib\@empty
\bibitem [{\citenamefont {Eliashberg}(1960)}]{Eliashberg1960}%
  \BibitemOpen
  \bibfield  {author} {\bibinfo {author} {\bibfnamefont {G.~M.}\ \bibnamefont
  {Eliashberg}},\ }\href@noop {} {\bibfield  {journal} {\bibinfo  {journal}
  {Sov. Phys. JETP}\ }\textbf {\bibinfo {volume} {11}},\ \bibinfo {pages} {696}
  (\bibinfo {year} {1960})}\BibitemShut {NoStop}%
\bibitem [{\citenamefont {Eliashberg}(1961)}]{Eliashberg1961}%
  \BibitemOpen
  \bibfield  {author} {\bibinfo {author} {\bibfnamefont {G.~M.}\ \bibnamefont
  {Eliashberg}},\ }\href@noop {} {\bibfield  {journal} {\bibinfo  {journal}
  {Sov. Phys. JETP}\ }\textbf {\bibinfo {volume} {12}},\ \bibinfo {pages}
  {1000} (\bibinfo {year} {1961})}\BibitemShut {NoStop}%
\bibitem [{\citenamefont {Rickayzen}(1965)}]{RickayzenBook}%
  \BibitemOpen
  \bibfield  {author} {\bibinfo {author} {\bibfnamefont {G.}~\bibnamefont
  {Rickayzen}},\ }\href@noop {} {\emph {\bibinfo {title} {Theory of
  Superconductivity}}}\ (\bibinfo  {publisher} {John Wiley and Sons Inc.},\
  \bibinfo {address} {New York},\ \bibinfo {year} {1965})\BibitemShut {NoStop}%
\bibitem [{\citenamefont {Carbotte}(1990)}]{RevModPhys.62.1027}%
  \BibitemOpen
  \bibfield  {author} {\bibinfo {author} {\bibfnamefont {J.~P.}\ \bibnamefont
  {Carbotte}},\ }\href {\doibase 10.1103/RevModPhys.62.1027} {\bibfield
  {journal} {\bibinfo  {journal} {Rev. Mod. Phys.}\ }\textbf {\bibinfo {volume}
  {62}},\ \bibinfo {pages} {1027} (\bibinfo {year} {1990})}\BibitemShut
  {NoStop}%
\bibitem [{\citenamefont {Marsiglio}(2020)}]{Marsiglio2020Jun}%
  \BibitemOpen
  \bibfield  {author} {\bibinfo {author} {\bibfnamefont {F.}~\bibnamefont
  {Marsiglio}},\ }\href {\doibase 10.1016/j.aop.2020.168102} {\bibfield
  {journal} {\bibinfo  {journal} {Ann. Phys.}\ }\textbf {\bibinfo {volume}
  {417}},\ \bibinfo {pages} {168102} (\bibinfo {year} {2020})}\BibitemShut
  {NoStop}%
\bibitem [{\citenamefont {Allen}\ and\ \citenamefont
  {Dynes}(1975)}]{Allen1975Aug}%
  \BibitemOpen
  \bibfield  {author} {\bibinfo {author} {\bibfnamefont {P.~B.}\ \bibnamefont
  {Allen}}\ and\ \bibinfo {author} {\bibfnamefont {R.~C.}\ \bibnamefont
  {Dynes}},\ }\href {\doibase 10.1103/PhysRevB.12.905} {\bibfield  {journal}
  {\bibinfo  {journal} {Phys. Rev. B}\ }\textbf {\bibinfo {volume} {12}},\
  \bibinfo {pages} {905} (\bibinfo {year} {1975})}\BibitemShut {NoStop}%
\bibitem [{\citenamefont {Leavens}\ and\ \citenamefont
  {Carbotte}(1971)}]{Leavens2011Feb}%
  \BibitemOpen
  \bibfield  {author} {\bibinfo {author} {\bibfnamefont {C.~R.}\ \bibnamefont
  {Leavens}}\ and\ \bibinfo {author} {\bibfnamefont {J.~P.}\ \bibnamefont
  {Carbotte}},\ }\href {\doibase 10.1139/p71-088} {\bibfield  {journal}
  {\bibinfo  {journal} {Can. J. Phys.}\ } (\bibinfo {year} {1971}),\
  10.1139/p71-088}\BibitemShut {NoStop}%
\bibitem [{\citenamefont {Schrieffer}\ \emph {et~al.}(1963)\citenamefont
  {Schrieffer}, \citenamefont {Scalapino},\ and\ \citenamefont
  {Wilkins}}]{Schrieffer1963Apr}%
  \BibitemOpen
  \bibfield  {author} {\bibinfo {author} {\bibfnamefont {J.~R.}\ \bibnamefont
  {Schrieffer}}, \bibinfo {author} {\bibfnamefont {D.~J.}\ \bibnamefont
  {Scalapino}}, \ and\ \bibinfo {author} {\bibfnamefont {J.~W.}\ \bibnamefont
  {Wilkins}},\ }\href {\doibase 10.1103/PhysRevLett.10.336} {\bibfield
  {journal} {\bibinfo  {journal} {Phys. Rev. Lett.}\ }\textbf {\bibinfo
  {volume} {10}},\ \bibinfo {pages} {336} (\bibinfo {year} {1963})}\BibitemShut
  {NoStop}%
\bibitem [{\citenamefont {Scalapino}(1969)}]{Parks1}%
  \BibitemOpen
  \bibfield  {author} {\bibinfo {author} {\bibfnamefont {D.~J.}\ \bibnamefont
  {Scalapino}},\ }in\ \href@noop {} {\emph {\bibinfo {booktitle}
  {Superconductivity: Part 1 (In Two Parts)}}},\ \bibinfo {editor} {edited by\
  \bibinfo {editor} {\bibfnamefont {R.}~\bibnamefont {Parks}}}\ (\bibinfo
  {publisher} {Marcel Dekker Inc., New York},\ \bibinfo {year} {1969})\ pp.\
  \bibinfo {pages} {449--560}\BibitemShut {NoStop}%
\bibitem [{\citenamefont {Scalapino}\ \emph {et~al.}(1965)\citenamefont
  {Scalapino}, \citenamefont {Wada},\ and\ \citenamefont
  {Swihart}}]{Scalapino1965Jan}%
  \BibitemOpen
  \bibfield  {author} {\bibinfo {author} {\bibfnamefont {D.~J.}\ \bibnamefont
  {Scalapino}}, \bibinfo {author} {\bibfnamefont {Y.}~\bibnamefont {Wada}}, \
  and\ \bibinfo {author} {\bibfnamefont {J.~C.}\ \bibnamefont {Swihart}},\
  }\href {\doibase 10.1103/PhysRevLett.14.102} {\bibfield  {journal} {\bibinfo
  {journal} {Phys. Rev. Lett.}\ }\textbf {\bibinfo {volume} {14}},\ \bibinfo
  {pages} {102} (\bibinfo {year} {1965})}\BibitemShut {NoStop}%
\bibitem [{\citenamefont {Marsiglio}\ \emph {et~al.}(1988)\citenamefont
  {Marsiglio}, \citenamefont {Schossmann},\ and\ \citenamefont
  {Carbotte}}]{Marsiglio1988Apr}%
  \BibitemOpen
  \bibfield  {author} {\bibinfo {author} {\bibfnamefont {F.}~\bibnamefont
  {Marsiglio}}, \bibinfo {author} {\bibfnamefont {M.}~\bibnamefont
  {Schossmann}}, \ and\ \bibinfo {author} {\bibfnamefont {J.~P.}\ \bibnamefont
  {Carbotte}},\ }\href {\doibase 10.1103/PhysRevB.37.4965} {\bibfield
  {journal} {\bibinfo  {journal} {Phys. Rev. B}\ }\textbf {\bibinfo {volume}
  {37}},\ \bibinfo {pages} {4965} (\bibinfo {year} {1988})}\BibitemShut
  {NoStop}%
\bibitem [{\citenamefont {Leavens}\ and\ \citenamefont
  {Carbotte}(1974)}]{Leavens1974Jan}%
  \BibitemOpen
  \bibfield  {author} {\bibinfo {author} {\bibfnamefont {C.~R.}\ \bibnamefont
  {Leavens}}\ and\ \bibinfo {author} {\bibfnamefont {J.~P.}\ \bibnamefont
  {Carbotte}},\ }\href {\doibase 10.1007/BF00654817} {\bibfield  {journal}
  {\bibinfo  {journal} {J. Low Temp. Phys.}\ }\textbf {\bibinfo {volume}
  {14}},\ \bibinfo {pages} {195} (\bibinfo {year} {1974})}\BibitemShut
  {NoStop}%
\bibitem [{\citenamefont {Travaglino}\ and\ \citenamefont
  {Zaccone}(2023)}]{Travaglino2023Jan}%
  \BibitemOpen
  \bibfield  {author} {\bibinfo {author} {\bibfnamefont {R.}~\bibnamefont
  {Travaglino}}\ and\ \bibinfo {author} {\bibfnamefont {A.}~\bibnamefont
  {Zaccone}},\ }\href {\doibase 10.1063/5.0132820} {\bibfield  {journal}
  {\bibinfo  {journal} {J. Appl. Phys.}\ }\textbf {\bibinfo {volume} {133}}
  (\bibinfo {year} {2023}),\ 10.1063/5.0132820}\BibitemShut {NoStop}%
\bibitem [{\citenamefont {Ummarino}\ and\ \citenamefont
  {Zaccone}(2024)}]{Ummarino2024Nov}%
  \BibitemOpen
  \bibfield  {author} {\bibinfo {author} {\bibfnamefont {G.~A.}\ \bibnamefont
  {Ummarino}}\ and\ \bibinfo {author} {\bibfnamefont {A.}~\bibnamefont
  {Zaccone}},\ }\href {\doibase 10.1088/1361-648X/ad92ed} {\bibfield  {journal}
  {\bibinfo  {journal} {J. Phys.: Condens. Matter}\ }\textbf {\bibinfo {volume}
  {37}},\ \bibinfo {pages} {065703} (\bibinfo {year} {2024})}\BibitemShut
  {NoStop}%
\bibitem [{\citenamefont {Simonato}\ \emph {et~al.}(2023)\citenamefont
  {Simonato}, \citenamefont {Katsnelson},\ and\ \citenamefont
  {R{\ifmmode\ddot{o}\else\"{o}\fi}sner}}]{Simonato2023Aug}%
  \BibitemOpen
  \bibfield  {author} {\bibinfo {author} {\bibfnamefont {M.}~\bibnamefont
  {Simonato}}, \bibinfo {author} {\bibfnamefont {M.~I.}\ \bibnamefont
  {Katsnelson}}, \ and\ \bibinfo {author} {\bibfnamefont {M.}~\bibnamefont
  {R{\ifmmode\ddot{o}\else\"{o}\fi}sner}},\ }\href {\doibase
  10.1103/PhysRevB.108.064513} {\bibfield  {journal} {\bibinfo  {journal}
  {Phys. Rev. B}\ }\textbf {\bibinfo {volume} {108}},\ \bibinfo {pages}
  {064513} (\bibinfo {year} {2023})}\BibitemShut {NoStop}%
\bibitem [{\citenamefont {Zaccone}\ and\ \citenamefont
  {Fomin}(2024)}]{Zaccone2024Apr}%
  \BibitemOpen
  \bibfield  {author} {\bibinfo {author} {\bibfnamefont {A.}~\bibnamefont
  {Zaccone}}\ and\ \bibinfo {author} {\bibfnamefont {V.~M.}\ \bibnamefont
  {Fomin}},\ }\href {\doibase 10.1103/PhysRevB.109.144520} {\bibfield
  {journal} {\bibinfo  {journal} {Phys. Rev. B}\ }\textbf {\bibinfo {volume}
  {109}},\ \bibinfo {pages} {144520} (\bibinfo {year} {2024})}\BibitemShut
  {NoStop}%
\bibitem [{\citenamefont {Feibelman}\ and\ \citenamefont
  {Hamann}(1984)}]{Feibelman1984Jun}%
  \BibitemOpen
  \bibfield  {author} {\bibinfo {author} {\bibfnamefont {P.~J.}\ \bibnamefont
  {Feibelman}}\ and\ \bibinfo {author} {\bibfnamefont {D.~R.}\ \bibnamefont
  {Hamann}},\ }\href {\doibase 10.1103/PhysRevB.29.6463} {\bibfield  {journal}
  {\bibinfo  {journal} {Phys. Rev. B}\ }\textbf {\bibinfo {volume} {29}},\
  \bibinfo {pages} {6463} (\bibinfo {year} {1984})}\BibitemShut {NoStop}%
\bibitem [{\citenamefont {Nguyen}\ \emph {et~al.}(2019)\citenamefont {Nguyen},
  \citenamefont {Wei},\ and\ \citenamefont {Chou}}]{Nguyen2019May}%
  \BibitemOpen
  \bibfield  {author} {\bibinfo {author} {\bibfnamefont {D.-L.}\ \bibnamefont
  {Nguyen}}, \bibinfo {author} {\bibfnamefont {C.-M.}\ \bibnamefont {Wei}}, \
  and\ \bibinfo {author} {\bibfnamefont {M.-Y.}\ \bibnamefont {Chou}},\ }\href
  {\doibase 10.1103/PhysRevB.99.205401} {\bibfield  {journal} {\bibinfo
  {journal} {Phys. Rev. B}\ }\textbf {\bibinfo {volume} {99}},\ \bibinfo
  {pages} {205401} (\bibinfo {year} {2019})}\BibitemShut {NoStop}%
\bibitem [{\citenamefont {Allen}(2024)}]{Allen2024Jun}%
  \BibitemOpen
  \bibfield  {author} {\bibinfo {author} {\bibfnamefont {P.~B.}\ \bibnamefont
  {Allen}},\ }\href {\doibase 10.48550/arXiv.2406.16197} {\bibfield  {journal}
  {\bibinfo  {journal} {ArXiv}\ } (\bibinfo {year} {2024}),\
  10.48550/arXiv.2406.16197},\ \Eprint {http://arxiv.org/abs/2406.16197}
  {2406.16197} \BibitemShut {NoStop}%
\bibitem [{\citenamefont {M{\ifmmode\ddot{u}\else\"{u}\fi}ller}\ \emph
  {et~al.}(2019)\citenamefont {M{\ifmmode\ddot{u}\else\"{u}\fi}ller},
  \citenamefont {Cole},\ and\ \citenamefont {Lisenfeld}}]{Muller2019Oct}%
  \BibitemOpen
  \bibfield  {author} {\bibinfo {author} {\bibfnamefont {C.}~\bibnamefont
  {M{\ifmmode\ddot{u}\else\"{u}\fi}ller}}, \bibinfo {author} {\bibfnamefont
  {J.~H.}\ \bibnamefont {Cole}}, \ and\ \bibinfo {author} {\bibfnamefont
  {J.}~\bibnamefont {Lisenfeld}},\ }\href {\doibase 10.1088/1361-6633/ab3a7e}
  {\bibfield  {journal} {\bibinfo  {journal} {Rep. Prog. Phys.}\ }\textbf
  {\bibinfo {volume} {82}},\ \bibinfo {pages} {124501} (\bibinfo {year}
  {2019})}\BibitemShut {NoStop}%
\bibitem [{\citenamefont {Martinis}\ \emph {et~al.}(2005)\citenamefont
  {Martinis}, \citenamefont {Cooper}, \citenamefont {McDermott}, \citenamefont
  {Steffen}, \citenamefont {Ansmann}, \citenamefont {Osborn}, \citenamefont
  {Cicak}, \citenamefont {Oh}, \citenamefont {Pappas}, \citenamefont
  {Simmonds},\ and\ \citenamefont {Yu}}]{Martinis2005Nov}%
  \BibitemOpen
  \bibfield  {author} {\bibinfo {author} {\bibfnamefont {J.~M.}\ \bibnamefont
  {Martinis}}, \bibinfo {author} {\bibfnamefont {K.~B.}\ \bibnamefont
  {Cooper}}, \bibinfo {author} {\bibfnamefont {R.}~\bibnamefont {McDermott}},
  \bibinfo {author} {\bibfnamefont {M.}~\bibnamefont {Steffen}}, \bibinfo
  {author} {\bibfnamefont {M.}~\bibnamefont {Ansmann}}, \bibinfo {author}
  {\bibfnamefont {K.~D.}\ \bibnamefont {Osborn}}, \bibinfo {author}
  {\bibfnamefont {K.}~\bibnamefont {Cicak}}, \bibinfo {author} {\bibfnamefont
  {S.}~\bibnamefont {Oh}}, \bibinfo {author} {\bibfnamefont {D.~P.}\
  \bibnamefont {Pappas}}, \bibinfo {author} {\bibfnamefont {R.~W.}\
  \bibnamefont {Simmonds}}, \ and\ \bibinfo {author} {\bibfnamefont {C.~C.}\
  \bibnamefont {Yu}},\ }\href {\doibase 10.1103/PhysRevLett.95.210503}
  {\bibfield  {journal} {\bibinfo  {journal} {Phys. Rev. Lett.}\ }\textbf
  {\bibinfo {volume} {95}},\ \bibinfo {pages} {210503} (\bibinfo {year}
  {2005})}\BibitemShut {NoStop}%
\bibitem [{\citenamefont {Schreier}\ \emph {et~al.}(2008)\citenamefont
  {Schreier}, \citenamefont {Houck}, \citenamefont {Koch}, \citenamefont
  {Schuster}, \citenamefont {Johnson}, \citenamefont {Chow}, \citenamefont
  {Gambetta}, \citenamefont {Majer}, \citenamefont {Frunzio}, \citenamefont
  {Devoret}, \citenamefont {Girvin},\ and\ \citenamefont
  {Schoelkopf}}]{Schreier2008May}%
  \BibitemOpen
  \bibfield  {author} {\bibinfo {author} {\bibfnamefont {J.~A.}\ \bibnamefont
  {Schreier}}, \bibinfo {author} {\bibfnamefont {A.~A.}\ \bibnamefont {Houck}},
  \bibinfo {author} {\bibfnamefont {J.}~\bibnamefont {Koch}}, \bibinfo {author}
  {\bibfnamefont {D.~I.}\ \bibnamefont {Schuster}}, \bibinfo {author}
  {\bibfnamefont {B.~R.}\ \bibnamefont {Johnson}}, \bibinfo {author}
  {\bibfnamefont {J.~M.}\ \bibnamefont {Chow}}, \bibinfo {author}
  {\bibfnamefont {J.~M.}\ \bibnamefont {Gambetta}}, \bibinfo {author}
  {\bibfnamefont {J.}~\bibnamefont {Majer}}, \bibinfo {author} {\bibfnamefont
  {L.}~\bibnamefont {Frunzio}}, \bibinfo {author} {\bibfnamefont {M.~H.}\
  \bibnamefont {Devoret}}, \bibinfo {author} {\bibfnamefont {S.~M.}\
  \bibnamefont {Girvin}}, \ and\ \bibinfo {author} {\bibfnamefont {R.~J.}\
  \bibnamefont {Schoelkopf}},\ }\href {\doibase 10.1103/PhysRevB.77.180502}
  {\bibfield  {journal} {\bibinfo  {journal} {Phys. Rev. B}\ }\textbf {\bibinfo
  {volume} {77}},\ \bibinfo {pages} {180502} (\bibinfo {year}
  {2008})}\BibitemShut {NoStop}%
\bibitem [{\citenamefont {Gunnarsson}\ \emph {et~al.}(2013)\citenamefont
  {Gunnarsson}, \citenamefont {Pirkkalainen}, \citenamefont {Li}, \citenamefont
  {Paraoanu}, \citenamefont {Hakonen}, \citenamefont
  {Sillanp{\ifmmode\ddot{a}\else\"{a}\fi}{\ifmmode\ddot{a}\else\"{a}\fi}},\
  and\ \citenamefont {Prunnila}}]{Gunnarsson2013Jul}%
  \BibitemOpen
  \bibfield  {author} {\bibinfo {author} {\bibfnamefont {D.}~\bibnamefont
  {Gunnarsson}}, \bibinfo {author} {\bibfnamefont {J.-M.}\ \bibnamefont
  {Pirkkalainen}}, \bibinfo {author} {\bibfnamefont {J.}~\bibnamefont {Li}},
  \bibinfo {author} {\bibfnamefont {G.~S.}\ \bibnamefont {Paraoanu}}, \bibinfo
  {author} {\bibfnamefont {P.}~\bibnamefont {Hakonen}}, \bibinfo {author}
  {\bibfnamefont {M.}~\bibnamefont
  {Sillanp{\ifmmode\ddot{a}\else\"{a}\fi}{\ifmmode\ddot{a}\else\"{a}\fi}}}, \
  and\ \bibinfo {author} {\bibfnamefont {M.}~\bibnamefont {Prunnila}},\ }\href
  {\doibase 10.1088/0953-2048/26/8/085010} {\bibfield  {journal} {\bibinfo
  {journal} {Supercond. Sci. Technol.}\ }\textbf {\bibinfo {volume} {26}},\
  \bibinfo {pages} {085010} (\bibinfo {year} {2013})}\BibitemShut {NoStop}%
\bibitem [{\citenamefont {Queen}\ \emph {et~al.}(2013)\citenamefont {Queen},
  \citenamefont {Liu}, \citenamefont {Karel}, \citenamefont {Metcalf},\ and\
  \citenamefont {Hellman}}]{Queen2013Mar}%
  \BibitemOpen
  \bibfield  {author} {\bibinfo {author} {\bibfnamefont {D.~R.}\ \bibnamefont
  {Queen}}, \bibinfo {author} {\bibfnamefont {X.}~\bibnamefont {Liu}}, \bibinfo
  {author} {\bibfnamefont {J.}~\bibnamefont {Karel}}, \bibinfo {author}
  {\bibfnamefont {T.~H.}\ \bibnamefont {Metcalf}}, \ and\ \bibinfo {author}
  {\bibfnamefont {F.}~\bibnamefont {Hellman}},\ }\href {\doibase
  10.1103/PhysRevLett.110.135901} {\bibfield  {journal} {\bibinfo  {journal}
  {Phys. Rev. Lett.}\ }\textbf {\bibinfo {volume} {110}},\ \bibinfo {pages}
  {135901} (\bibinfo {year} {2013})}\BibitemShut {NoStop}%
\bibitem [{\citenamefont {Wang}\ \emph {et~al.}(2025)\citenamefont {Wang},
  \citenamefont {Yu},\ and\ \citenamefont {Wu}}]{Wang2025Feb}%
  \BibitemOpen
  \bibfield  {author} {\bibinfo {author} {\bibfnamefont {Z.}~\bibnamefont
  {Wang}}, \bibinfo {author} {\bibfnamefont {C.~C.}\ \bibnamefont {Yu}}, \ and\
  \bibinfo {author} {\bibfnamefont {R.}~\bibnamefont {Wu}},\ }\href {\doibase
  10.1103/PhysRevApplied.23.024017} {\bibfield  {journal} {\bibinfo  {journal}
  {Phys. Rev. Appl.}\ }\textbf {\bibinfo {volume} {23}},\ \bibinfo {pages}
  {024017} (\bibinfo {year} {2025})}\BibitemShut {NoStop}%
\bibitem [{\citenamefont {Bergmann}\ and\ \citenamefont
  {Rainer}(1973)}]{Bergmann1973Aug}%
  \BibitemOpen
  \bibfield  {author} {\bibinfo {author} {\bibfnamefont {G.}~\bibnamefont
  {Bergmann}}\ and\ \bibinfo {author} {\bibfnamefont {D.}~\bibnamefont
  {Rainer}},\ }\href {\doibase 10.1007/BF02351862} {\bibfield  {journal}
  {\bibinfo  {journal} {Z. Phys.}\ }\textbf {\bibinfo {volume} {263}},\
  \bibinfo {pages} {59} (\bibinfo {year} {1973})}\BibitemShut {NoStop}%
\bibitem [{\citenamefont {Anderson}(1959)}]{Anderson1959Sep}%
  \BibitemOpen
  \bibfield  {author} {\bibinfo {author} {\bibfnamefont {P.~W.}\ \bibnamefont
  {Anderson}},\ }\href {\doibase 10.1016/0022-3697(59)90036-8} {\bibfield
  {journal} {\bibinfo  {journal} {J. Phys. Chem. Solids}\ }\textbf {\bibinfo
  {volume} {11}},\ \bibinfo {pages} {26} (\bibinfo {year} {1959})}\BibitemShut
  {NoStop}%
\bibitem [{\citenamefont {Giannozzi}\ \emph {et~al.}(2020)\citenamefont
  {Giannozzi}, \citenamefont {Baseggio}, \citenamefont {Bonfà}, \citenamefont
  {Brunato}, \citenamefont {Car}, \citenamefont {Carnimeo}, \citenamefont
  {Cavazzoni}, \citenamefont {de~Gironcoli}, \citenamefont {Delugas},
  \citenamefont {Ferrari~Ruffino}, \citenamefont {Ferretti}, \citenamefont
  {Marzari}, \citenamefont {Timrov}, \citenamefont {Urru},\ and\ \citenamefont
  {Baroni}}]{QE-2020}%
  \BibitemOpen
  \bibfield  {author} {\bibinfo {author} {\bibfnamefont {P.}~\bibnamefont
  {Giannozzi}}, \bibinfo {author} {\bibfnamefont {O.}~\bibnamefont {Baseggio}},
  \bibinfo {author} {\bibfnamefont {P.}~\bibnamefont {Bonfà}}, \bibinfo
  {author} {\bibfnamefont {D.}~\bibnamefont {Brunato}}, \bibinfo {author}
  {\bibfnamefont {R.}~\bibnamefont {Car}}, \bibinfo {author} {\bibfnamefont
  {I.}~\bibnamefont {Carnimeo}}, \bibinfo {author} {\bibfnamefont
  {C.}~\bibnamefont {Cavazzoni}}, \bibinfo {author} {\bibfnamefont
  {S.}~\bibnamefont {de~Gironcoli}}, \bibinfo {author} {\bibfnamefont
  {P.}~\bibnamefont {Delugas}}, \bibinfo {author} {\bibfnamefont
  {F.}~\bibnamefont {Ferrari~Ruffino}}, \bibinfo {author} {\bibfnamefont
  {A.}~\bibnamefont {Ferretti}}, \bibinfo {author} {\bibfnamefont
  {N.}~\bibnamefont {Marzari}}, \bibinfo {author} {\bibfnamefont
  {I.}~\bibnamefont {Timrov}}, \bibinfo {author} {\bibfnamefont
  {A.}~\bibnamefont {Urru}}, \ and\ \bibinfo {author} {\bibfnamefont
  {S.}~\bibnamefont {Baroni}},\ }\href {\doibase 10.1063/5.0005082} {\bibfield
  {journal} {\bibinfo  {journal} {J. Chem. Phys.}\ }\textbf {\bibinfo {volume}
  {152}},\ \bibinfo {pages} {154105} (\bibinfo {year} {2020})}\BibitemShut
  {NoStop}%
\bibitem [{\citenamefont {Kresse}\ and\ \citenamefont
  {Joubert}(1999)}]{kresse1999VASP}%
  \BibitemOpen
  \bibfield  {author} {\bibinfo {author} {\bibfnamefont {G.}~\bibnamefont
  {Kresse}}\ and\ \bibinfo {author} {\bibfnamefont {D.}~\bibnamefont
  {Joubert}},\ }\href {\doibase 10.1103/PhysRevB.59.1758} {\bibfield  {journal}
  {\bibinfo  {journal} {Phys. Rev. B}\ }\textbf {\bibinfo {volume} {59}},\
  \bibinfo {pages} {1758} (\bibinfo {year} {1999})}\BibitemShut {NoStop}%
\bibitem [{\citenamefont {Perdew}\ \emph {et~al.}(1996)\citenamefont {Perdew},
  \citenamefont {Burke},\ and\ \citenamefont {Ernzerhof}}]{Perdew1996}%
  \BibitemOpen
  \bibfield  {author} {\bibinfo {author} {\bibfnamefont {J.~P.}\ \bibnamefont
  {Perdew}}, \bibinfo {author} {\bibfnamefont {K.}~\bibnamefont {Burke}}, \
  and\ \bibinfo {author} {\bibfnamefont {M.}~\bibnamefont {Ernzerhof}},\ }\href
  {\doibase 10.1103/PhysRevLett.77.3865} {\bibfield  {journal} {\bibinfo
  {journal} {Phys. Rev. Lett.}\ }\textbf {\bibinfo {volume} {77}},\ \bibinfo
  {pages} {3865} (\bibinfo {year} {1996})}\BibitemShut {NoStop}%
\bibitem [{\citenamefont {Van~Setten}\ \emph {et~al.}(2018)\citenamefont
  {Van~Setten}, \citenamefont {Giantomassi}, \citenamefont {Bousquet},
  \citenamefont {Verstraete}, \citenamefont {Hamann}, \citenamefont {Gonze},\
  and\ \citenamefont {Rignanese}}]{van2018pseudodojo}%
  \BibitemOpen
  \bibfield  {author} {\bibinfo {author} {\bibfnamefont {M.~J.}\ \bibnamefont
  {Van~Setten}}, \bibinfo {author} {\bibfnamefont {M.}~\bibnamefont
  {Giantomassi}}, \bibinfo {author} {\bibfnamefont {E.}~\bibnamefont
  {Bousquet}}, \bibinfo {author} {\bibfnamefont {M.~J.}\ \bibnamefont
  {Verstraete}}, \bibinfo {author} {\bibfnamefont {D.~R.}\ \bibnamefont
  {Hamann}}, \bibinfo {author} {\bibfnamefont {X.}~\bibnamefont {Gonze}}, \
  and\ \bibinfo {author} {\bibfnamefont {G.-M.}\ \bibnamefont {Rignanese}},\
  }\href {\doibase https://doi.org/10.1016/j.cpc.2018.01.012} {\bibfield
  {journal} {\bibinfo  {journal} {Comput. Phys. Commun.}\ }\textbf {\bibinfo
  {volume} {226}},\ \bibinfo {pages} {39} (\bibinfo {year} {2018})}\BibitemShut
  {NoStop}%
\bibitem [{\citenamefont {Monkhorst}\ and\ \citenamefont
  {Pack}(1976)}]{PhysRevB.13.5188}%
  \BibitemOpen
  \bibfield  {author} {\bibinfo {author} {\bibfnamefont {H.~J.}\ \bibnamefont
  {Monkhorst}}\ and\ \bibinfo {author} {\bibfnamefont {J.~D.}\ \bibnamefont
  {Pack}},\ }\href {\doibase 10.1103/PhysRevB.13.5188} {\bibfield  {journal}
  {\bibinfo  {journal} {Phys. Rev. B}\ }\textbf {\bibinfo {volume} {13}},\
  \bibinfo {pages} {5188} (\bibinfo {year} {1976})}\BibitemShut {NoStop}%
\bibitem [{\citenamefont {Noffsinger}\ \emph {et~al.}(2010)\citenamefont
  {Noffsinger}, \citenamefont {Giustino}, \citenamefont {Malone}, \citenamefont
  {Park}, \citenamefont {Louie},\ and\ \citenamefont
  {Cohen}}]{Noffsinger2010Dec}%
  \BibitemOpen
  \bibfield  {author} {\bibinfo {author} {\bibfnamefont {J.}~\bibnamefont
  {Noffsinger}}, \bibinfo {author} {\bibfnamefont {F.}~\bibnamefont
  {Giustino}}, \bibinfo {author} {\bibfnamefont {B.~D.}\ \bibnamefont
  {Malone}}, \bibinfo {author} {\bibfnamefont {C.-H.}\ \bibnamefont {Park}},
  \bibinfo {author} {\bibfnamefont {S.~G.}\ \bibnamefont {Louie}}, \ and\
  \bibinfo {author} {\bibfnamefont {M.~L.}\ \bibnamefont {Cohen}},\ }\href
  {\doibase 10.1016/j.cpc.2010.08.027} {\bibfield  {journal} {\bibinfo
  {journal} {Comput. Phys. Commun.}\ }\textbf {\bibinfo {volume} {181}},\
  \bibinfo {pages} {2140} (\bibinfo {year} {2010})}\BibitemShut {NoStop}%
\bibitem [{\citenamefont {Lee}\ \emph {et~al.}(2023)\citenamefont {Lee},
  \citenamefont {Ponc{\'e}}, \citenamefont {Bushick}, \citenamefont
  {Hajinazar}, \citenamefont {Lafuente-Bartolome}, \citenamefont {Leveillee},
  \citenamefont {Lian}, \citenamefont {Lihm}, \citenamefont {Macheda},
  \citenamefont {Mori} \emph {et~al.}}]{lee2023electron}%
  \BibitemOpen
  \bibfield  {author} {\bibinfo {author} {\bibfnamefont {H.}~\bibnamefont
  {Lee}}, \bibinfo {author} {\bibfnamefont {S.}~\bibnamefont {Ponc{\'e}}},
  \bibinfo {author} {\bibfnamefont {K.}~\bibnamefont {Bushick}}, \bibinfo
  {author} {\bibfnamefont {S.}~\bibnamefont {Hajinazar}}, \bibinfo {author}
  {\bibfnamefont {J.}~\bibnamefont {Lafuente-Bartolome}}, \bibinfo {author}
  {\bibfnamefont {J.}~\bibnamefont {Leveillee}}, \bibinfo {author}
  {\bibfnamefont {C.}~\bibnamefont {Lian}}, \bibinfo {author} {\bibfnamefont
  {J.-M.}\ \bibnamefont {Lihm}}, \bibinfo {author} {\bibfnamefont
  {F.}~\bibnamefont {Macheda}}, \bibinfo {author} {\bibfnamefont
  {H.}~\bibnamefont {Mori}},  \emph {et~al.},\ }\href {\doibase
  https://doi.org/10.1038/s41524-023-01107-3} {\bibfield  {journal} {\bibinfo
  {journal} {npj Comput. Mater.}\ }\textbf {\bibinfo {volume} {9}},\ \bibinfo
  {pages} {156} (\bibinfo {year} {2023})}\BibitemShut {NoStop}%
\bibitem [{\citenamefont {Pizzi}\ \emph {et~al.}(2020)\citenamefont {Pizzi},
  \citenamefont {Vitale}, \citenamefont {Arita}, \citenamefont {Blugel},
  \citenamefont {Freimuth}, \citenamefont {G{\'{e}}ranton}, \citenamefont
  {Gibertini}, \citenamefont {Gresch}, \citenamefont {Johnson}, \citenamefont
  {Koretsune}, \citenamefont {Iba{\~{n}}ez-Azpiroz}, \citenamefont {Lee},
  \citenamefont {Lihm}, \citenamefont {Marchand}, \citenamefont {Marrazzo},
  \citenamefont {Mokrousov}, \citenamefont {Mustafa}, \citenamefont {Nohara},
  \citenamefont {Nomura}, \citenamefont {Paulatto}, \citenamefont
  {Ponc{\'{e}}}, \citenamefont {Ponweiser}, \citenamefont {Qiao}, \citenamefont
  {Thole}, \citenamefont {Tsirkin}, \citenamefont {Wierzbowska}, \citenamefont
  {Marzari}, \citenamefont {Vanderbilt}, \citenamefont {Souza}, \citenamefont
  {Mostofi},\ and\ \citenamefont {Yates}}]{Pizzi2020}%
  \BibitemOpen
  \bibfield  {author} {\bibinfo {author} {\bibfnamefont {G.}~\bibnamefont
  {Pizzi}}, \bibinfo {author} {\bibfnamefont {V.}~\bibnamefont {Vitale}},
  \bibinfo {author} {\bibfnamefont {R.}~\bibnamefont {Arita}}, \bibinfo
  {author} {\bibfnamefont {S.}~\bibnamefont {Blugel}}, \bibinfo {author}
  {\bibfnamefont {F.}~\bibnamefont {Freimuth}}, \bibinfo {author}
  {\bibfnamefont {G.}~\bibnamefont {G{\'{e}}ranton}}, \bibinfo {author}
  {\bibfnamefont {M.}~\bibnamefont {Gibertini}}, \bibinfo {author}
  {\bibfnamefont {D.}~\bibnamefont {Gresch}}, \bibinfo {author} {\bibfnamefont
  {C.}~\bibnamefont {Johnson}}, \bibinfo {author} {\bibfnamefont
  {T.}~\bibnamefont {Koretsune}}, \bibinfo {author} {\bibfnamefont
  {J.}~\bibnamefont {Iba{\~{n}}ez-Azpiroz}}, \bibinfo {author} {\bibfnamefont
  {H.}~\bibnamefont {Lee}}, \bibinfo {author} {\bibfnamefont {J.-M.}\
  \bibnamefont {Lihm}}, \bibinfo {author} {\bibfnamefont {D.}~\bibnamefont
  {Marchand}}, \bibinfo {author} {\bibfnamefont {A.}~\bibnamefont {Marrazzo}},
  \bibinfo {author} {\bibfnamefont {Y.}~\bibnamefont {Mokrousov}}, \bibinfo
  {author} {\bibfnamefont {J.~I.}\ \bibnamefont {Mustafa}}, \bibinfo {author}
  {\bibfnamefont {Y.}~\bibnamefont {Nohara}}, \bibinfo {author} {\bibfnamefont
  {Y.}~\bibnamefont {Nomura}}, \bibinfo {author} {\bibfnamefont
  {L.}~\bibnamefont {Paulatto}}, \bibinfo {author} {\bibfnamefont
  {S.}~\bibnamefont {Ponc{\'{e}}}}, \bibinfo {author} {\bibfnamefont
  {T.}~\bibnamefont {Ponweiser}}, \bibinfo {author} {\bibfnamefont
  {J.}~\bibnamefont {Qiao}}, \bibinfo {author} {\bibfnamefont {F.}~\bibnamefont
  {Thole}}, \bibinfo {author} {\bibfnamefont {S.~S.}\ \bibnamefont {Tsirkin}},
  \bibinfo {author} {\bibfnamefont {M.}~\bibnamefont {Wierzbowska}}, \bibinfo
  {author} {\bibfnamefont {N.}~\bibnamefont {Marzari}}, \bibinfo {author}
  {\bibfnamefont {D.}~\bibnamefont {Vanderbilt}}, \bibinfo {author}
  {\bibfnamefont {I.}~\bibnamefont {Souza}}, \bibinfo {author} {\bibfnamefont
  {A.~A.}\ \bibnamefont {Mostofi}}, \ and\ \bibinfo {author} {\bibfnamefont
  {J.~R.}\ \bibnamefont {Yates}},\ }\href {\doibase 10.1088/1361-648x/ab51ff}
  {\bibfield  {journal} {\bibinfo  {journal} {J. Phys. Condens. Matter}\
  }\textbf {\bibinfo {volume} {32}},\ \bibinfo {pages} {165902} (\bibinfo
  {year} {2020})}\BibitemShut {NoStop}%
\bibitem [{\citenamefont {Strongin}\ \emph {et~al.}(1968)\citenamefont
  {Strongin}, \citenamefont {Kammerer}, \citenamefont {Crow}, \citenamefont
  {Parks}, \citenamefont {Douglass},\ and\ \citenamefont
  {Jensen}}]{Strongin1968Oct}%
  \BibitemOpen
  \bibfield  {author} {\bibinfo {author} {\bibfnamefont {M.}~\bibnamefont
  {Strongin}}, \bibinfo {author} {\bibfnamefont {O.~F.}\ \bibnamefont
  {Kammerer}}, \bibinfo {author} {\bibfnamefont {J.~E.}\ \bibnamefont {Crow}},
  \bibinfo {author} {\bibfnamefont {R.~D.}\ \bibnamefont {Parks}}, \bibinfo
  {author} {\bibfnamefont {D.~H.}\ \bibnamefont {Douglass}}, \ and\ \bibinfo
  {author} {\bibfnamefont {M.~A.}\ \bibnamefont {Jensen}},\ }\href {\doibase
  10.1103/PhysRevLett.21.1320} {\bibfield  {journal} {\bibinfo  {journal}
  {Phys. Rev. Lett.}\ }\textbf {\bibinfo {volume} {21}},\ \bibinfo {pages}
  {1320} (\bibinfo {year} {1968})}\BibitemShut {NoStop}%
\bibitem [{\citenamefont {Allen}(1974)}]{Allen1974May}%
  \BibitemOpen
  \bibfield  {author} {\bibinfo {author} {\bibfnamefont {P.~B.}\ \bibnamefont
  {Allen}},\ }\href {\doibase 10.1016/0038-1098(74)90397-4} {\bibfield
  {journal} {\bibinfo  {journal} {Solid State Commun.}\ }\textbf {\bibinfo
  {volume} {14}},\ \bibinfo {pages} {937} (\bibinfo {year} {1974})}\BibitemShut
  {NoStop}%
\bibitem [{\citenamefont {Leavens}\ and\ \citenamefont
  {Fenton}(1981)}]{Leavens1981Nov}%
  \BibitemOpen
  \bibfield  {author} {\bibinfo {author} {\bibfnamefont {C.~R.}\ \bibnamefont
  {Leavens}}\ and\ \bibinfo {author} {\bibfnamefont {E.~W.}\ \bibnamefont
  {Fenton}},\ }\href {\doibase 10.1103/PhysRevB.24.5086} {\bibfield  {journal}
  {\bibinfo  {journal} {Phys. Rev. B}\ }\textbf {\bibinfo {volume} {24}},\
  \bibinfo {pages} {5086} (\bibinfo {year} {1981})}\BibitemShut {NoStop}%
\bibitem [{\citenamefont {Tamura}(1993)}]{Tamura1993Mar}%
  \BibitemOpen
  \bibfield  {author} {\bibinfo {author} {\bibfnamefont {A.}~\bibnamefont
  {Tamura}},\ }\href {\doibase 10.1007/BF01425677} {\bibfield  {journal}
  {\bibinfo  {journal} {Z. Phys. D: At. Mol. Clusters}\ }\textbf {\bibinfo
  {volume} {26}},\ \bibinfo {pages} {240} (\bibinfo {year} {1993})}\BibitemShut
  {NoStop}%
\bibitem [{\citenamefont {Bose}\ \emph {et~al.}(2009)\citenamefont {Bose},
  \citenamefont {Galande}, \citenamefont {Chockalingam}, \citenamefont
  {Banerjee}, \citenamefont {Raychaudhuri},\ and\ \citenamefont
  {Ayyub}}]{Bose2009Apr}%
  \BibitemOpen
  \bibfield  {author} {\bibinfo {author} {\bibfnamefont {S.}~\bibnamefont
  {Bose}}, \bibinfo {author} {\bibfnamefont {C.}~\bibnamefont {Galande}},
  \bibinfo {author} {\bibfnamefont {S.~P.}\ \bibnamefont {Chockalingam}},
  \bibinfo {author} {\bibfnamefont {R.}~\bibnamefont {Banerjee}}, \bibinfo
  {author} {\bibfnamefont {P.}~\bibnamefont {Raychaudhuri}}, \ and\ \bibinfo
  {author} {\bibfnamefont {P.}~\bibnamefont {Ayyub}},\ }\href {\doibase
  10.1088/0953-8984/21/20/205702} {\bibfield  {journal} {\bibinfo  {journal}
  {J. Phys.: Condens. Matter}\ }\textbf {\bibinfo {volume} {21}},\ \bibinfo
  {pages} {205702} (\bibinfo {year} {2009})}\BibitemShut {NoStop}%
\bibitem [{\citenamefont {Bozyigit}\ \emph {et~al.}(2016)\citenamefont
  {Bozyigit}, \citenamefont {Yazdani}, \citenamefont {Yarema}, \citenamefont
  {Yarema}, \citenamefont {Lin}, \citenamefont {Volk}, \citenamefont
  {Vuttivorakulchai}, \citenamefont {Luisier}, \citenamefont {Juranyi},\ and\
  \citenamefont {Wood}}]{bozyigit2016soft}%
  \BibitemOpen
  \bibfield  {author} {\bibinfo {author} {\bibfnamefont {D.}~\bibnamefont
  {Bozyigit}}, \bibinfo {author} {\bibfnamefont {N.}~\bibnamefont {Yazdani}},
  \bibinfo {author} {\bibfnamefont {M.}~\bibnamefont {Yarema}}, \bibinfo
  {author} {\bibfnamefont {O.}~\bibnamefont {Yarema}}, \bibinfo {author}
  {\bibfnamefont {W.~M.~M.}\ \bibnamefont {Lin}}, \bibinfo {author}
  {\bibfnamefont {S.}~\bibnamefont {Volk}}, \bibinfo {author} {\bibfnamefont
  {K.}~\bibnamefont {Vuttivorakulchai}}, \bibinfo {author} {\bibfnamefont
  {M.}~\bibnamefont {Luisier}}, \bibinfo {author} {\bibfnamefont
  {F.}~\bibnamefont {Juranyi}}, \ and\ \bibinfo {author} {\bibfnamefont
  {V.}~\bibnamefont {Wood}},\ }\href {\doibase
  https://doi.org/10.1038/nature16977} {\bibfield  {journal} {\bibinfo
  {journal} {Nature}\ }\textbf {\bibinfo {volume} {531}},\ \bibinfo {pages}
  {618} (\bibinfo {year} {2016})}\BibitemShut {NoStop}%
\bibitem [{\citenamefont {Lozano}\ \emph {et~al.}(2019)\citenamefont {Lozano},
  \citenamefont {Couet}, \citenamefont {Petermann}, \citenamefont {Hamoir},
  \citenamefont {Jochum}, \citenamefont {Picot}, \citenamefont
  {Men{\ifmmode\acute{e}\else\'{e}\fi}ndez}, \citenamefont {Houben},
  \citenamefont {Joly}, \citenamefont {Antohe}, \citenamefont {Hu},
  \citenamefont {Leu}, \citenamefont {Alatas}, \citenamefont {Said},
  \citenamefont {Roelants}, \citenamefont {Partoens}, \citenamefont
  {Milo{\ifmmode\check{s}\else\v{s}\fi}evi{\ifmmode\acute{c}\else\'{c}\fi}},
  \citenamefont {Peeters}, \citenamefont {Piraux}, \citenamefont {Van~de
  Vondel}, \citenamefont {Vantomme}, \citenamefont {Temst},\ and\ \citenamefont
  {Van~Bael}}]{Lozano2019Feb}%
  \BibitemOpen
  \bibfield  {author} {\bibinfo {author} {\bibfnamefont {D.~P.}\ \bibnamefont
  {Lozano}}, \bibinfo {author} {\bibfnamefont {S.}~\bibnamefont {Couet}},
  \bibinfo {author} {\bibfnamefont {C.}~\bibnamefont {Petermann}}, \bibinfo
  {author} {\bibfnamefont {G.}~\bibnamefont {Hamoir}}, \bibinfo {author}
  {\bibfnamefont {J.~K.}\ \bibnamefont {Jochum}}, \bibinfo {author}
  {\bibfnamefont {T.}~\bibnamefont {Picot}}, \bibinfo {author} {\bibfnamefont
  {E.}~\bibnamefont {Men{\ifmmode\acute{e}\else\'{e}\fi}ndez}}, \bibinfo
  {author} {\bibfnamefont {K.}~\bibnamefont {Houben}}, \bibinfo {author}
  {\bibfnamefont {V.}~\bibnamefont {Joly}}, \bibinfo {author} {\bibfnamefont
  {V.~A.}\ \bibnamefont {Antohe}}, \bibinfo {author} {\bibfnamefont {M.~Y.}\
  \bibnamefont {Hu}}, \bibinfo {author} {\bibfnamefont {B.~M.}\ \bibnamefont
  {Leu}}, \bibinfo {author} {\bibfnamefont {A.}~\bibnamefont {Alatas}},
  \bibinfo {author} {\bibfnamefont {A.~H.}\ \bibnamefont {Said}}, \bibinfo
  {author} {\bibfnamefont {S.}~\bibnamefont {Roelants}}, \bibinfo {author}
  {\bibfnamefont {B.}~\bibnamefont {Partoens}}, \bibinfo {author}
  {\bibfnamefont {M.~V.}\ \bibnamefont
  {Milo{\ifmmode\check{s}\else\v{s}\fi}evi{\ifmmode\acute{c}\else\'{c}\fi}}},
  \bibinfo {author} {\bibfnamefont {F.~M.}\ \bibnamefont {Peeters}}, \bibinfo
  {author} {\bibfnamefont {L.}~\bibnamefont {Piraux}}, \bibinfo {author}
  {\bibfnamefont {J.}~\bibnamefont {Van~de Vondel}}, \bibinfo {author}
  {\bibfnamefont {A.}~\bibnamefont {Vantomme}}, \bibinfo {author}
  {\bibfnamefont {K.}~\bibnamefont {Temst}}, \ and\ \bibinfo {author}
  {\bibfnamefont {M.~J.}\ \bibnamefont {Van~Bael}},\ }\href {\doibase
  10.1103/PhysRevB.99.064512} {\bibfield  {journal} {\bibinfo  {journal} {Phys.
  Rev. B}\ }\textbf {\bibinfo {volume} {99}},\ \bibinfo {pages} {064512}
  (\bibinfo {year} {2019})}\BibitemShut {NoStop}%
\bibitem [{\citenamefont {Houben}\ \emph {et~al.}(2020)\citenamefont {Houben},
  \citenamefont {Jochum}, \citenamefont {Couet}, \citenamefont
  {Men{\ifmmode\acute{e}\else\'{e}\fi}ndez}, \citenamefont {Picot},
  \citenamefont {Hu}, \citenamefont {Zhao}, \citenamefont {Alp}, \citenamefont
  {Vantomme}, \citenamefont {Temst},\ and\ \citenamefont
  {Van~Bael}}]{Houben2020Mar}%
  \BibitemOpen
  \bibfield  {author} {\bibinfo {author} {\bibfnamefont {K.}~\bibnamefont
  {Houben}}, \bibinfo {author} {\bibfnamefont {J.~K.}\ \bibnamefont {Jochum}},
  \bibinfo {author} {\bibfnamefont {S.}~\bibnamefont {Couet}}, \bibinfo
  {author} {\bibfnamefont {E.}~\bibnamefont
  {Men{\ifmmode\acute{e}\else\'{e}\fi}ndez}}, \bibinfo {author} {\bibfnamefont
  {T.}~\bibnamefont {Picot}}, \bibinfo {author} {\bibfnamefont {M.~Y.}\
  \bibnamefont {Hu}}, \bibinfo {author} {\bibfnamefont {J.~Y.}\ \bibnamefont
  {Zhao}}, \bibinfo {author} {\bibfnamefont {E.~E.}\ \bibnamefont {Alp}},
  \bibinfo {author} {\bibfnamefont {A.}~\bibnamefont {Vantomme}}, \bibinfo
  {author} {\bibfnamefont {K.}~\bibnamefont {Temst}}, \ and\ \bibinfo {author}
  {\bibfnamefont {M.~J.}\ \bibnamefont {Van~Bael}},\ }\href {\doibase
  10.1038/s41598-020-62617-4} {\bibfield  {journal} {\bibinfo  {journal} {Sci.
  Rep.}\ }\textbf {\bibinfo {volume} {10}},\ \bibinfo {pages} {1} (\bibinfo
  {year} {2020})}\BibitemShut {NoStop}%
\bibitem [{\citenamefont {Fibich}(1965)}]{Fibich1965Apr}%
  \BibitemOpen
  \bibfield  {author} {\bibinfo {author} {\bibfnamefont {M.}~\bibnamefont
  {Fibich}},\ }\href {\doibase 10.1103/PhysRevLett.14.561} {\bibfield
  {journal} {\bibinfo  {journal} {Phys. Rev. Lett.}\ }\textbf {\bibinfo
  {volume} {14}},\ \bibinfo {pages} {561} (\bibinfo {year} {1965})}\BibitemShut
  {NoStop}%
\bibitem [{\citenamefont {Marsiglio}\ and\ \citenamefont
  {Carbotte}(1991)}]{Marsiglio1991Mar}%
  \BibitemOpen
  \bibfield  {author} {\bibinfo {author} {\bibfnamefont {F.}~\bibnamefont
  {Marsiglio}}\ and\ \bibinfo {author} {\bibfnamefont {J.~P.}\ \bibnamefont
  {Carbotte}},\ }\href {\doibase 10.1103/PhysRevB.43.5355} {\bibfield
  {journal} {\bibinfo  {journal} {Phys. Rev. B}\ }\textbf {\bibinfo {volume}
  {43}},\ \bibinfo {pages} {5355} (\bibinfo {year} {1991})}\BibitemShut
  {NoStop}%
\bibitem [{\citenamefont {Chubukov}\ \emph {et~al.}(2020)\citenamefont
  {Chubukov}, \citenamefont {Abanov}, \citenamefont {Esterlis},\ and\
  \citenamefont {Kivelson}}]{Chubukov2020Jun}%
  \BibitemOpen
  \bibfield  {author} {\bibinfo {author} {\bibfnamefont {A.~V.}\ \bibnamefont
  {Chubukov}}, \bibinfo {author} {\bibfnamefont {A.}~\bibnamefont {Abanov}},
  \bibinfo {author} {\bibfnamefont {I.}~\bibnamefont {Esterlis}}, \ and\
  \bibinfo {author} {\bibfnamefont {S.~A.}\ \bibnamefont {Kivelson}},\ }\href
  {\doibase 10.1016/j.aop.2020.168190} {\bibfield  {journal} {\bibinfo
  {journal} {Ann. Phys.}\ }\textbf {\bibinfo {volume} {417}},\ \bibinfo {pages}
  {168190} (\bibinfo {year} {2020})}\BibitemShut {NoStop}%
\bibitem [{\citenamefont {Owen}\ and\ \citenamefont
  {Scalapino}(1971)}]{Owen1971Oct}%
  \BibitemOpen
  \bibfield  {author} {\bibinfo {author} {\bibfnamefont {C.~S.}\ \bibnamefont
  {Owen}}\ and\ \bibinfo {author} {\bibfnamefont {D.~J.}\ \bibnamefont
  {Scalapino}},\ }\href {\doibase 10.1016/0031-8914(71)90320-X} {\bibfield
  {journal} {\bibinfo  {journal} {Physica}\ }\textbf {\bibinfo {volume} {55}},\
  \bibinfo {pages} {691} (\bibinfo {year} {1971})}\BibitemShut {NoStop}%
\bibitem [{\citenamefont {Kresin}(1987)}]{Kresin1987Jun}%
  \BibitemOpen
  \bibfield  {author} {\bibinfo {author} {\bibfnamefont {V.~Z.}\ \bibnamefont
  {Kresin}},\ }\href {\doibase 10.1016/0375-9601(87)90744-4} {\bibfield
  {journal} {\bibinfo  {journal} {Phys. Lett. A}\ }\textbf {\bibinfo {volume}
  {122}},\ \bibinfo {pages} {434} (\bibinfo {year} {1987})}\BibitemShut
  {NoStop}%
\bibitem [{\citenamefont {Tajik}(2018)}]{Tajik2018}%
  \BibitemOpen
  \bibfield  {author} {\bibinfo {author} {\bibfnamefont {S.}~\bibnamefont
  {Tajik}},\ }\emph {\bibinfo {title} {Effect of Rattling Phonons on
  Superconductivity of KOs$_2$O$_6$}},\ \href
  {https://brocku.scholaris.ca/items/9dadd2fc-a49d-4d0c-a5b1-365f91636fb4/full}
  {\bibinfo {type} {{M.Sc. Thesis}}},\ \bibinfo  {school} {Brock University},
  \bibinfo {address} {St. Catharines, Ontario, Canada} (\bibinfo {year}
  {2018})\BibitemShut {NoStop}%
\end{thebibliography}%


\begin{thebibliography}{123}%
\makeatletter
\providecommand \@ifxundefined [1]{%
 \@ifx{#1\undefined}
}%
\providecommand \@ifnum [1]{%
 \ifnum #1\expandafter \@firstoftwo
 \else \expandafter \@secondoftwo
 \fi
}%
\providecommand \@ifx [1]{%
 \ifx #1\expandafter \@firstoftwo
 \else \expandafter \@secondoftwo
 \fi
}%
\providecommand \natexlab [1]{#1}%
\providecommand \enquote  [1]{``#1''}%
\providecommand \bibnamefont  [1]{#1}%
\providecommand \bibfnamefont [1]{#1}%
\providecommand \citenamefont [1]{#1}%
\providecommand \href@noop [0]{\@secondoftwo}%
\providecommand \href [0]{\begingroup \@sanitize@url \@href}%
\providecommand \@href[1]{\@@startlink{#1}\@@href}%
\providecommand \@@href[1]{\endgroup#1\@@endlink}%
\providecommand \@sanitize@url [0]{\catcode `\\12\catcode `\$12\catcode
  `\&12\catcode `\#12\catcode `\^12\catcode `\_12\catcode `\%12\relax}%
\providecommand \@@startlink[1]{}%
\providecommand \@@endlink[0]{}%
\providecommand \url  [0]{\begingroup\@sanitize@url \@url }%
\providecommand \@url [1]{\endgroup\@href {#1}{\urlprefix }}%
\providecommand \urlprefix  [0]{URL }%
\providecommand \Eprint [0]{\href }%
\providecommand \doibase [0]{http://dx.doi.org/}%
\providecommand \selectlanguage [0]{\@gobble}%
\providecommand \bibinfo  [0]{\@secondoftwo}%
\providecommand \bibfield  [0]{\@secondoftwo}%
\providecommand \translation [1]{[#1]}%
\providecommand \BibitemOpen [0]{}%
\providecommand \bibitemStop [0]{}%
\providecommand \bibitemNoStop [0]{.\EOS\space}%
\providecommand \EOS [0]{\spacefactor3000\relax}%
\providecommand \BibitemShut  [1]{\csname bibitem#1\endcsname}%
\let\auto@bib@innerbib\@empty
\bibitem [{\citenamefont {Bardeen}\ \emph
  {et~al.}(1957{\natexlab{a}})\citenamefont {Bardeen}, \citenamefont {Cooper},\
  and\ \citenamefont {Schrieffer}}]{Bardeen1957}%
  \BibitemOpen
  \bibfield  {author} {\bibinfo {author} {\bibfnamefont {J.}~\bibnamefont
  {Bardeen}}, \bibinfo {author} {\bibfnamefont {L.~N.}\ \bibnamefont {Cooper}},
  \ and\ \bibinfo {author} {\bibfnamefont {J.~R.}\ \bibnamefont {Schrieffer}},\
  }\href {\doibase 10.1103/PhysRev.108.1175} {\bibfield  {journal} {\bibinfo
  {journal} {Phys. Rev.}\ }\textbf {\bibinfo {volume} {108}},\ \bibinfo {pages}
  {1175} (\bibinfo {year} {1957}{\natexlab{a}})}\BibitemShut {NoStop}%
\bibitem [{\citenamefont {McMillan}(1968)}]{McMillan1968Mar}%
  \BibitemOpen
  \bibfield  {author} {\bibinfo {author} {\bibfnamefont {W.~L.}\ \bibnamefont
  {McMillan}},\ }\href {\doibase 10.1103/PhysRev.167.331} {\bibfield  {journal}
  {\bibinfo  {journal} {Phys. Rev.}\ }\textbf {\bibinfo {volume} {167}},\
  \bibinfo {pages} {331} (\bibinfo {year} {1968})}\BibitemShut {NoStop}%
\bibitem [{\citenamefont {Rickayzen}(1965)}]{RickayzenBook}%
  \BibitemOpen
  \bibfield  {author} {\bibinfo {author} {\bibfnamefont {G.}~\bibnamefont
  {Rickayzen}},\ }\href@noop {} {\emph {\bibinfo {title} {Theory of
  Superconductivity}}}\ (\bibinfo  {publisher} {John Wiley and Sons Inc.},\
  \bibinfo {address} {New York},\ \bibinfo {year} {1965})\BibitemShut {NoStop}%
\bibitem [{\citenamefont {Allen}\ and\ \citenamefont
  {Dynes}(1975)}]{Allen1975Aug}%
  \BibitemOpen
  \bibfield  {author} {\bibinfo {author} {\bibfnamefont {P.~B.}\ \bibnamefont
  {Allen}}\ and\ \bibinfo {author} {\bibfnamefont {R.~C.}\ \bibnamefont
  {Dynes}},\ }\href {\doibase 10.1103/PhysRevB.12.905} {\bibfield  {journal}
  {\bibinfo  {journal} {Phys. Rev. B}\ }\textbf {\bibinfo {volume} {12}},\
  \bibinfo {pages} {905} (\bibinfo {year} {1975})}\BibitemShut {NoStop}%
\bibitem [{\citenamefont {Carbotte}(1990)}]{RevModPhys.62.1027}%
  \BibitemOpen
  \bibfield  {author} {\bibinfo {author} {\bibfnamefont {J.~P.}\ \bibnamefont
  {Carbotte}},\ }\href {\doibase 10.1103/RevModPhys.62.1027} {\bibfield
  {journal} {\bibinfo  {journal} {Rev. Mod. Phys.}\ }\textbf {\bibinfo {volume}
  {62}},\ \bibinfo {pages} {1027} (\bibinfo {year} {1990})}\BibitemShut
  {NoStop}%
\bibitem [{\citenamefont {Garland}\ \emph {et~al.}(1968)\citenamefont
  {Garland}, \citenamefont {Bennemann},\ and\ \citenamefont
  {Mueller}}]{Garland1968Oct}%
  \BibitemOpen
  \bibfield  {author} {\bibinfo {author} {\bibfnamefont {J.~W.}\ \bibnamefont
  {Garland}}, \bibinfo {author} {\bibfnamefont {K.~H.}\ \bibnamefont
  {Bennemann}}, \ and\ \bibinfo {author} {\bibfnamefont {F.~M.}\ \bibnamefont
  {Mueller}},\ }\href {\doibase 10.1103/PhysRevLett.21.1315} {\bibfield
  {journal} {\bibinfo  {journal} {Phys. Rev. Lett.}\ }\textbf {\bibinfo
  {volume} {21}},\ \bibinfo {pages} {1315} (\bibinfo {year}
  {1968})}\BibitemShut {NoStop}%
\bibitem [{\citenamefont {Mayoh}\ and\ \citenamefont
  {Garc{\ifmmode\acute{\imath}\else\'{\i}\fi}a-Garc{\ifmmode\acute{\imath}\else\'{\i}\fi}a}(2014)}]{Mayoh2014Oct}%
  \BibitemOpen
  \bibfield  {author} {\bibinfo {author} {\bibfnamefont {J.}~\bibnamefont
  {Mayoh}}\ and\ \bibinfo {author} {\bibfnamefont {A.~M.}\ \bibnamefont
  {Garc{\ifmmode\acute{\imath}\else\'{\i}\fi}a-Garc{\ifmmode\acute{\imath}\else\'{\i}\fi}a}},\
  }\href {\doibase 10.1103/PhysRevB.90.134513} {\bibfield  {journal} {\bibinfo
  {journal} {Phys. Rev. B}\ }\textbf {\bibinfo {volume} {90}},\ \bibinfo
  {pages} {134513} (\bibinfo {year} {2014})}\BibitemShut {NoStop}%
\bibitem [{\citenamefont {Pracht}\ \emph {et~al.}(2016)\citenamefont {Pracht},
  \citenamefont {Bachar}, \citenamefont {Benfatto}, \citenamefont {Deutscher},
  \citenamefont {Farber}, \citenamefont {Dressel},\ and\ \citenamefont
  {Scheffler}}]{Pracht2016Mar}%
  \BibitemOpen
  \bibfield  {author} {\bibinfo {author} {\bibfnamefont {U.~S.}\ \bibnamefont
  {Pracht}}, \bibinfo {author} {\bibfnamefont {N.}~\bibnamefont {Bachar}},
  \bibinfo {author} {\bibfnamefont {L.}~\bibnamefont {Benfatto}}, \bibinfo
  {author} {\bibfnamefont {G.}~\bibnamefont {Deutscher}}, \bibinfo {author}
  {\bibfnamefont {E.}~\bibnamefont {Farber}}, \bibinfo {author} {\bibfnamefont
  {M.}~\bibnamefont {Dressel}}, \ and\ \bibinfo {author} {\bibfnamefont
  {M.}~\bibnamefont {Scheffler}},\ }\href {\doibase 10.1103/PhysRevB.93.100503}
  {\bibfield  {journal} {\bibinfo  {journal} {Phys. Rev. B}\ }\textbf {\bibinfo
  {volume} {93}},\ \bibinfo {pages} {100503} (\bibinfo {year}
  {2016})}\BibitemShut {NoStop}%
\bibitem [{\citenamefont {Leavens}\ and\ \citenamefont
  {Fenton}(1981)}]{Leavens1981Nov}%
  \BibitemOpen
  \bibfield  {author} {\bibinfo {author} {\bibfnamefont {C.~R.}\ \bibnamefont
  {Leavens}}\ and\ \bibinfo {author} {\bibfnamefont {E.~W.}\ \bibnamefont
  {Fenton}},\ }\href {\doibase 10.1103/PhysRevB.24.5086} {\bibfield  {journal}
  {\bibinfo  {journal} {Phys. Rev. B}\ }\textbf {\bibinfo {volume} {24}},\
  \bibinfo {pages} {5086} (\bibinfo {year} {1981})}\BibitemShut {NoStop}%
\bibitem [{\citenamefont {Tamura}(1993)}]{Tamura1993Mar}%
  \BibitemOpen
  \bibfield  {author} {\bibinfo {author} {\bibfnamefont {A.}~\bibnamefont
  {Tamura}},\ }\href {\doibase 10.1007/BF01425677} {\bibfield  {journal}
  {\bibinfo  {journal} {Z. Phys. D: At. Mol. Clusters}\ }\textbf {\bibinfo
  {volume} {26}},\ \bibinfo {pages} {240} (\bibinfo {year} {1993})}\BibitemShut
  {NoStop}%
\bibitem [{\citenamefont {Croitoru}\ \emph {et~al.}(2016)\citenamefont
  {Croitoru}, \citenamefont {Shanenko}, \citenamefont {Vagov}, \citenamefont
  {Vasenko}, \citenamefont
  {Milo{\ifmmode\check{s}\else\v{s}\fi}evi{\ifmmode\acute{c}\else\'{c}\fi}},
  \citenamefont {Axt},\ and\ \citenamefont {Peeters}}]{Croitoru2016Mar}%
  \BibitemOpen
  \bibfield  {author} {\bibinfo {author} {\bibfnamefont {M.~D.}\ \bibnamefont
  {Croitoru}}, \bibinfo {author} {\bibfnamefont {A.~A.}\ \bibnamefont
  {Shanenko}}, \bibinfo {author} {\bibfnamefont {A.}~\bibnamefont {Vagov}},
  \bibinfo {author} {\bibfnamefont {A.~S.}\ \bibnamefont {Vasenko}}, \bibinfo
  {author} {\bibfnamefont {M.~V.}\ \bibnamefont
  {Milo{\ifmmode\check{s}\else\v{s}\fi}evi{\ifmmode\acute{c}\else\'{c}\fi}}},
  \bibinfo {author} {\bibfnamefont {V.~M.}\ \bibnamefont {Axt}}, \ and\
  \bibinfo {author} {\bibfnamefont {F.~M.}\ \bibnamefont {Peeters}},\ }\href
  {\doibase 10.1007/s10948-015-3319-8} {\bibfield  {journal} {\bibinfo
  {journal} {J. Supercond. Novel Magn.}\ }\textbf {\bibinfo {volume} {29}},\
  \bibinfo {pages} {605} (\bibinfo {year} {2016})}\BibitemShut {NoStop}%
\bibitem [{\citenamefont {Grankin}\ \emph {et~al.}(2024)\citenamefont
  {Grankin}, \citenamefont {Hafezi},\ and\ \citenamefont
  {Galitski}}]{Grankin2024Aug}%
  \BibitemOpen
  \bibfield  {author} {\bibinfo {author} {\bibfnamefont {A.}~\bibnamefont
  {Grankin}}, \bibinfo {author} {\bibfnamefont {M.}~\bibnamefont {Hafezi}}, \
  and\ \bibinfo {author} {\bibfnamefont {V.}~\bibnamefont {Galitski}},\ }\href
  {\doibase 10.48550/arXiv.2408.03927} {\bibfield  {journal} {\bibinfo
  {journal} {arXiv}\ } (\bibinfo {year} {2024}),\ 10.48550/arXiv.2408.03927},\
  \Eprint {http://arxiv.org/abs/2408.03927} {2408.03927} \BibitemShut {NoStop}%
\bibitem [{\citenamefont {Lee}\ \emph {et~al.}(2025)\citenamefont {Lee},
  \citenamefont {Lee}, \citenamefont {Yun}, \citenamefont {Zapata},
  \citenamefont {Su{\ifmmode\acute{a}\else\'{a}\fi}rez}, \citenamefont
  {Sirena}, \citenamefont {Kim},\ and\ \citenamefont {Haberkorn}}]{Lee2025Nov}%
  \BibitemOpen
  \bibfield  {author} {\bibinfo {author} {\bibfnamefont {C.}~\bibnamefont
  {Lee}}, \bibinfo {author} {\bibfnamefont {Y.}~\bibnamefont {Lee}}, \bibinfo
  {author} {\bibfnamefont {J.}~\bibnamefont {Yun}}, \bibinfo {author}
  {\bibfnamefont {J.~C.}\ \bibnamefont {Zapata}}, \bibinfo {author}
  {\bibfnamefont {S.}~\bibnamefont {Su{\ifmmode\acute{a}\else\'{a}\fi}rez}},
  \bibinfo {author} {\bibfnamefont {M.}~\bibnamefont {Sirena}}, \bibinfo
  {author} {\bibfnamefont {J.}~\bibnamefont {Kim}}, \ and\ \bibinfo {author}
  {\bibfnamefont {N.}~\bibnamefont {Haberkorn}},\ }\href {\doibase
  10.1016/j.matlet.2025.138982} {\bibfield  {journal} {\bibinfo  {journal}
  {Mater. Lett.}\ }\textbf {\bibinfo {volume} {398}},\ \bibinfo {pages}
  {138982} (\bibinfo {year} {2025})}\BibitemShut {NoStop}%
\bibitem [{\citenamefont {Douglass}\ and\ \citenamefont
  {Meservey}(1964)}]{Douglass1964Jul}%
  \BibitemOpen
  \bibfield  {author} {\bibinfo {author} {\bibfnamefont {D.~H.}\ \bibnamefont
  {Douglass}}\ and\ \bibinfo {author} {\bibfnamefont {R.}~\bibnamefont
  {Meservey}},\ }\href {\doibase 10.1103/PhysRev.135.A19} {\bibfield  {journal}
  {\bibinfo  {journal} {Phys. Rev.}\ }\textbf {\bibinfo {volume} {135}},\
  \bibinfo {pages} {A19} (\bibinfo {year} {1964})}\BibitemShut {NoStop}%
\bibitem [{\citenamefont {Strongin}\ \emph {et~al.}(1965)\citenamefont
  {Strongin}, \citenamefont {Kammerer},\ and\ \citenamefont
  {Paskin}}]{Strongin1965Jun}%
  \BibitemOpen
  \bibfield  {author} {\bibinfo {author} {\bibfnamefont {M.}~\bibnamefont
  {Strongin}}, \bibinfo {author} {\bibfnamefont {O.~F.}\ \bibnamefont
  {Kammerer}}, \ and\ \bibinfo {author} {\bibfnamefont {A.}~\bibnamefont
  {Paskin}},\ }\href {\doibase 10.1103/PhysRevLett.14.949} {\bibfield
  {journal} {\bibinfo  {journal} {Phys. Rev. Lett.}\ }\textbf {\bibinfo
  {volume} {14}},\ \bibinfo {pages} {949} (\bibinfo {year} {1965})}\BibitemShut
  {NoStop}%
\bibitem [{\citenamefont {Abeles}\ \emph {et~al.}(1966)\citenamefont {Abeles},
  \citenamefont {Cohen},\ and\ \citenamefont {Cullen}}]{Abeles1966Sep}%
  \BibitemOpen
  \bibfield  {author} {\bibinfo {author} {\bibfnamefont {B.}~\bibnamefont
  {Abeles}}, \bibinfo {author} {\bibfnamefont {R.~W.}\ \bibnamefont {Cohen}}, \
  and\ \bibinfo {author} {\bibfnamefont {G.~W.}\ \bibnamefont {Cullen}},\
  }\href {\doibase 10.1103/PhysRevLett.17.632} {\bibfield  {journal} {\bibinfo
  {journal} {Phys. Rev. Lett.}\ }\textbf {\bibinfo {volume} {17}},\ \bibinfo
  {pages} {632} (\bibinfo {year} {1966})}\BibitemShut {NoStop}%
\bibitem [{\citenamefont {Cohen}\ and\ \citenamefont
  {Abeles}(1968)}]{Cohen1968Apr}%
  \BibitemOpen
  \bibfield  {author} {\bibinfo {author} {\bibfnamefont {R.~W.}\ \bibnamefont
  {Cohen}}\ and\ \bibinfo {author} {\bibfnamefont {B.}~\bibnamefont {Abeles}},\
  }\href {\doibase 10.1103/PhysRev.168.444} {\bibfield  {journal} {\bibinfo
  {journal} {Phys. Rev.}\ }\textbf {\bibinfo {volume} {168}},\ \bibinfo {pages}
  {444} (\bibinfo {year} {1968})}\BibitemShut {NoStop}%
\bibitem [{\citenamefont {Strongin}\ \emph {et~al.}(1968)\citenamefont
  {Strongin}, \citenamefont {Kammerer}, \citenamefont {Crow}, \citenamefont
  {Parks}, \citenamefont {Douglass},\ and\ \citenamefont
  {Jensen}}]{Strongin1968Oct}%
  \BibitemOpen
  \bibfield  {author} {\bibinfo {author} {\bibfnamefont {M.}~\bibnamefont
  {Strongin}}, \bibinfo {author} {\bibfnamefont {O.~F.}\ \bibnamefont
  {Kammerer}}, \bibinfo {author} {\bibfnamefont {J.~E.}\ \bibnamefont {Crow}},
  \bibinfo {author} {\bibfnamefont {R.~D.}\ \bibnamefont {Parks}}, \bibinfo
  {author} {\bibfnamefont {D.~H.}\ \bibnamefont {Douglass}}, \ and\ \bibinfo
  {author} {\bibfnamefont {M.~A.}\ \bibnamefont {Jensen}},\ }\href {\doibase
  10.1103/PhysRevLett.21.1320} {\bibfield  {journal} {\bibinfo  {journal}
  {Phys. Rev. Lett.}\ }\textbf {\bibinfo {volume} {21}},\ \bibinfo {pages}
  {1320} (\bibinfo {year} {1968})}\BibitemShut {NoStop}%
\bibitem [{\citenamefont {Chubov}\ \emph {et~al.}(1969)\citenamefont {Chubov},
  \citenamefont {Eremenko},\ and\ \citenamefont
  {Pilipenko}}]{chubov1969dependence}%
  \BibitemOpen
  \bibfield  {author} {\bibinfo {author} {\bibfnamefont {P.}~\bibnamefont
  {Chubov}}, \bibinfo {author} {\bibfnamefont {V.}~\bibnamefont {Eremenko}}, \
  and\ \bibinfo {author} {\bibfnamefont {Y.~A.}\ \bibnamefont {Pilipenko}},\
  }\href@noop {} {\bibfield  {journal} {\bibinfo  {journal} {Sov Phys JETP}\
  }\textbf {\bibinfo {volume} {28}},\ \bibinfo {pages} {389} (\bibinfo {year}
  {1969})}\BibitemShut {NoStop}%
\bibitem [{\citenamefont {Meservey}\ and\ \citenamefont
  {Tedrow}(1971)}]{Meservey1971Jan}%
  \BibitemOpen
  \bibfield  {author} {\bibinfo {author} {\bibfnamefont {R.}~\bibnamefont
  {Meservey}}\ and\ \bibinfo {author} {\bibfnamefont {P.~M.}\ \bibnamefont
  {Tedrow}},\ }\href {\doibase 10.1063/1.1659648} {\bibfield  {journal}
  {\bibinfo  {journal} {J. Appl. Phys.}\ }\textbf {\bibinfo {volume} {42}},\
  \bibinfo {pages} {51} (\bibinfo {year} {1971})}\BibitemShut {NoStop}%
\bibitem [{\citenamefont {Townsend}\ \emph {et~al.}(1972)\citenamefont
  {Townsend}, \citenamefont {Gregory},\ and\ \citenamefont
  {Taylor}}]{Townsend1972Jan}%
  \BibitemOpen
  \bibfield  {author} {\bibinfo {author} {\bibfnamefont {P.}~\bibnamefont
  {Townsend}}, \bibinfo {author} {\bibfnamefont {S.}~\bibnamefont {Gregory}}, \
  and\ \bibinfo {author} {\bibfnamefont {R.~G.}\ \bibnamefont {Taylor}},\
  }\href {\doibase 10.1103/PhysRevB.5.54} {\bibfield  {journal} {\bibinfo
  {journal} {Phys. Rev. B}\ }\textbf {\bibinfo {volume} {5}},\ \bibinfo {pages}
  {54} (\bibinfo {year} {1972})}\BibitemShut {NoStop}%
\bibitem [{\citenamefont {Allen}(1974)}]{Allen1974May}%
  \BibitemOpen
  \bibfield  {author} {\bibinfo {author} {\bibfnamefont {P.~B.}\ \bibnamefont
  {Allen}},\ }\href {\doibase 10.1016/0038-1098(74)90397-4} {\bibfield
  {journal} {\bibinfo  {journal} {Solid State Commun.}\ }\textbf {\bibinfo
  {volume} {14}},\ \bibinfo {pages} {937} (\bibinfo {year} {1974})}\BibitemShut
  {NoStop}%
\bibitem [{\citenamefont {Pettit}\ and\ \citenamefont
  {Silcox}(1976)}]{Pettit1976Apr}%
  \BibitemOpen
  \bibfield  {author} {\bibinfo {author} {\bibfnamefont {R.~B.}\ \bibnamefont
  {Pettit}}\ and\ \bibinfo {author} {\bibfnamefont {J.}~\bibnamefont
  {Silcox}},\ }\href {\doibase 10.1103/PhysRevB.13.2865} {\bibfield  {journal}
  {\bibinfo  {journal} {Phys. Rev. B}\ }\textbf {\bibinfo {volume} {13}},\
  \bibinfo {pages} {2865} (\bibinfo {year} {1976})}\BibitemShut {NoStop}%
\bibitem [{\citenamefont {Feibelman}\ and\ \citenamefont
  {Hamann}(1984)}]{Feibelman1984Jun}%
  \BibitemOpen
  \bibfield  {author} {\bibinfo {author} {\bibfnamefont {P.~J.}\ \bibnamefont
  {Feibelman}}\ and\ \bibinfo {author} {\bibfnamefont {D.~R.}\ \bibnamefont
  {Hamann}},\ }\href {\doibase 10.1103/PhysRevB.29.6463} {\bibfield  {journal}
  {\bibinfo  {journal} {Phys. Rev. B}\ }\textbf {\bibinfo {volume} {29}},\
  \bibinfo {pages} {6463} (\bibinfo {year} {1984})}\BibitemShut {NoStop}%
\bibitem [{\citenamefont {Bose}\ \emph {et~al.}(2009)\citenamefont {Bose},
  \citenamefont {Galande}, \citenamefont {Chockalingam}, \citenamefont
  {Banerjee}, \citenamefont {Raychaudhuri},\ and\ \citenamefont
  {Ayyub}}]{Bose2009Apr}%
  \BibitemOpen
  \bibfield  {author} {\bibinfo {author} {\bibfnamefont {S.}~\bibnamefont
  {Bose}}, \bibinfo {author} {\bibfnamefont {C.}~\bibnamefont {Galande}},
  \bibinfo {author} {\bibfnamefont {S.~P.}\ \bibnamefont {Chockalingam}},
  \bibinfo {author} {\bibfnamefont {R.}~\bibnamefont {Banerjee}}, \bibinfo
  {author} {\bibfnamefont {P.}~\bibnamefont {Raychaudhuri}}, \ and\ \bibinfo
  {author} {\bibfnamefont {P.}~\bibnamefont {Ayyub}},\ }\href {\doibase
  10.1088/0953-8984/21/20/205702} {\bibfield  {journal} {\bibinfo  {journal}
  {J. Phys.: Condens. Matter}\ }\textbf {\bibinfo {volume} {21}},\ \bibinfo
  {pages} {205702} (\bibinfo {year} {2009})}\BibitemShut {NoStop}%
\bibitem [{\citenamefont {Walmsley}\ \emph {et~al.}(2011)\citenamefont
  {Walmsley}, \citenamefont {Campbell},\ and\ \citenamefont
  {Dynes}}]{Walmsley2011Feb}%
  \BibitemOpen
  \bibfield  {author} {\bibinfo {author} {\bibfnamefont {D.~G.}\ \bibnamefont
  {Walmsley}}, \bibinfo {author} {\bibfnamefont {C.~K.}\ \bibnamefont
  {Campbell}}, \ and\ \bibinfo {author} {\bibfnamefont {R.~C.}\ \bibnamefont
  {Dynes}},\ }\href {\doibase 10.1139/p68-141} {\bibfield  {journal} {\bibinfo
  {journal} {Can. J. Phys.}\ } (\bibinfo {year} {2011}),\
  10.1139/p68-141}\BibitemShut {NoStop}%
\bibitem [{\citenamefont {Cherney}\ and\ \citenamefont
  {Shewchun}(2011)}]{Cherney2011Feb}%
  \BibitemOpen
  \bibfield  {author} {\bibinfo {author} {\bibfnamefont {O.~A.~E.}\
  \bibnamefont {Cherney}}\ and\ \bibinfo {author} {\bibfnamefont
  {J.}~\bibnamefont {Shewchun}},\ }\href {\doibase 10.1139/p69-138} {\bibfield
  {journal} {\bibinfo  {journal} {Can. J. Phys.}\ } (\bibinfo {year} {2011}),\
  10.1139/p69-138}\BibitemShut {NoStop}%
\bibitem [{\citenamefont {Lozano}\ \emph {et~al.}(2019)\citenamefont {Lozano},
  \citenamefont {Couet}, \citenamefont {Petermann}, \citenamefont {Hamoir},
  \citenamefont {Jochum}, \citenamefont {Picot}, \citenamefont
  {Men{\ifmmode\acute{e}\else\'{e}\fi}ndez}, \citenamefont {Houben},
  \citenamefont {Joly}, \citenamefont {Antohe}, \citenamefont {Hu},
  \citenamefont {Leu}, \citenamefont {Alatas}, \citenamefont {Said},
  \citenamefont {Roelants}, \citenamefont {Partoens}, \citenamefont
  {Milo{\ifmmode\check{s}\else\v{s}\fi}evi{\ifmmode\acute{c}\else\'{c}\fi}},
  \citenamefont {Peeters}, \citenamefont {Piraux}, \citenamefont {Van~de
  Vondel}, \citenamefont {Vantomme}, \citenamefont {Temst},\ and\ \citenamefont
  {Van~Bael}}]{Lozano2019Feb}%
  \BibitemOpen
  \bibfield  {author} {\bibinfo {author} {\bibfnamefont {D.~P.}\ \bibnamefont
  {Lozano}}, \bibinfo {author} {\bibfnamefont {S.}~\bibnamefont {Couet}},
  \bibinfo {author} {\bibfnamefont {C.}~\bibnamefont {Petermann}}, \bibinfo
  {author} {\bibfnamefont {G.}~\bibnamefont {Hamoir}}, \bibinfo {author}
  {\bibfnamefont {J.~K.}\ \bibnamefont {Jochum}}, \bibinfo {author}
  {\bibfnamefont {T.}~\bibnamefont {Picot}}, \bibinfo {author} {\bibfnamefont
  {E.}~\bibnamefont {Men{\ifmmode\acute{e}\else\'{e}\fi}ndez}}, \bibinfo
  {author} {\bibfnamefont {K.}~\bibnamefont {Houben}}, \bibinfo {author}
  {\bibfnamefont {V.}~\bibnamefont {Joly}}, \bibinfo {author} {\bibfnamefont
  {V.~A.}\ \bibnamefont {Antohe}}, \bibinfo {author} {\bibfnamefont {M.~Y.}\
  \bibnamefont {Hu}}, \bibinfo {author} {\bibfnamefont {B.~M.}\ \bibnamefont
  {Leu}}, \bibinfo {author} {\bibfnamefont {A.}~\bibnamefont {Alatas}},
  \bibinfo {author} {\bibfnamefont {A.~H.}\ \bibnamefont {Said}}, \bibinfo
  {author} {\bibfnamefont {S.}~\bibnamefont {Roelants}}, \bibinfo {author}
  {\bibfnamefont {B.}~\bibnamefont {Partoens}}, \bibinfo {author}
  {\bibfnamefont {M.~V.}\ \bibnamefont
  {Milo{\ifmmode\check{s}\else\v{s}\fi}evi{\ifmmode\acute{c}\else\'{c}\fi}}},
  \bibinfo {author} {\bibfnamefont {F.~M.}\ \bibnamefont {Peeters}}, \bibinfo
  {author} {\bibfnamefont {L.}~\bibnamefont {Piraux}}, \bibinfo {author}
  {\bibfnamefont {J.}~\bibnamefont {Van~de Vondel}}, \bibinfo {author}
  {\bibfnamefont {A.}~\bibnamefont {Vantomme}}, \bibinfo {author}
  {\bibfnamefont {K.}~\bibnamefont {Temst}}, \ and\ \bibinfo {author}
  {\bibfnamefont {M.~J.}\ \bibnamefont {Van~Bael}},\ }\href {\doibase
  10.1103/PhysRevB.99.064512} {\bibfield  {journal} {\bibinfo  {journal} {Phys.
  Rev. B}\ }\textbf {\bibinfo {volume} {99}},\ \bibinfo {pages} {064512}
  (\bibinfo {year} {2019})}\BibitemShut {NoStop}%
\bibitem [{\citenamefont {Houben}\ \emph {et~al.}(2020)\citenamefont {Houben},
  \citenamefont {Jochum}, \citenamefont {Couet}, \citenamefont
  {Men{\ifmmode\acute{e}\else\'{e}\fi}ndez}, \citenamefont {Picot},
  \citenamefont {Hu}, \citenamefont {Zhao}, \citenamefont {Alp}, \citenamefont
  {Vantomme}, \citenamefont {Temst},\ and\ \citenamefont
  {Van~Bael}}]{Houben2020Mar}%
  \BibitemOpen
  \bibfield  {author} {\bibinfo {author} {\bibfnamefont {K.}~\bibnamefont
  {Houben}}, \bibinfo {author} {\bibfnamefont {J.~K.}\ \bibnamefont {Jochum}},
  \bibinfo {author} {\bibfnamefont {S.}~\bibnamefont {Couet}}, \bibinfo
  {author} {\bibfnamefont {E.}~\bibnamefont
  {Men{\ifmmode\acute{e}\else\'{e}\fi}ndez}}, \bibinfo {author} {\bibfnamefont
  {T.}~\bibnamefont {Picot}}, \bibinfo {author} {\bibfnamefont {M.~Y.}\
  \bibnamefont {Hu}}, \bibinfo {author} {\bibfnamefont {J.~Y.}\ \bibnamefont
  {Zhao}}, \bibinfo {author} {\bibfnamefont {E.~E.}\ \bibnamefont {Alp}},
  \bibinfo {author} {\bibfnamefont {A.}~\bibnamefont {Vantomme}}, \bibinfo
  {author} {\bibfnamefont {K.}~\bibnamefont {Temst}}, \ and\ \bibinfo {author}
  {\bibfnamefont {M.~J.}\ \bibnamefont {Van~Bael}},\ }\href {\doibase
  10.1038/s41598-020-62617-4} {\bibfield  {journal} {\bibinfo  {journal} {Sci.
  Rep.}\ }\textbf {\bibinfo {volume} {10}},\ \bibinfo {pages} {1} (\bibinfo
  {year} {2020})}\BibitemShut {NoStop}%
\bibitem [{\citenamefont {Guo}\ \emph {et~al.}(2004)\citenamefont {Guo},
  \citenamefont {Zhang}, \citenamefont {Bao}, \citenamefont {Han},
  \citenamefont {Tang}, \citenamefont {Zhang}, \citenamefont {Zhu},
  \citenamefont {Wang}, \citenamefont {Niu}, \citenamefont {Qiu}, \citenamefont
  {Jia}, \citenamefont {Zhao},\ and\ \citenamefont {Xue}}]{Guo2004Dec}%
  \BibitemOpen
  \bibfield  {author} {\bibinfo {author} {\bibfnamefont {Y.}~\bibnamefont
  {Guo}}, \bibinfo {author} {\bibfnamefont {Y.-F.}\ \bibnamefont {Zhang}},
  \bibinfo {author} {\bibfnamefont {X.-Y.}\ \bibnamefont {Bao}}, \bibinfo
  {author} {\bibfnamefont {T.-Z.}\ \bibnamefont {Han}}, \bibinfo {author}
  {\bibfnamefont {Z.}~\bibnamefont {Tang}}, \bibinfo {author} {\bibfnamefont
  {L.-X.}\ \bibnamefont {Zhang}}, \bibinfo {author} {\bibfnamefont {W.-G.}\
  \bibnamefont {Zhu}}, \bibinfo {author} {\bibfnamefont {E.~G.}\ \bibnamefont
  {Wang}}, \bibinfo {author} {\bibfnamefont {Q.}~\bibnamefont {Niu}}, \bibinfo
  {author} {\bibfnamefont {Z.~Q.}\ \bibnamefont {Qiu}}, \bibinfo {author}
  {\bibfnamefont {J.-F.}\ \bibnamefont {Jia}}, \bibinfo {author} {\bibfnamefont
  {Z.-X.}\ \bibnamefont {Zhao}}, \ and\ \bibinfo {author} {\bibfnamefont
  {Q.-K.}\ \bibnamefont {Xue}},\ }\href {\doibase 10.1126/science.1105130}
  {\bibfield  {journal} {\bibinfo  {journal} {Science}\ }\textbf {\bibinfo
  {volume} {306}},\ \bibinfo {pages} {1915} (\bibinfo {year}
  {2004})}\BibitemShut {NoStop}%
\bibitem [{\citenamefont {Shanenko}\ \emph {et~al.}(2006)\citenamefont
  {Shanenko}, \citenamefont {Croitoru},\ and\ \citenamefont
  {Peeters}}]{Shanenko2006Sep}%
  \BibitemOpen
  \bibfield  {author} {\bibinfo {author} {\bibfnamefont {A.~A.}\ \bibnamefont
  {Shanenko}}, \bibinfo {author} {\bibfnamefont {M.~D.}\ \bibnamefont
  {Croitoru}}, \ and\ \bibinfo {author} {\bibfnamefont {F.~M.}\ \bibnamefont
  {Peeters}},\ }\href {\doibase 10.1209/epl/i2006-10274-6} {\bibfield
  {journal} {\bibinfo  {journal} {Europhys. Lett.}\ }\textbf {\bibinfo {volume}
  {76}},\ \bibinfo {pages} {498} (\bibinfo {year} {2006})}\BibitemShut
  {NoStop}%
\bibitem [{\citenamefont {Adams}\ \emph {et~al.}(2017)\citenamefont {Adams},
  \citenamefont {Nam}, \citenamefont {Shih},\ and\ \citenamefont
  {Catelani}}]{Adams2017Mar}%
  \BibitemOpen
  \bibfield  {author} {\bibinfo {author} {\bibfnamefont {P.~W.}\ \bibnamefont
  {Adams}}, \bibinfo {author} {\bibfnamefont {H.}~\bibnamefont {Nam}}, \bibinfo
  {author} {\bibfnamefont {C.~K.}\ \bibnamefont {Shih}}, \ and\ \bibinfo
  {author} {\bibfnamefont {G.}~\bibnamefont {Catelani}},\ }\href {\doibase
  10.1103/PhysRevB.95.094520} {\bibfield  {journal} {\bibinfo  {journal} {Phys.
  Rev. B}\ }\textbf {\bibinfo {volume} {95}},\ \bibinfo {pages} {094520}
  (\bibinfo {year} {2017})}\BibitemShut {NoStop}%
\bibitem [{\citenamefont {Nguyen}\ \emph {et~al.}(2019)\citenamefont {Nguyen},
  \citenamefont {Wei},\ and\ \citenamefont {Chou}}]{Nguyen2019May}%
  \BibitemOpen
  \bibfield  {author} {\bibinfo {author} {\bibfnamefont {D.-L.}\ \bibnamefont
  {Nguyen}}, \bibinfo {author} {\bibfnamefont {C.-M.}\ \bibnamefont {Wei}}, \
  and\ \bibinfo {author} {\bibfnamefont {M.-Y.}\ \bibnamefont {Chou}},\ }\href
  {\doibase 10.1103/PhysRevB.99.205401} {\bibfield  {journal} {\bibinfo
  {journal} {Phys. Rev. B}\ }\textbf {\bibinfo {volume} {99}},\ \bibinfo
  {pages} {205401} (\bibinfo {year} {2019})}\BibitemShut {NoStop}%
\bibitem [{\citenamefont {Yu}\ \emph {et~al.}(2022)\citenamefont {Yu},
  \citenamefont {Yang}, \citenamefont {Baggioli}, \citenamefont {Phillips},
  \citenamefont {Zaccone}, \citenamefont {Zhang}, \citenamefont {Kajimoto},
  \citenamefont {Nakamura}, \citenamefont {Yu},\ and\ \citenamefont
  {Hong}}]{Yu2022Jun}%
  \BibitemOpen
  \bibfield  {author} {\bibinfo {author} {\bibfnamefont {Y.}~\bibnamefont
  {Yu}}, \bibinfo {author} {\bibfnamefont {C.}~\bibnamefont {Yang}}, \bibinfo
  {author} {\bibfnamefont {M.}~\bibnamefont {Baggioli}}, \bibinfo {author}
  {\bibfnamefont {A.~E.}\ \bibnamefont {Phillips}}, \bibinfo {author}
  {\bibfnamefont {A.}~\bibnamefont {Zaccone}}, \bibinfo {author} {\bibfnamefont
  {L.}~\bibnamefont {Zhang}}, \bibinfo {author} {\bibfnamefont
  {R.}~\bibnamefont {Kajimoto}}, \bibinfo {author} {\bibfnamefont
  {M.}~\bibnamefont {Nakamura}}, \bibinfo {author} {\bibfnamefont
  {D.}~\bibnamefont {Yu}}, \ and\ \bibinfo {author} {\bibfnamefont
  {L.}~\bibnamefont {Hong}},\ }\href {\doibase 10.1038/s41467-022-31349-6}
  {\bibfield  {journal} {\bibinfo  {journal} {Nat. Commun.}\ }\textbf {\bibinfo
  {volume} {13}},\ \bibinfo {pages} {1} (\bibinfo {year} {2022})}\BibitemShut
  {NoStop}%
\bibitem [{\citenamefont {van Weerdenburg}\ \emph {et~al.}(2023)\citenamefont
  {van Weerdenburg}, \citenamefont {Kamlapure}, \citenamefont {Fyhn},
  \citenamefont {Huang}, \citenamefont {van Mullekom}, \citenamefont
  {Steinbrecher}, \citenamefont {Krogstrup}, \citenamefont {Linder},\ and\
  \citenamefont {Khajetoorians}}]{vanWeerdenburg2023Mar}%
  \BibitemOpen
  \bibfield  {author} {\bibinfo {author} {\bibfnamefont {W.~M.~J.}\
  \bibnamefont {van Weerdenburg}}, \bibinfo {author} {\bibfnamefont
  {A.}~\bibnamefont {Kamlapure}}, \bibinfo {author} {\bibfnamefont {E.~H.}\
  \bibnamefont {Fyhn}}, \bibinfo {author} {\bibfnamefont {X.}~\bibnamefont
  {Huang}}, \bibinfo {author} {\bibfnamefont {N.~P.~E.}\ \bibnamefont {van
  Mullekom}}, \bibinfo {author} {\bibfnamefont {M.}~\bibnamefont
  {Steinbrecher}}, \bibinfo {author} {\bibfnamefont {P.}~\bibnamefont
  {Krogstrup}}, \bibinfo {author} {\bibfnamefont {J.}~\bibnamefont {Linder}}, \
  and\ \bibinfo {author} {\bibfnamefont {A.~A.}\ \bibnamefont
  {Khajetoorians}},\ }\href {\doibase 10.1126/sciadv.adf5500} {\bibfield
  {journal} {\bibinfo  {journal} {Sci. Adv.}\ }\textbf {\bibinfo {volume} {9}}
  (\bibinfo {year} {2023}),\ 10.1126/sciadv.adf5500}\BibitemShut {NoStop}%
\bibitem [{\citenamefont {Allen}(2024)}]{Allen2024Jun}%
  \BibitemOpen
  \bibfield  {author} {\bibinfo {author} {\bibfnamefont {P.~B.}\ \bibnamefont
  {Allen}},\ }\href {\doibase 10.48550/arXiv.2406.16197} {\bibfield  {journal}
  {\bibinfo  {journal} {ArXiv}\ } (\bibinfo {year} {2024}),\
  10.48550/arXiv.2406.16197},\ \Eprint {http://arxiv.org/abs/2406.16197}
  {2406.16197} \BibitemShut {NoStop}%
\bibitem [{\citenamefont {Deshpande}\ \emph {et~al.}(2025)\citenamefont
  {Deshpande}, \citenamefont {Pusskeiler}, \citenamefont {Prange},
  \citenamefont {Rogge}, \citenamefont {Dressel},\ and\ \citenamefont
  {Scheffler}}]{Deshpande2025Jan}%
  \BibitemOpen
  \bibfield  {author} {\bibinfo {author} {\bibfnamefont {A.}~\bibnamefont
  {Deshpande}}, \bibinfo {author} {\bibfnamefont {J.}~\bibnamefont
  {Pusskeiler}}, \bibinfo {author} {\bibfnamefont {C.}~\bibnamefont {Prange}},
  \bibinfo {author} {\bibfnamefont {U.}~\bibnamefont {Rogge}}, \bibinfo
  {author} {\bibfnamefont {M.}~\bibnamefont {Dressel}}, \ and\ \bibinfo
  {author} {\bibfnamefont {M.}~\bibnamefont {Scheffler}},\ }\href {\doibase
  10.1063/5.0250146} {\bibfield  {journal} {\bibinfo  {journal} {J. Appl.
  Phys.}\ }\textbf {\bibinfo {volume} {137}} (\bibinfo {year} {2025}),\
  10.1063/5.0250146}\BibitemShut {NoStop}%
\bibitem [{\citenamefont {Ginzburg}\ and\ \citenamefont
  {Kirzhnits}(1964)}]{ginzburg1964superconductivity}%
  \BibitemOpen
  \bibfield  {author} {\bibinfo {author} {\bibfnamefont {V.}~\bibnamefont
  {Ginzburg}}\ and\ \bibinfo {author} {\bibfnamefont {D.}~\bibnamefont
  {Kirzhnits}},\ }\href {http://jetp.ras.ru/cgi-bin/e/index/e/19/1/p269?a=list}
  {\bibfield  {journal} {\bibinfo  {journal} {Zh. Eksp. Teor. Fiz}\ }\textbf
  {\bibinfo {volume} {46}},\ \bibinfo {pages} {397} (\bibinfo {year}
  {1964})}\BibitemShut {NoStop}%
\bibitem [{\citenamefont {Ginzburg}(1964)}]{Ginzburg1964Nov}%
  \BibitemOpen
  \bibfield  {author} {\bibinfo {author} {\bibfnamefont {V.~L.}\ \bibnamefont
  {Ginzburg}},\ }\href {\doibase 10.1016/0031-9163(64)90672-9} {\bibfield
  {journal} {\bibinfo  {journal} {Physics Letters}\ }\textbf {\bibinfo {volume}
  {13}},\ \bibinfo {pages} {101} (\bibinfo {year} {1964})}\BibitemShut
  {NoStop}%
\bibitem [{\citenamefont {Little}(1964)}]{Little1964Jun}%
  \BibitemOpen
  \bibfield  {author} {\bibinfo {author} {\bibfnamefont {W.~A.}\ \bibnamefont
  {Little}},\ }\href {\doibase 10.1103/PhysRev.134.A1416} {\bibfield  {journal}
  {\bibinfo  {journal} {Phys. Rev.}\ }\textbf {\bibinfo {volume} {134}},\
  \bibinfo {pages} {A1416} (\bibinfo {year} {1964})}\BibitemShut {NoStop}%
\bibitem [{\citenamefont {Little}(1987)}]{Little1987}%
  \BibitemOpen
  \bibfield  {author} {\bibinfo {author} {\bibfnamefont {W.~A.}\ \bibnamefont
  {Little}},\ }in\ \href {\doibase 10.1007/978-1-4613-1937-5_37} {\emph
  {\bibinfo {booktitle} {{Novel Superconductivity}}}}\ (\bibinfo  {publisher}
  {Springer, Boston, MA},\ \bibinfo {address} {Boston, MA, USA},\ \bibinfo
  {year} {1987})\ pp.\ \bibinfo {pages} {341--353}\BibitemShut {NoStop}%
\bibitem [{\citenamefont {Little}(1996)}]{Little1996}%
  \BibitemOpen
  \bibfield  {author} {\bibinfo {author} {\bibfnamefont {W.~A.}\ \bibnamefont
  {Little}},\ }in\ \href {\doibase 10.1007/978-1-4613-0411-1_5} {\emph
  {\bibinfo {booktitle} {{From High-Temperature Superconductivity to
  Microminiature Refrigeration}}}}\ (\bibinfo  {publisher} {Springer, Boston,
  MA},\ \bibinfo {address} {Boston, MA, USA},\ \bibinfo {year} {1996})\ pp.\
  \bibinfo {pages} {35--43}\BibitemShut {NoStop}%
\bibitem [{\citenamefont {Ginzburg}(1989)}]{Ginzburg1989Jan}%
  \BibitemOpen
  \bibfield  {author} {\bibinfo {author} {\bibfnamefont {V.~L.}\ \bibnamefont
  {Ginzburg}},\ }in\ \href {\doibase 10.1016/S0079-6417(08)60040-2} {\emph
  {\bibinfo {booktitle} {{Progress in Low Temperature Physics}}}},\
  Vol.~\bibinfo {volume} {12}\ (\bibinfo  {publisher} {Elsevier},\ \bibinfo
  {address} {Walthm, MA, USA},\ \bibinfo {year} {1989})\ pp.\ \bibinfo {pages}
  {1--44}\BibitemShut {NoStop}%
\bibitem [{\citenamefont {Baggioli}\ \emph {et~al.}(2020)\citenamefont
  {Baggioli}, \citenamefont {Setty},\ and\ \citenamefont
  {Zaccone}}]{Baggioli2020Jun}%
  \BibitemOpen
  \bibfield  {author} {\bibinfo {author} {\bibfnamefont {M.}~\bibnamefont
  {Baggioli}}, \bibinfo {author} {\bibfnamefont {C.}~\bibnamefont {Setty}}, \
  and\ \bibinfo {author} {\bibfnamefont {A.}~\bibnamefont {Zaccone}},\ }\href
  {\doibase 10.1103/PhysRevB.101.214502} {\bibfield  {journal} {\bibinfo
  {journal} {Phys. Rev. B}\ }\textbf {\bibinfo {volume} {101}},\ \bibinfo
  {pages} {214502} (\bibinfo {year} {2020})}\BibitemShut {NoStop}%
\bibitem [{\citenamefont {Travaglino}\ and\ \citenamefont
  {Zaccone}(2023)}]{Travaglino2023Jan}%
  \BibitemOpen
  \bibfield  {author} {\bibinfo {author} {\bibfnamefont {R.}~\bibnamefont
  {Travaglino}}\ and\ \bibinfo {author} {\bibfnamefont {A.}~\bibnamefont
  {Zaccone}},\ }\href {\doibase 10.1063/5.0132820} {\bibfield  {journal}
  {\bibinfo  {journal} {J. Appl. Phys.}\ }\textbf {\bibinfo {volume} {133}}
  (\bibinfo {year} {2023}),\ 10.1063/5.0132820}\BibitemShut {NoStop}%
\bibitem [{\citenamefont {Zaccone}\ and\ \citenamefont
  {Fomin}(2024)}]{Zaccone2024Apr}%
  \BibitemOpen
  \bibfield  {author} {\bibinfo {author} {\bibfnamefont {A.}~\bibnamefont
  {Zaccone}}\ and\ \bibinfo {author} {\bibfnamefont {V.~M.}\ \bibnamefont
  {Fomin}},\ }\href {\doibase 10.1103/PhysRevB.109.144520} {\bibfield
  {journal} {\bibinfo  {journal} {Phys. Rev. B}\ }\textbf {\bibinfo {volume}
  {109}},\ \bibinfo {pages} {144520} (\bibinfo {year} {2024})}\BibitemShut
  {NoStop}%
\bibitem [{\citenamefont {Zaccone}(2025)}]{Zaccone2025May}%
  \BibitemOpen
  \bibfield  {author} {\bibinfo {author} {\bibfnamefont {A.}~\bibnamefont
  {Zaccone}},\ }\href {\doibase 10.1088/2515-7639/adc83f} {\bibfield  {journal}
  {\bibinfo  {journal} {J. Phys.: Mater.}\ }\textbf {\bibinfo {volume} {8}},\
  \bibinfo {pages} {031001} (\bibinfo {year} {2025})}\BibitemShut {NoStop}%
\bibitem [{\citenamefont {Phillips}(1987)}]{Phillips1987Dec}%
  \BibitemOpen
  \bibfield  {author} {\bibinfo {author} {\bibfnamefont {W.~A.}\ \bibnamefont
  {Phillips}},\ }\href {\doibase 10.1088/0034-4885/50/12/003} {\bibfield
  {journal} {\bibinfo  {journal} {Rep. Prog. Phys.}\ }\textbf {\bibinfo
  {volume} {50}},\ \bibinfo {pages} {1657} (\bibinfo {year}
  {1987})}\BibitemShut {NoStop}%
\bibitem [{\citenamefont {W.~Anderson}\ \emph {et~al.}(1972)\citenamefont
  {W.~Anderson}, \citenamefont {Halperin},\ and\ \citenamefont
  {Varma}}]{W.Anderson1972Jan}%
  \BibitemOpen
  \bibfield  {author} {\bibinfo {author} {\bibfnamefont {P.}~\bibnamefont
  {W.~Anderson}}, \bibinfo {author} {\bibfnamefont {B.~I.}\ \bibnamefont
  {Halperin}}, \ and\ \bibinfo {author} {\bibfnamefont {C.~M.}\ \bibnamefont
  {Varma}},\ }\href
  {https://www.tandfonline.com/doi/abs/10.1080/14786437208229210} {\bibfield
  {journal} {\bibinfo  {journal} {Philos. Mag.}\ } (\bibinfo {year}
  {1972})}\BibitemShut {NoStop}%
\bibitem [{\citenamefont {Wisbey}\ \emph {et~al.}(2010)\citenamefont {Wisbey},
  \citenamefont {Gao}, \citenamefont {Vissers}, \citenamefont {da~Silva},
  \citenamefont {Kline}, \citenamefont {Vale},\ and\ \citenamefont
  {Pappas}}]{Wisbey2010Nov}%
  \BibitemOpen
  \bibfield  {author} {\bibinfo {author} {\bibfnamefont {D.~S.}\ \bibnamefont
  {Wisbey}}, \bibinfo {author} {\bibfnamefont {J.}~\bibnamefont {Gao}},
  \bibinfo {author} {\bibfnamefont {M.~R.}\ \bibnamefont {Vissers}}, \bibinfo
  {author} {\bibfnamefont {F.~C.~S.}\ \bibnamefont {da~Silva}}, \bibinfo
  {author} {\bibfnamefont {J.~S.}\ \bibnamefont {Kline}}, \bibinfo {author}
  {\bibfnamefont {L.}~\bibnamefont {Vale}}, \ and\ \bibinfo {author}
  {\bibfnamefont {D.~P.}\ \bibnamefont {Pappas}},\ }\href {\doibase
  10.1063/1.3499608} {\bibfield  {journal} {\bibinfo  {journal} {J. Appl.
  Phys.}\ }\textbf {\bibinfo {volume} {108}},\ \bibinfo {pages} {093918}
  (\bibinfo {year} {2010})}\BibitemShut {NoStop}%
\bibitem [{\citenamefont {Wenner}\ \emph {et~al.}(2011)\citenamefont {Wenner},
  \citenamefont {Barends}, \citenamefont {Bialczak}, \citenamefont {Chen},
  \citenamefont {Kelly}, \citenamefont {Lucero}, \citenamefont {Mariantoni},
  \citenamefont {Megrant}, \citenamefont {O{'}Malley}, \citenamefont {Sank},
  \citenamefont {Vainsencher}, \citenamefont {Wang}, \citenamefont {White},
  \citenamefont {Yin}, \citenamefont {Zhao}, \citenamefont {Cleland},\ and\
  \citenamefont {Martinis}}]{Wenner2011Sep}%
  \BibitemOpen
  \bibfield  {author} {\bibinfo {author} {\bibfnamefont {J.}~\bibnamefont
  {Wenner}}, \bibinfo {author} {\bibfnamefont {R.}~\bibnamefont {Barends}},
  \bibinfo {author} {\bibfnamefont {R.~C.}\ \bibnamefont {Bialczak}}, \bibinfo
  {author} {\bibfnamefont {Y.}~\bibnamefont {Chen}}, \bibinfo {author}
  {\bibfnamefont {J.}~\bibnamefont {Kelly}}, \bibinfo {author} {\bibfnamefont
  {E.}~\bibnamefont {Lucero}}, \bibinfo {author} {\bibfnamefont
  {M.}~\bibnamefont {Mariantoni}}, \bibinfo {author} {\bibfnamefont
  {A.}~\bibnamefont {Megrant}}, \bibinfo {author} {\bibfnamefont {P.~J.~J.}\
  \bibnamefont {O{'}Malley}}, \bibinfo {author} {\bibfnamefont
  {D.}~\bibnamefont {Sank}}, \bibinfo {author} {\bibfnamefont {A.}~\bibnamefont
  {Vainsencher}}, \bibinfo {author} {\bibfnamefont {H.}~\bibnamefont {Wang}},
  \bibinfo {author} {\bibfnamefont {T.~C.}\ \bibnamefont {White}}, \bibinfo
  {author} {\bibfnamefont {Y.}~\bibnamefont {Yin}}, \bibinfo {author}
  {\bibfnamefont {J.}~\bibnamefont {Zhao}}, \bibinfo {author} {\bibfnamefont
  {A.~N.}\ \bibnamefont {Cleland}}, \ and\ \bibinfo {author} {\bibfnamefont
  {J.~M.}\ \bibnamefont {Martinis}},\ }\href {\doibase 10.1063/1.3637047}
  {\bibfield  {journal} {\bibinfo  {journal} {Appl. Phys. Lett.}\ }\textbf
  {\bibinfo {volume} {99}},\ \bibinfo {pages} {113513} (\bibinfo {year}
  {2011})}\BibitemShut {NoStop}%
\bibitem [{\citenamefont {DuBois}\ \emph {et~al.}(2013)\citenamefont {DuBois},
  \citenamefont {Per}, \citenamefont {Russo},\ and\ \citenamefont
  {Cole}}]{DuBois2013Feb}%
  \BibitemOpen
  \bibfield  {author} {\bibinfo {author} {\bibfnamefont {T.~C.}\ \bibnamefont
  {DuBois}}, \bibinfo {author} {\bibfnamefont {M.~C.}\ \bibnamefont {Per}},
  \bibinfo {author} {\bibfnamefont {S.~P.}\ \bibnamefont {Russo}}, \ and\
  \bibinfo {author} {\bibfnamefont {J.~H.}\ \bibnamefont {Cole}},\ }\href
  {\doibase 10.1103/PhysRevLett.110.077002} {\bibfield  {journal} {\bibinfo
  {journal} {Phys. Rev. Lett.}\ }\textbf {\bibinfo {volume} {110}},\ \bibinfo
  {pages} {077002} (\bibinfo {year} {2013})}\BibitemShut {NoStop}%
\bibitem [{\citenamefont {Earnest}\ \emph {et~al.}(2018)\citenamefont
  {Earnest}, \citenamefont {B{\ifmmode\acute{e}\else\'{e}\fi}janin},
  \citenamefont {McConkey}, \citenamefont {Peters}, \citenamefont {Korinek},
  \citenamefont {Yuan},\ and\ \citenamefont {Mariantoni}}]{Earnest2018Nov}%
  \BibitemOpen
  \bibfield  {author} {\bibinfo {author} {\bibfnamefont {C.~T.}\ \bibnamefont
  {Earnest}}, \bibinfo {author} {\bibfnamefont {J.~H.}\ \bibnamefont
  {B{\ifmmode\acute{e}\else\'{e}\fi}janin}}, \bibinfo {author} {\bibfnamefont
  {T.~G.}\ \bibnamefont {McConkey}}, \bibinfo {author} {\bibfnamefont {E.~A.}\
  \bibnamefont {Peters}}, \bibinfo {author} {\bibfnamefont {A.}~\bibnamefont
  {Korinek}}, \bibinfo {author} {\bibfnamefont {H.}~\bibnamefont {Yuan}}, \
  and\ \bibinfo {author} {\bibfnamefont {M.}~\bibnamefont {Mariantoni}},\
  }\href {\doibase 10.1088/1361-6668/aae548} {\bibfield  {journal} {\bibinfo
  {journal} {Supercond. Sci. Technol.}\ }\textbf {\bibinfo {volume} {31}},\
  \bibinfo {pages} {125013} (\bibinfo {year} {2018})}\BibitemShut {NoStop}%
\bibitem [{\citenamefont {M{\ifmmode\ddot{u}\else\"{u}\fi}ller}\ \emph
  {et~al.}(2019)\citenamefont {M{\ifmmode\ddot{u}\else\"{u}\fi}ller},
  \citenamefont {Cole},\ and\ \citenamefont {Lisenfeld}}]{Muller2019Oct}%
  \BibitemOpen
  \bibfield  {author} {\bibinfo {author} {\bibfnamefont {C.}~\bibnamefont
  {M{\ifmmode\ddot{u}\else\"{u}\fi}ller}}, \bibinfo {author} {\bibfnamefont
  {J.~H.}\ \bibnamefont {Cole}}, \ and\ \bibinfo {author} {\bibfnamefont
  {J.}~\bibnamefont {Lisenfeld}},\ }\href {\doibase 10.1088/1361-6633/ab3a7e}
  {\bibfield  {journal} {\bibinfo  {journal} {Rep. Prog. Phys.}\ }\textbf
  {\bibinfo {volume} {82}},\ \bibinfo {pages} {124501} (\bibinfo {year}
  {2019})}\BibitemShut {NoStop}%
\bibitem [{\citenamefont {Chayanun}\ \emph {et~al.}(2024)\citenamefont
  {Chayanun}, \citenamefont
  {Bizn{\ifmmode\acute{a}\else\'{a}\fi}rov{\ifmmode\acute{a}\else\'{a}\fi}},
  \citenamefont {Zeng}, \citenamefont {Malmberg}, \citenamefont {Nylander},
  \citenamefont {Osman}, \citenamefont {Rommel}, \citenamefont {Tam},
  \citenamefont {Olsson}, \citenamefont {Delsing}, \citenamefont {Yurgens},
  \citenamefont {Bylander},\ and\ \citenamefont
  {Fadavi~Roudsari}}]{Chayanun2024Jun}%
  \BibitemOpen
  \bibfield  {author} {\bibinfo {author} {\bibfnamefont {L.}~\bibnamefont
  {Chayanun}}, \bibinfo {author} {\bibfnamefont {J.}~\bibnamefont
  {Bizn{\ifmmode\acute{a}\else\'{a}\fi}rov{\ifmmode\acute{a}\else\'{a}\fi}}},
  \bibinfo {author} {\bibfnamefont {L.}~\bibnamefont {Zeng}}, \bibinfo {author}
  {\bibfnamefont {P.}~\bibnamefont {Malmberg}}, \bibinfo {author}
  {\bibfnamefont {A.}~\bibnamefont {Nylander}}, \bibinfo {author}
  {\bibfnamefont {A.}~\bibnamefont {Osman}}, \bibinfo {author} {\bibfnamefont
  {M.}~\bibnamefont {Rommel}}, \bibinfo {author} {\bibfnamefont {P.~L.}\
  \bibnamefont {Tam}}, \bibinfo {author} {\bibfnamefont {E.}~\bibnamefont
  {Olsson}}, \bibinfo {author} {\bibfnamefont {P.}~\bibnamefont {Delsing}},
  \bibinfo {author} {\bibfnamefont {A.}~\bibnamefont {Yurgens}}, \bibinfo
  {author} {\bibfnamefont {J.}~\bibnamefont {Bylander}}, \ and\ \bibinfo
  {author} {\bibfnamefont {A.}~\bibnamefont {Fadavi~Roudsari}},\ }\href
  {\doibase 10.1063/5.0208140} {\bibfield  {journal} {\bibinfo  {journal} {APL
  Quantum}\ }\textbf {\bibinfo {volume} {1}} (\bibinfo {year} {2024}),\
  10.1063/5.0208140}\BibitemShut {NoStop}%
\bibitem [{\citenamefont {Simmonds}\ \emph {et~al.}(2004)\citenamefont
  {Simmonds}, \citenamefont {Lang}, \citenamefont {Hite}, \citenamefont {Nam},
  \citenamefont {Pappas},\ and\ \citenamefont {Martinis}}]{Simmonds2004Aug}%
  \BibitemOpen
  \bibfield  {author} {\bibinfo {author} {\bibfnamefont {R.~W.}\ \bibnamefont
  {Simmonds}}, \bibinfo {author} {\bibfnamefont {K.~M.}\ \bibnamefont {Lang}},
  \bibinfo {author} {\bibfnamefont {D.~A.}\ \bibnamefont {Hite}}, \bibinfo
  {author} {\bibfnamefont {S.}~\bibnamefont {Nam}}, \bibinfo {author}
  {\bibfnamefont {D.~P.}\ \bibnamefont {Pappas}}, \ and\ \bibinfo {author}
  {\bibfnamefont {J.~M.}\ \bibnamefont {Martinis}},\ }\href {\doibase
  10.1103/PhysRevLett.93.077003} {\bibfield  {journal} {\bibinfo  {journal}
  {Phys. Rev. Lett.}\ }\textbf {\bibinfo {volume} {93}},\ \bibinfo {pages}
  {077003} (\bibinfo {year} {2004})}\BibitemShut {NoStop}%
\bibitem [{\citenamefont {Shnirman}\ \emph {et~al.}(2005)\citenamefont
  {Shnirman}, \citenamefont {Sch{\ifmmode\ddot{o}\else\"{o}\fi}n},
  \citenamefont {Martin},\ and\ \citenamefont {Makhlin}}]{Shnirman2005Apr}%
  \BibitemOpen
  \bibfield  {author} {\bibinfo {author} {\bibfnamefont {A.}~\bibnamefont
  {Shnirman}}, \bibinfo {author} {\bibfnamefont {G.}~\bibnamefont
  {Sch{\ifmmode\ddot{o}\else\"{o}\fi}n}}, \bibinfo {author} {\bibfnamefont
  {I.}~\bibnamefont {Martin}}, \ and\ \bibinfo {author} {\bibfnamefont
  {Y.}~\bibnamefont {Makhlin}},\ }\href {\doibase
  10.1103/PhysRevLett.94.127002} {\bibfield  {journal} {\bibinfo  {journal}
  {Phys. Rev. Lett.}\ }\textbf {\bibinfo {volume} {94}},\ \bibinfo {pages}
  {127002} (\bibinfo {year} {2005})}\BibitemShut {NoStop}%
\bibitem [{\citenamefont {Ku}\ and\ \citenamefont {Yu}(2005)}]{Ku2005Jul}%
  \BibitemOpen
  \bibfield  {author} {\bibinfo {author} {\bibfnamefont {L.-C.}\ \bibnamefont
  {Ku}}\ and\ \bibinfo {author} {\bibfnamefont {C.~C.}\ \bibnamefont {Yu}},\
  }\href {\doibase 10.1103/PhysRevB.72.024526} {\bibfield  {journal} {\bibinfo
  {journal} {Phys. Rev. B}\ }\textbf {\bibinfo {volume} {72}},\ \bibinfo
  {pages} {024526} (\bibinfo {year} {2005})}\BibitemShut {NoStop}%
\bibitem [{\citenamefont {Martinis}\ \emph {et~al.}(2005)\citenamefont
  {Martinis}, \citenamefont {Cooper}, \citenamefont {McDermott}, \citenamefont
  {Steffen}, \citenamefont {Ansmann}, \citenamefont {Osborn}, \citenamefont
  {Cicak}, \citenamefont {Oh}, \citenamefont {Pappas}, \citenamefont
  {Simmonds},\ and\ \citenamefont {Yu}}]{Martinis2005Nov}%
  \BibitemOpen
  \bibfield  {author} {\bibinfo {author} {\bibfnamefont {J.~M.}\ \bibnamefont
  {Martinis}}, \bibinfo {author} {\bibfnamefont {K.~B.}\ \bibnamefont
  {Cooper}}, \bibinfo {author} {\bibfnamefont {R.}~\bibnamefont {McDermott}},
  \bibinfo {author} {\bibfnamefont {M.}~\bibnamefont {Steffen}}, \bibinfo
  {author} {\bibfnamefont {M.}~\bibnamefont {Ansmann}}, \bibinfo {author}
  {\bibfnamefont {K.~D.}\ \bibnamefont {Osborn}}, \bibinfo {author}
  {\bibfnamefont {K.}~\bibnamefont {Cicak}}, \bibinfo {author} {\bibfnamefont
  {S.}~\bibnamefont {Oh}}, \bibinfo {author} {\bibfnamefont {D.~P.}\
  \bibnamefont {Pappas}}, \bibinfo {author} {\bibfnamefont {R.~W.}\
  \bibnamefont {Simmonds}}, \ and\ \bibinfo {author} {\bibfnamefont {C.~C.}\
  \bibnamefont {Yu}},\ }\href {\doibase 10.1103/PhysRevLett.95.210503}
  {\bibfield  {journal} {\bibinfo  {journal} {Phys. Rev. Lett.}\ }\textbf
  {\bibinfo {volume} {95}},\ \bibinfo {pages} {210503} (\bibinfo {year}
  {2005})}\BibitemShut {NoStop}%
\bibitem [{\citenamefont {Hung}\ \emph {et~al.}(2023)\citenamefont {Hung},
  \citenamefont {Kohler},\ and\ \citenamefont {Osborn}}]{Hung2023Feb}%
  \BibitemOpen
  \bibfield  {author} {\bibinfo {author} {\bibfnamefont {C.-C.}\ \bibnamefont
  {Hung}}, \bibinfo {author} {\bibfnamefont {T.}~\bibnamefont {Kohler}}, \ and\
  \bibinfo {author} {\bibfnamefont {K.~D.}\ \bibnamefont {Osborn}},\ }\href
  {\doibase 10.48550/arXiv.2302.00318} {\bibfield  {journal} {\bibinfo
  {journal} {arXiv}\ } (\bibinfo {year} {2023}),\ 10.48550/arXiv.2302.00318},\
  \Eprint {http://arxiv.org/abs/2302.00318} {2302.00318} \BibitemShut {NoStop}%
\bibitem [{\citenamefont {Zhang}\ \emph {et~al.}(2024)\citenamefont {Zhang},
  \citenamefont {Godeneli}, \citenamefont {He}, \citenamefont {Odeh},
  \citenamefont {Zhou}, \citenamefont {Meesala},\ and\ \citenamefont
  {Sipahigil}}]{Zhang2024Oct}%
  \BibitemOpen
  \bibfield  {author} {\bibinfo {author} {\bibfnamefont {Z.-H.}\ \bibnamefont
  {Zhang}}, \bibinfo {author} {\bibfnamefont {K.}~\bibnamefont {Godeneli}},
  \bibinfo {author} {\bibfnamefont {J.}~\bibnamefont {He}}, \bibinfo {author}
  {\bibfnamefont {M.}~\bibnamefont {Odeh}}, \bibinfo {author} {\bibfnamefont
  {H.}~\bibnamefont {Zhou}}, \bibinfo {author} {\bibfnamefont {S.}~\bibnamefont
  {Meesala}}, \ and\ \bibinfo {author} {\bibfnamefont {A.}~\bibnamefont
  {Sipahigil}},\ }\href {\doibase 10.1103/PhysRevX.14.041022} {\bibfield
  {journal} {\bibinfo  {journal} {Phys. Rev. X}\ }\textbf {\bibinfo {volume}
  {14}},\ \bibinfo {pages} {041022} (\bibinfo {year} {2024})}\BibitemShut
  {NoStop}%
\bibitem [{\citenamefont {Oh}\ \emph {et~al.}(2006)\citenamefont {Oh},
  \citenamefont {Cicak}, \citenamefont {Kline}, \citenamefont
  {Sillanp{\ifmmode\ddot{a}\else\"{a}\fi}{\ifmmode\ddot{a}\else\"{a}\fi}},
  \citenamefont {Osborn}, \citenamefont {Whittaker}, \citenamefont {Simmonds},\
  and\ \citenamefont {Pappas}}]{Oh2006Sep}%
  \BibitemOpen
  \bibfield  {author} {\bibinfo {author} {\bibfnamefont {S.}~\bibnamefont
  {Oh}}, \bibinfo {author} {\bibfnamefont {K.}~\bibnamefont {Cicak}}, \bibinfo
  {author} {\bibfnamefont {J.~S.}\ \bibnamefont {Kline}}, \bibinfo {author}
  {\bibfnamefont {M.~A.}\ \bibnamefont
  {Sillanp{\ifmmode\ddot{a}\else\"{a}\fi}{\ifmmode\ddot{a}\else\"{a}\fi}}},
  \bibinfo {author} {\bibfnamefont {K.~D.}\ \bibnamefont {Osborn}}, \bibinfo
  {author} {\bibfnamefont {J.~D.}\ \bibnamefont {Whittaker}}, \bibinfo {author}
  {\bibfnamefont {R.~W.}\ \bibnamefont {Simmonds}}, \ and\ \bibinfo {author}
  {\bibfnamefont {D.~P.}\ \bibnamefont {Pappas}},\ }\href {\doibase
  10.1103/PhysRevB.74.100502} {\bibfield  {journal} {\bibinfo  {journal} {Phys.
  Rev. B}\ }\textbf {\bibinfo {volume} {74}},\ \bibinfo {pages} {100502}
  (\bibinfo {year} {2006})}\BibitemShut {NoStop}%
\bibitem [{\citenamefont {Oliver}\ and\ \citenamefont
  {Welander}(2013)}]{Oliver2013Oct}%
  \BibitemOpen
  \bibfield  {author} {\bibinfo {author} {\bibfnamefont {W.~D.}\ \bibnamefont
  {Oliver}}\ and\ \bibinfo {author} {\bibfnamefont {P.~B.}\ \bibnamefont
  {Welander}},\ }\href {\doibase 10.1557/mrs.2013.229} {\bibfield  {journal}
  {\bibinfo  {journal} {MRS Bull.}\ }\textbf {\bibinfo {volume} {38}},\
  \bibinfo {pages} {816} (\bibinfo {year} {2013})}\BibitemShut {NoStop}%
\bibitem [{\citenamefont {Tyner}\ \emph {et~al.}(2025)\citenamefont {Tyner},
  \citenamefont {Heath}, \citenamefont {Thann}, \citenamefont {Michal},
  \citenamefont {Krogstrup}, \citenamefont {Svendsen},\ and\ \citenamefont
  {Balatsky}}]{Tyner2025Jul}%
  \BibitemOpen
  \bibfield  {author} {\bibinfo {author} {\bibfnamefont {A.~C.}\ \bibnamefont
  {Tyner}}, \bibinfo {author} {\bibfnamefont {J.~T.}\ \bibnamefont {Heath}},
  \bibinfo {author} {\bibfnamefont {T.~C.}\ \bibnamefont {Thann}}, \bibinfo
  {author} {\bibfnamefont {V.~P.}\ \bibnamefont {Michal}}, \bibinfo {author}
  {\bibfnamefont {P.}~\bibnamefont {Krogstrup}}, \bibinfo {author}
  {\bibfnamefont {M.~K.}\ \bibnamefont {Svendsen}}, \ and\ \bibinfo {author}
  {\bibfnamefont {A.~V.}\ \bibnamefont {Balatsky}},\ }\href {\doibase
  10.1002/qute.202500170} {\bibfield  {journal} {\bibinfo  {journal} {Adv.
  Quantum Technol.}\ }\textbf {\bibinfo {volume} {n/a}},\ \bibinfo {pages}
  {e2500170} (\bibinfo {year} {2025})}\BibitemShut {NoStop}%
\bibitem [{\citenamefont {Hung}\ \emph {et~al.}(2022)\citenamefont {Hung},
  \citenamefont {Yu}, \citenamefont {Foroozani}, \citenamefont {Fritz},
  \citenamefont {Gerthsen},\ and\ \citenamefont {Osborn}}]{Hung2022Mar}%
  \BibitemOpen
  \bibfield  {author} {\bibinfo {author} {\bibfnamefont {C.-C.}\ \bibnamefont
  {Hung}}, \bibinfo {author} {\bibfnamefont {L.}~\bibnamefont {Yu}}, \bibinfo
  {author} {\bibfnamefont {N.}~\bibnamefont {Foroozani}}, \bibinfo {author}
  {\bibfnamefont {S.}~\bibnamefont {Fritz}}, \bibinfo {author} {\bibfnamefont
  {D.}~\bibnamefont {Gerthsen}}, \ and\ \bibinfo {author} {\bibfnamefont
  {K.~D.}\ \bibnamefont {Osborn}},\ }\href {\doibase
  10.1103/PhysRevApplied.17.034025} {\bibfield  {journal} {\bibinfo  {journal}
  {Phys. Rev. Appl.}\ }\textbf {\bibinfo {volume} {17}},\ \bibinfo {pages}
  {034025} (\bibinfo {year} {2022})}\BibitemShut {NoStop}%
\bibitem [{\citenamefont {Balatsky}\ \emph {et~al.}(2006)\citenamefont
  {Balatsky}, \citenamefont {Vekhter},\ and\ \citenamefont
  {Zhu}}]{Balatsky2006May}%
  \BibitemOpen
  \bibfield  {author} {\bibinfo {author} {\bibfnamefont {A.~V.}\ \bibnamefont
  {Balatsky}}, \bibinfo {author} {\bibfnamefont {I.}~\bibnamefont {Vekhter}}, \
  and\ \bibinfo {author} {\bibfnamefont {J.-X.}\ \bibnamefont {Zhu}},\ }\href
  {\doibase 10.1103/RevModPhys.78.373} {\bibfield  {journal} {\bibinfo
  {journal} {Rev. Mod. Phys.}\ }\textbf {\bibinfo {volume} {78}},\ \bibinfo
  {pages} {373} (\bibinfo {year} {2006})}\BibitemShut {NoStop}%
\bibitem [{\citenamefont {Heath}\ \emph {et~al.}(2025)\citenamefont {Heath},
  \citenamefont {Tyner}, \citenamefont {Thann}, \citenamefont {Michal},
  \citenamefont {Krogstrup}, \citenamefont {Svendsen},\ and\ \citenamefont
  {Balatsky}}]{Heath2025Mar}%
  \BibitemOpen
  \bibfield  {author} {\bibinfo {author} {\bibfnamefont {J.~T.}\ \bibnamefont
  {Heath}}, \bibinfo {author} {\bibfnamefont {A.~C.}\ \bibnamefont {Tyner}},
  \bibinfo {author} {\bibfnamefont {T.~C.}\ \bibnamefont {Thann}}, \bibinfo
  {author} {\bibfnamefont {V.~P.}\ \bibnamefont {Michal}}, \bibinfo {author}
  {\bibfnamefont {P.}~\bibnamefont {Krogstrup}}, \bibinfo {author}
  {\bibfnamefont {M.~K.}\ \bibnamefont {Svendsen}}, \ and\ \bibinfo {author}
  {\bibfnamefont {A.~V.}\ \bibnamefont {Balatsky}},\ }\href {\doibase
  10.48550/arXiv.2503.08767} {\bibfield  {journal} {\bibinfo  {journal}
  {arXiv}\ } (\bibinfo {year} {2025}),\ 10.48550/arXiv.2503.08767},\ \Eprint
  {http://arxiv.org/abs/2503.08767} {2503.08767} \BibitemShut {NoStop}%
\bibitem [{\citenamefont {Fibich}(1965)}]{Fibich1965Apr}%
  \BibitemOpen
  \bibfield  {author} {\bibinfo {author} {\bibfnamefont {M.}~\bibnamefont
  {Fibich}},\ }\href {\doibase 10.1103/PhysRevLett.14.561} {\bibfield
  {journal} {\bibinfo  {journal} {Phys. Rev. Lett.}\ }\textbf {\bibinfo
  {volume} {14}},\ \bibinfo {pages} {561} (\bibinfo {year} {1965})}\BibitemShut
  {NoStop}%
\bibitem [{\citenamefont {Eliashberg}(1960)}]{Eliashberg1960}%
  \BibitemOpen
  \bibfield  {author} {\bibinfo {author} {\bibfnamefont {G.~M.}\ \bibnamefont
  {Eliashberg}},\ }\href@noop {} {\bibfield  {journal} {\bibinfo  {journal}
  {Sov. Phys. JETP}\ }\textbf {\bibinfo {volume} {11}},\ \bibinfo {pages} {696}
  (\bibinfo {year} {1960})}\BibitemShut {NoStop}%
\bibitem [{\citenamefont {Eliashberg}(1961)}]{Eliashberg1961}%
  \BibitemOpen
  \bibfield  {author} {\bibinfo {author} {\bibfnamefont {G.~M.}\ \bibnamefont
  {Eliashberg}},\ }\href@noop {} {\bibfield  {journal} {\bibinfo  {journal}
  {Sov. Phys. JETP}\ }\textbf {\bibinfo {volume} {12}},\ \bibinfo {pages}
  {1000} (\bibinfo {year} {1961})}\BibitemShut {NoStop}%
\bibitem [{\citenamefont {Scalapino}(1969)}]{Parks1}%
  \BibitemOpen
  \bibfield  {author} {\bibinfo {author} {\bibfnamefont {D.~J.}\ \bibnamefont
  {Scalapino}},\ }in\ \href@noop {} {\emph {\bibinfo {booktitle}
  {Superconductivity: Part 1 (In Two Parts)}}},\ \bibinfo {editor} {edited by\
  \bibinfo {editor} {\bibfnamefont {R.}~\bibnamefont {Parks}}}\ (\bibinfo
  {publisher} {Marcel Dekker Inc., New York},\ \bibinfo {year} {1969})\ pp.\
  \bibinfo {pages} {449--560}\BibitemShut {NoStop}%
\bibitem [{\citenamefont {Scalapino}\ \emph {et~al.}(1965)\citenamefont
  {Scalapino}, \citenamefont {Wada},\ and\ \citenamefont
  {Swihart}}]{Scalapino1965Jan}%
  \BibitemOpen
  \bibfield  {author} {\bibinfo {author} {\bibfnamefont {D.~J.}\ \bibnamefont
  {Scalapino}}, \bibinfo {author} {\bibfnamefont {Y.}~\bibnamefont {Wada}}, \
  and\ \bibinfo {author} {\bibfnamefont {J.~C.}\ \bibnamefont {Swihart}},\
  }\href {\doibase 10.1103/PhysRevLett.14.102} {\bibfield  {journal} {\bibinfo
  {journal} {Phys. Rev. Lett.}\ }\textbf {\bibinfo {volume} {14}},\ \bibinfo
  {pages} {102} (\bibinfo {year} {1965})}\BibitemShut {NoStop}%
\bibitem [{\citenamefont {Marsiglio}\ and\ \citenamefont
  {Carbotte}(1991)}]{Marsiglio1991Mar}%
  \BibitemOpen
  \bibfield  {author} {\bibinfo {author} {\bibfnamefont {F.}~\bibnamefont
  {Marsiglio}}\ and\ \bibinfo {author} {\bibfnamefont {J.~P.}\ \bibnamefont
  {Carbotte}},\ }\href {\doibase 10.1103/PhysRevB.43.5355} {\bibfield
  {journal} {\bibinfo  {journal} {Phys. Rev. B}\ }\textbf {\bibinfo {volume}
  {43}},\ \bibinfo {pages} {5355} (\bibinfo {year} {1991})}\BibitemShut
  {NoStop}%
\bibitem [{\citenamefont {Chubukov}\ \emph {et~al.}(2020)\citenamefont
  {Chubukov}, \citenamefont {Abanov}, \citenamefont {Esterlis},\ and\
  \citenamefont {Kivelson}}]{Chubukov2020Jun}%
  \BibitemOpen
  \bibfield  {author} {\bibinfo {author} {\bibfnamefont {A.~V.}\ \bibnamefont
  {Chubukov}}, \bibinfo {author} {\bibfnamefont {A.}~\bibnamefont {Abanov}},
  \bibinfo {author} {\bibfnamefont {I.}~\bibnamefont {Esterlis}}, \ and\
  \bibinfo {author} {\bibfnamefont {S.~A.}\ \bibnamefont {Kivelson}},\ }\href
  {\doibase 10.1016/j.aop.2020.168190} {\bibfield  {journal} {\bibinfo
  {journal} {Ann. Phys.}\ }\textbf {\bibinfo {volume} {417}},\ \bibinfo {pages}
  {168190} (\bibinfo {year} {2020})}\BibitemShut {NoStop}%
\bibitem [{\citenamefont {Ummarino}\ and\ \citenamefont
  {Zaccone}(2024)}]{Ummarino2024Nov}%
  \BibitemOpen
  \bibfield  {author} {\bibinfo {author} {\bibfnamefont {G.~A.}\ \bibnamefont
  {Ummarino}}\ and\ \bibinfo {author} {\bibfnamefont {A.}~\bibnamefont
  {Zaccone}},\ }\href {\doibase 10.1088/1361-648X/ad92ed} {\bibfield  {journal}
  {\bibinfo  {journal} {J. Phys.: Condens. Matter}\ }\textbf {\bibinfo {volume}
  {37}},\ \bibinfo {pages} {065703} (\bibinfo {year} {2024})}\BibitemShut
  {NoStop}%
\bibitem [{\citenamefont {Lisenfeld}\ \emph {et~al.}(2023)\citenamefont
  {Lisenfeld}, \citenamefont {Bilmes},\ and\ \citenamefont
  {Ustinov}}]{Lisenfeld2023Jan}%
  \BibitemOpen
  \bibfield  {author} {\bibinfo {author} {\bibfnamefont {J.}~\bibnamefont
  {Lisenfeld}}, \bibinfo {author} {\bibfnamefont {A.}~\bibnamefont {Bilmes}}, \
  and\ \bibinfo {author} {\bibfnamefont {A.~V.}\ \bibnamefont {Ustinov}},\
  }\href {\doibase 10.1038/s41534-023-00678-9} {\bibfield  {journal} {\bibinfo
  {journal} {npj Quantum Inf.}\ }\textbf {\bibinfo {volume} {9}},\ \bibinfo
  {pages} {1} (\bibinfo {year} {2023})}\BibitemShut {NoStop}%
\bibitem [{\citenamefont {Pappas}\ \emph {et~al.}(2024)\citenamefont {Pappas},
  \citenamefont {Field}, \citenamefont {Kopas}, \citenamefont {Howard},
  \citenamefont {Wang}, \citenamefont {Lachman}, \citenamefont {Oh},
  \citenamefont {Zhou}, \citenamefont {Gold}, \citenamefont {Stiehl},
  \citenamefont {Yadavalli}, \citenamefont {Sete}, \citenamefont {Bestwick},
  \citenamefont {Kramer},\ and\ \citenamefont {Mutus}}]{Pappas2024Aug}%
  \BibitemOpen
  \bibfield  {author} {\bibinfo {author} {\bibfnamefont {D.~P.}\ \bibnamefont
  {Pappas}}, \bibinfo {author} {\bibfnamefont {M.}~\bibnamefont {Field}},
  \bibinfo {author} {\bibfnamefont {C.~J.}\ \bibnamefont {Kopas}}, \bibinfo
  {author} {\bibfnamefont {J.~A.}\ \bibnamefont {Howard}}, \bibinfo {author}
  {\bibfnamefont {X.}~\bibnamefont {Wang}}, \bibinfo {author} {\bibfnamefont
  {E.}~\bibnamefont {Lachman}}, \bibinfo {author} {\bibfnamefont
  {J.}~\bibnamefont {Oh}}, \bibinfo {author} {\bibfnamefont {L.}~\bibnamefont
  {Zhou}}, \bibinfo {author} {\bibfnamefont {A.}~\bibnamefont {Gold}}, \bibinfo
  {author} {\bibfnamefont {G.~M.}\ \bibnamefont {Stiehl}}, \bibinfo {author}
  {\bibfnamefont {K.}~\bibnamefont {Yadavalli}}, \bibinfo {author}
  {\bibfnamefont {E.~A.}\ \bibnamefont {Sete}}, \bibinfo {author}
  {\bibfnamefont {A.}~\bibnamefont {Bestwick}}, \bibinfo {author}
  {\bibfnamefont {M.~J.}\ \bibnamefont {Kramer}}, \ and\ \bibinfo {author}
  {\bibfnamefont {J.~Y.}\ \bibnamefont {Mutus}},\ }\href {\doibase
  10.1038/s43246-024-00596-z} {\bibfield  {journal} {\bibinfo  {journal}
  {Commun. Mater.}\ }\textbf {\bibinfo {volume} {5}},\ \bibinfo {pages} {1}
  (\bibinfo {year} {2024})}\BibitemShut {NoStop}%
\bibitem [{\citenamefont {Chen}\ \emph {et~al.}(2024)\citenamefont {Chen},
  \citenamefont {Owens}, \citenamefont {Putterman}, \citenamefont
  {Sch{\ifmmode\ddot{a}\else\"{a}\fi}fer},\ and\ \citenamefont
  {Painter}}]{Chen2024Sep}%
  \BibitemOpen
  \bibfield  {author} {\bibinfo {author} {\bibfnamefont {M.}~\bibnamefont
  {Chen}}, \bibinfo {author} {\bibfnamefont {J.~C.}\ \bibnamefont {Owens}},
  \bibinfo {author} {\bibfnamefont {H.}~\bibnamefont {Putterman}}, \bibinfo
  {author} {\bibfnamefont {M.}~\bibnamefont
  {Sch{\ifmmode\ddot{a}\else\"{a}\fi}fer}}, \ and\ \bibinfo {author}
  {\bibfnamefont {O.}~\bibnamefont {Painter}},\ }\href {\doibase
  10.1126/sciadv.ado6240} {\bibfield  {journal} {\bibinfo  {journal} {Sci.
  Adv.}\ }\textbf {\bibinfo {volume} {10}} (\bibinfo {year} {2024}),\
  10.1126/sciadv.ado6240}\BibitemShut {NoStop}%
\bibitem [{\citenamefont {Chen}\ \emph {et~al.}(2025)\citenamefont {Chen},
  \citenamefont {Lee}, \citenamefont {Liu}, \citenamefont {Marinelli},
  \citenamefont {Naik}, \citenamefont {Kang}, \citenamefont {Goss},
  \citenamefont {Kim}, \citenamefont {Santiago},\ and\ \citenamefont
  {Siddiqi}}]{Chen2025Mar}%
  \BibitemOpen
  \bibfield  {author} {\bibinfo {author} {\bibfnamefont {L.}~\bibnamefont
  {Chen}}, \bibinfo {author} {\bibfnamefont {K.-H.}\ \bibnamefont {Lee}},
  \bibinfo {author} {\bibfnamefont {C.-H.}\ \bibnamefont {Liu}}, \bibinfo
  {author} {\bibfnamefont {B.}~\bibnamefont {Marinelli}}, \bibinfo {author}
  {\bibfnamefont {R.~K.}\ \bibnamefont {Naik}}, \bibinfo {author}
  {\bibfnamefont {Z.}~\bibnamefont {Kang}}, \bibinfo {author} {\bibfnamefont
  {N.}~\bibnamefont {Goss}}, \bibinfo {author} {\bibfnamefont {H.}~\bibnamefont
  {Kim}}, \bibinfo {author} {\bibfnamefont {D.~I.}\ \bibnamefont {Santiago}}, \
  and\ \bibinfo {author} {\bibfnamefont {I.}~\bibnamefont {Siddiqi}},\ }\href
  {\doibase 10.48550/arXiv.2503.04702} {\bibfield  {journal} {\bibinfo
  {journal} {arXiv}\ } (\bibinfo {year} {2025}),\ 10.48550/arXiv.2503.04702},\
  \Eprint {http://arxiv.org/abs/2503.04702} {2503.04702} \BibitemShut {NoStop}%
\bibitem [{\citenamefont {Leavens}\ and\ \citenamefont
  {Carbotte}(1974)}]{Leavens1974Jan}%
  \BibitemOpen
  \bibfield  {author} {\bibinfo {author} {\bibfnamefont {C.~R.}\ \bibnamefont
  {Leavens}}\ and\ \bibinfo {author} {\bibfnamefont {J.~P.}\ \bibnamefont
  {Carbotte}},\ }\href {\doibase 10.1007/BF00654817} {\bibfield  {journal}
  {\bibinfo  {journal} {J. Low Temp. Phys.}\ }\textbf {\bibinfo {volume}
  {14}},\ \bibinfo {pages} {195} (\bibinfo {year} {1974})}\BibitemShut
  {NoStop}%
\bibitem [{\citenamefont {Mitrovi{\ifmmode\acute{c}\else\'{c}\fi}}\ \emph
  {et~al.}(1984)\citenamefont {Mitrovi{\ifmmode\acute{c}\else\'{c}\fi}},
  \citenamefont {Zarate},\ and\ \citenamefont {Carbotte}}]{Mitrovic1984Jan}%
  \BibitemOpen
  \bibfield  {author} {\bibinfo {author} {\bibfnamefont {B.}~\bibnamefont
  {Mitrovi{\ifmmode\acute{c}\else\'{c}\fi}}}, \bibinfo {author} {\bibfnamefont
  {H.~G.}\ \bibnamefont {Zarate}}, \ and\ \bibinfo {author} {\bibfnamefont
  {J.~P.}\ \bibnamefont {Carbotte}},\ }\href {\doibase 10.1103/PhysRevB.29.184}
  {\bibfield  {journal} {\bibinfo  {journal} {Phys. Rev. B}\ }\textbf {\bibinfo
  {volume} {29}},\ \bibinfo {pages} {184} (\bibinfo {year} {1984})}\BibitemShut
  {NoStop}%
\bibitem [{\citenamefont {Leavens}\ and\ \citenamefont
  {Carbotte}(1971)}]{Leavens2011Feb}%
  \BibitemOpen
  \bibfield  {author} {\bibinfo {author} {\bibfnamefont {C.~R.}\ \bibnamefont
  {Leavens}}\ and\ \bibinfo {author} {\bibfnamefont {J.~P.}\ \bibnamefont
  {Carbotte}},\ }\href {\doibase 10.1139/p71-088} {\bibfield  {journal}
  {\bibinfo  {journal} {Can. J. Phys.}\ } (\bibinfo {year} {1971}),\
  10.1139/p71-088}\BibitemShut {NoStop}%
\bibitem [{\citenamefont {Marsiglio}(2020)}]{Marsiglio2020Jun}%
  \BibitemOpen
  \bibfield  {author} {\bibinfo {author} {\bibfnamefont {F.}~\bibnamefont
  {Marsiglio}},\ }\href {\doibase 10.1016/j.aop.2020.168102} {\bibfield
  {journal} {\bibinfo  {journal} {Ann. Phys.}\ }\textbf {\bibinfo {volume}
  {417}},\ \bibinfo {pages} {168102} (\bibinfo {year} {2020})}\BibitemShut
  {NoStop}%
\bibitem [{\citenamefont {Balatsky}\ and\ \citenamefont
  {Zhu}(2006)}]{Balatsky2006Sep}%
  \BibitemOpen
  \bibfield  {author} {\bibinfo {author} {\bibfnamefont {A.~V.}\ \bibnamefont
  {Balatsky}}\ and\ \bibinfo {author} {\bibfnamefont {J.-X.}\ \bibnamefont
  {Zhu}},\ }\href {\doibase 10.1103/PhysRevB.74.094517} {\bibfield  {journal}
  {\bibinfo  {journal} {Phys. Rev. B}\ }\textbf {\bibinfo {volume} {74}},\
  \bibinfo {pages} {094517} (\bibinfo {year} {2006})}\BibitemShut {NoStop}%
\bibitem [{\citenamefont {Bergmann}\ and\ \citenamefont
  {Rainer}(1973)}]{Bergmann1973Aug}%
  \BibitemOpen
  \bibfield  {author} {\bibinfo {author} {\bibfnamefont {G.}~\bibnamefont
  {Bergmann}}\ and\ \bibinfo {author} {\bibfnamefont {D.}~\bibnamefont
  {Rainer}},\ }\href {\doibase 10.1007/BF02351862} {\bibfield  {journal}
  {\bibinfo  {journal} {Z. Phys.}\ }\textbf {\bibinfo {volume} {263}},\
  \bibinfo {pages} {59} (\bibinfo {year} {1973})}\BibitemShut {NoStop}%
\bibitem [{\citenamefont {Wang}\ \emph {et~al.}(2025)\citenamefont {Wang},
  \citenamefont {Yu},\ and\ \citenamefont {Wu}}]{Wang2025Feb}%
  \BibitemOpen
  \bibfield  {author} {\bibinfo {author} {\bibfnamefont {Z.}~\bibnamefont
  {Wang}}, \bibinfo {author} {\bibfnamefont {C.~C.}\ \bibnamefont {Yu}}, \ and\
  \bibinfo {author} {\bibfnamefont {R.}~\bibnamefont {Wu}},\ }\href {\doibase
  10.1103/PhysRevApplied.23.024017} {\bibfield  {journal} {\bibinfo  {journal}
  {Phys. Rev. Appl.}\ }\textbf {\bibinfo {volume} {23}},\ \bibinfo {pages}
  {024017} (\bibinfo {year} {2025})}\BibitemShut {NoStop}%
\bibitem [{\citenamefont {Lee}\ \emph {et~al.}(1995)\citenamefont {Lee},
  \citenamefont {Chang},\ and\ \citenamefont {Cohen}}]{Lee1995Jul}%
  \BibitemOpen
  \bibfield  {author} {\bibinfo {author} {\bibfnamefont {K.-H.}\ \bibnamefont
  {Lee}}, \bibinfo {author} {\bibfnamefont {K.~J.}\ \bibnamefont {Chang}}, \
  and\ \bibinfo {author} {\bibfnamefont {M.~L.}\ \bibnamefont {Cohen}},\ }\href
  {\doibase 10.1103/PhysRevB.52.1425} {\bibfield  {journal} {\bibinfo
  {journal} {Phys. Rev. B}\ }\textbf {\bibinfo {volume} {52}},\ \bibinfo
  {pages} {1425} (\bibinfo {year} {1995})}\BibitemShut {NoStop}%
\bibitem [{\citenamefont {Simonato}\ \emph {et~al.}(2023)\citenamefont
  {Simonato}, \citenamefont {Katsnelson},\ and\ \citenamefont
  {R{\ifmmode\ddot{o}\else\"{o}\fi}sner}}]{Simonato2023Aug}%
  \BibitemOpen
  \bibfield  {author} {\bibinfo {author} {\bibfnamefont {M.}~\bibnamefont
  {Simonato}}, \bibinfo {author} {\bibfnamefont {M.~I.}\ \bibnamefont
  {Katsnelson}}, \ and\ \bibinfo {author} {\bibfnamefont {M.}~\bibnamefont
  {R{\ifmmode\ddot{o}\else\"{o}\fi}sner}},\ }\href {\doibase
  10.1103/PhysRevB.108.064513} {\bibfield  {journal} {\bibinfo  {journal}
  {Phys. Rev. B}\ }\textbf {\bibinfo {volume} {108}},\ \bibinfo {pages}
  {064513} (\bibinfo {year} {2023})}\BibitemShut {NoStop}%
\bibitem [{\citenamefont {Perdew}\ \emph {et~al.}(1996)\citenamefont {Perdew},
  \citenamefont {Burke},\ and\ \citenamefont {Ernzerhof}}]{Perdew1996}%
  \BibitemOpen
  \bibfield  {author} {\bibinfo {author} {\bibfnamefont {J.~P.}\ \bibnamefont
  {Perdew}}, \bibinfo {author} {\bibfnamefont {K.}~\bibnamefont {Burke}}, \
  and\ \bibinfo {author} {\bibfnamefont {M.}~\bibnamefont {Ernzerhof}},\ }\href
  {\doibase 10.1103/PhysRevLett.77.3865} {\bibfield  {journal} {\bibinfo
  {journal} {Phys. Rev. Lett.}\ }\textbf {\bibinfo {volume} {77}},\ \bibinfo
  {pages} {3865} (\bibinfo {year} {1996})}\BibitemShut {NoStop}%
\bibitem [{\citenamefont {Giannozzi}\ \emph {et~al.}(2020)\citenamefont
  {Giannozzi}, \citenamefont {Baseggio}, \citenamefont
  {Bonf{\ifmmode\grave{a}\else\`{a}\fi}}, \citenamefont {Brunato},
  \citenamefont {Car}, \citenamefont {Carnimeo}, \citenamefont {Cavazzoni},
  \citenamefont {de~Gironcoli}, \citenamefont {Delugas}, \citenamefont
  {Ferrari~Ruffino}, \citenamefont {Ferretti}, \citenamefont {Marzari},
  \citenamefont {Timrov}, \citenamefont {Urru},\ and\ \citenamefont
  {Baroni}}]{Giannozzi2020Apr}%
  \BibitemOpen
  \bibfield  {author} {\bibinfo {author} {\bibfnamefont {P.}~\bibnamefont
  {Giannozzi}}, \bibinfo {author} {\bibfnamefont {O.}~\bibnamefont {Baseggio}},
  \bibinfo {author} {\bibfnamefont {P.}~\bibnamefont
  {Bonf{\ifmmode\grave{a}\else\`{a}\fi}}}, \bibinfo {author} {\bibfnamefont
  {D.}~\bibnamefont {Brunato}}, \bibinfo {author} {\bibfnamefont
  {R.}~\bibnamefont {Car}}, \bibinfo {author} {\bibfnamefont {I.}~\bibnamefont
  {Carnimeo}}, \bibinfo {author} {\bibfnamefont {C.}~\bibnamefont {Cavazzoni}},
  \bibinfo {author} {\bibfnamefont {S.}~\bibnamefont {de~Gironcoli}}, \bibinfo
  {author} {\bibfnamefont {P.}~\bibnamefont {Delugas}}, \bibinfo {author}
  {\bibfnamefont {F.}~\bibnamefont {Ferrari~Ruffino}}, \bibinfo {author}
  {\bibfnamefont {A.}~\bibnamefont {Ferretti}}, \bibinfo {author}
  {\bibfnamefont {N.}~\bibnamefont {Marzari}}, \bibinfo {author} {\bibfnamefont
  {I.}~\bibnamefont {Timrov}}, \bibinfo {author} {\bibfnamefont
  {A.}~\bibnamefont {Urru}}, \ and\ \bibinfo {author} {\bibfnamefont
  {S.}~\bibnamefont {Baroni}},\ }\href {\doibase 10.1063/5.0005082} {\bibfield
  {journal} {\bibinfo  {journal} {J. Chem. Phys.}\ }\textbf {\bibinfo {volume}
  {152}},\ \bibinfo {pages} {154105} (\bibinfo {year} {2020})}\BibitemShut
  {NoStop}%
\bibitem [{\citenamefont {Kresse}\ and\ \citenamefont
  {Joubert}(1999)}]{kresse1999VASP}%
  \BibitemOpen
  \bibfield  {author} {\bibinfo {author} {\bibfnamefont {G.}~\bibnamefont
  {Kresse}}\ and\ \bibinfo {author} {\bibfnamefont {D.}~\bibnamefont
  {Joubert}},\ }\href {\doibase 10.1103/PhysRevB.59.1758} {\bibfield  {journal}
  {\bibinfo  {journal} {Phys. Rev. B}\ }\textbf {\bibinfo {volume} {59}},\
  \bibinfo {pages} {1758} (\bibinfo {year} {1999})}\BibitemShut {NoStop}%
\bibitem [{\citenamefont {Van~Setten}\ \emph {et~al.}(2018)\citenamefont
  {Van~Setten}, \citenamefont {Giantomassi}, \citenamefont {Bousquet},
  \citenamefont {Verstraete}, \citenamefont {Hamann}, \citenamefont {Gonze},\
  and\ \citenamefont {Rignanese}}]{van2018pseudodojo}%
  \BibitemOpen
  \bibfield  {author} {\bibinfo {author} {\bibfnamefont {M.~J.}\ \bibnamefont
  {Van~Setten}}, \bibinfo {author} {\bibfnamefont {M.}~\bibnamefont
  {Giantomassi}}, \bibinfo {author} {\bibfnamefont {E.}~\bibnamefont
  {Bousquet}}, \bibinfo {author} {\bibfnamefont {M.~J.}\ \bibnamefont
  {Verstraete}}, \bibinfo {author} {\bibfnamefont {D.~R.}\ \bibnamefont
  {Hamann}}, \bibinfo {author} {\bibfnamefont {X.}~\bibnamefont {Gonze}}, \
  and\ \bibinfo {author} {\bibfnamefont {G.-M.}\ \bibnamefont {Rignanese}},\
  }\href {\doibase https://doi.org/10.1016/j.cpc.2018.01.012} {\bibfield
  {journal} {\bibinfo  {journal} {Comput. Phys. Commun.}\ }\textbf {\bibinfo
  {volume} {226}},\ \bibinfo {pages} {39} (\bibinfo {year} {2018})}\BibitemShut
  {NoStop}%
\bibitem [{\citenamefont {Monkhorst}\ and\ \citenamefont
  {Pack}(1976)}]{PhysRevB.13.5188}%
  \BibitemOpen
  \bibfield  {author} {\bibinfo {author} {\bibfnamefont {H.~J.}\ \bibnamefont
  {Monkhorst}}\ and\ \bibinfo {author} {\bibfnamefont {J.~D.}\ \bibnamefont
  {Pack}},\ }\href {\doibase 10.1103/PhysRevB.13.5188} {\bibfield  {journal}
  {\bibinfo  {journal} {Phys. Rev. B}\ }\textbf {\bibinfo {volume} {13}},\
  \bibinfo {pages} {5188} (\bibinfo {year} {1976})}\BibitemShut {NoStop}%
\bibitem [{\citenamefont {Noffsinger}\ \emph {et~al.}(2010)\citenamefont
  {Noffsinger}, \citenamefont {Giustino}, \citenamefont {Malone}, \citenamefont
  {Park}, \citenamefont {Louie},\ and\ \citenamefont
  {Cohen}}]{Noffsinger2010Dec}%
  \BibitemOpen
  \bibfield  {author} {\bibinfo {author} {\bibfnamefont {J.}~\bibnamefont
  {Noffsinger}}, \bibinfo {author} {\bibfnamefont {F.}~\bibnamefont
  {Giustino}}, \bibinfo {author} {\bibfnamefont {B.~D.}\ \bibnamefont
  {Malone}}, \bibinfo {author} {\bibfnamefont {C.-H.}\ \bibnamefont {Park}},
  \bibinfo {author} {\bibfnamefont {S.~G.}\ \bibnamefont {Louie}}, \ and\
  \bibinfo {author} {\bibfnamefont {M.~L.}\ \bibnamefont {Cohen}},\ }\href
  {\doibase 10.1016/j.cpc.2010.08.027} {\bibfield  {journal} {\bibinfo
  {journal} {Comput. Phys. Commun.}\ }\textbf {\bibinfo {volume} {181}},\
  \bibinfo {pages} {2140} (\bibinfo {year} {2010})}\BibitemShut {NoStop}%
\bibitem [{\citenamefont {Lee}\ \emph {et~al.}(2023)\citenamefont {Lee},
  \citenamefont {Ponc{\'e}}, \citenamefont {Bushick}, \citenamefont
  {Hajinazar}, \citenamefont {Lafuente-Bartolome}, \citenamefont {Leveillee},
  \citenamefont {Lian}, \citenamefont {Lihm}, \citenamefont {Macheda},
  \citenamefont {Mori} \emph {et~al.}}]{lee2023electron}%
  \BibitemOpen
  \bibfield  {author} {\bibinfo {author} {\bibfnamefont {H.}~\bibnamefont
  {Lee}}, \bibinfo {author} {\bibfnamefont {S.}~\bibnamefont {Ponc{\'e}}},
  \bibinfo {author} {\bibfnamefont {K.}~\bibnamefont {Bushick}}, \bibinfo
  {author} {\bibfnamefont {S.}~\bibnamefont {Hajinazar}}, \bibinfo {author}
  {\bibfnamefont {J.}~\bibnamefont {Lafuente-Bartolome}}, \bibinfo {author}
  {\bibfnamefont {J.}~\bibnamefont {Leveillee}}, \bibinfo {author}
  {\bibfnamefont {C.}~\bibnamefont {Lian}}, \bibinfo {author} {\bibfnamefont
  {J.-M.}\ \bibnamefont {Lihm}}, \bibinfo {author} {\bibfnamefont
  {F.}~\bibnamefont {Macheda}}, \bibinfo {author} {\bibfnamefont
  {H.}~\bibnamefont {Mori}},  \emph {et~al.},\ }\href {\doibase
  https://doi.org/10.1038/s41524-023-01107-3} {\bibfield  {journal} {\bibinfo
  {journal} {npj Comput. Mater.}\ }\textbf {\bibinfo {volume} {9}},\ \bibinfo
  {pages} {156} (\bibinfo {year} {2023})}\BibitemShut {NoStop}%
\bibitem [{\citenamefont {Pizzi}\ \emph {et~al.}(2020)\citenamefont {Pizzi},
  \citenamefont {Vitale}, \citenamefont {Arita}, \citenamefont {Blugel},
  \citenamefont {Freimuth}, \citenamefont {G{\'{e}}ranton}, \citenamefont
  {Gibertini}, \citenamefont {Gresch}, \citenamefont {Johnson}, \citenamefont
  {Koretsune}, \citenamefont {Iba{\~{n}}ez-Azpiroz}, \citenamefont {Lee},
  \citenamefont {Lihm}, \citenamefont {Marchand}, \citenamefont {Marrazzo},
  \citenamefont {Mokrousov}, \citenamefont {Mustafa}, \citenamefont {Nohara},
  \citenamefont {Nomura}, \citenamefont {Paulatto}, \citenamefont
  {Ponc{\'{e}}}, \citenamefont {Ponweiser}, \citenamefont {Qiao}, \citenamefont
  {Thole}, \citenamefont {Tsirkin}, \citenamefont {Wierzbowska}, \citenamefont
  {Marzari}, \citenamefont {Vanderbilt}, \citenamefont {Souza}, \citenamefont
  {Mostofi},\ and\ \citenamefont {Yates}}]{Pizzi2020}%
  \BibitemOpen
  \bibfield  {author} {\bibinfo {author} {\bibfnamefont {G.}~\bibnamefont
  {Pizzi}}, \bibinfo {author} {\bibfnamefont {V.}~\bibnamefont {Vitale}},
  \bibinfo {author} {\bibfnamefont {R.}~\bibnamefont {Arita}}, \bibinfo
  {author} {\bibfnamefont {S.}~\bibnamefont {Blugel}}, \bibinfo {author}
  {\bibfnamefont {F.}~\bibnamefont {Freimuth}}, \bibinfo {author}
  {\bibfnamefont {G.}~\bibnamefont {G{\'{e}}ranton}}, \bibinfo {author}
  {\bibfnamefont {M.}~\bibnamefont {Gibertini}}, \bibinfo {author}
  {\bibfnamefont {D.}~\bibnamefont {Gresch}}, \bibinfo {author} {\bibfnamefont
  {C.}~\bibnamefont {Johnson}}, \bibinfo {author} {\bibfnamefont
  {T.}~\bibnamefont {Koretsune}}, \bibinfo {author} {\bibfnamefont
  {J.}~\bibnamefont {Iba{\~{n}}ez-Azpiroz}}, \bibinfo {author} {\bibfnamefont
  {H.}~\bibnamefont {Lee}}, \bibinfo {author} {\bibfnamefont {J.-M.}\
  \bibnamefont {Lihm}}, \bibinfo {author} {\bibfnamefont {D.}~\bibnamefont
  {Marchand}}, \bibinfo {author} {\bibfnamefont {A.}~\bibnamefont {Marrazzo}},
  \bibinfo {author} {\bibfnamefont {Y.}~\bibnamefont {Mokrousov}}, \bibinfo
  {author} {\bibfnamefont {J.~I.}\ \bibnamefont {Mustafa}}, \bibinfo {author}
  {\bibfnamefont {Y.}~\bibnamefont {Nohara}}, \bibinfo {author} {\bibfnamefont
  {Y.}~\bibnamefont {Nomura}}, \bibinfo {author} {\bibfnamefont
  {L.}~\bibnamefont {Paulatto}}, \bibinfo {author} {\bibfnamefont
  {S.}~\bibnamefont {Ponc{\'{e}}}}, \bibinfo {author} {\bibfnamefont
  {T.}~\bibnamefont {Ponweiser}}, \bibinfo {author} {\bibfnamefont
  {J.}~\bibnamefont {Qiao}}, \bibinfo {author} {\bibfnamefont {F.}~\bibnamefont
  {Thole}}, \bibinfo {author} {\bibfnamefont {S.~S.}\ \bibnamefont {Tsirkin}},
  \bibinfo {author} {\bibfnamefont {M.}~\bibnamefont {Wierzbowska}}, \bibinfo
  {author} {\bibfnamefont {N.}~\bibnamefont {Marzari}}, \bibinfo {author}
  {\bibfnamefont {D.}~\bibnamefont {Vanderbilt}}, \bibinfo {author}
  {\bibfnamefont {I.}~\bibnamefont {Souza}}, \bibinfo {author} {\bibfnamefont
  {A.~A.}\ \bibnamefont {Mostofi}}, \ and\ \bibinfo {author} {\bibfnamefont
  {J.~R.}\ \bibnamefont {Yates}},\ }\href {\doibase 10.1088/1361-648x/ab51ff}
  {\bibfield  {journal} {\bibinfo  {journal} {J. Phys. Condens. Matter}\
  }\textbf {\bibinfo {volume} {32}},\ \bibinfo {pages} {165902} (\bibinfo
  {year} {2020})}\BibitemShut {NoStop}%
\bibitem [{\citenamefont {Bozyigit}\ \emph {et~al.}(2016)\citenamefont
  {Bozyigit}, \citenamefont {Yazdani}, \citenamefont {Yarema}, \citenamefont
  {Yarema}, \citenamefont {Lin}, \citenamefont {Volk}, \citenamefont
  {Vuttivorakulchai}, \citenamefont {Luisier}, \citenamefont {Juranyi},\ and\
  \citenamefont {Wood}}]{bozyigit2016soft}%
  \BibitemOpen
  \bibfield  {author} {\bibinfo {author} {\bibfnamefont {D.}~\bibnamefont
  {Bozyigit}}, \bibinfo {author} {\bibfnamefont {N.}~\bibnamefont {Yazdani}},
  \bibinfo {author} {\bibfnamefont {M.}~\bibnamefont {Yarema}}, \bibinfo
  {author} {\bibfnamefont {O.}~\bibnamefont {Yarema}}, \bibinfo {author}
  {\bibfnamefont {W.~M.~M.}\ \bibnamefont {Lin}}, \bibinfo {author}
  {\bibfnamefont {S.}~\bibnamefont {Volk}}, \bibinfo {author} {\bibfnamefont
  {K.}~\bibnamefont {Vuttivorakulchai}}, \bibinfo {author} {\bibfnamefont
  {M.}~\bibnamefont {Luisier}}, \bibinfo {author} {\bibfnamefont
  {F.}~\bibnamefont {Juranyi}}, \ and\ \bibinfo {author} {\bibfnamefont
  {V.}~\bibnamefont {Wood}},\ }\href {\doibase
  https://doi.org/10.1038/nature16977} {\bibfield  {journal} {\bibinfo
  {journal} {Nature}\ }\textbf {\bibinfo {volume} {531}},\ \bibinfo {pages}
  {618} (\bibinfo {year} {2016})}\BibitemShut {NoStop}%
\bibitem [{\citenamefont {Yeh}\ \emph {et~al.}(2023)\citenamefont {Yeh},
  \citenamefont {Do}, \citenamefont {Liao}, \citenamefont {Hsu}, \citenamefont
  {Tu}, \citenamefont {Lin}, \citenamefont {Chang}, \citenamefont {Wang},
  \citenamefont {Gao}, \citenamefont {Wu}, \citenamefont {Wu}, \citenamefont
  {Lai}, \citenamefont {Martin}, \citenamefont {Lin}, \citenamefont
  {Panagopoulos},\ and\ \citenamefont {Liang}}]{Yeh2023Nov}%
  \BibitemOpen
  \bibfield  {author} {\bibinfo {author} {\bibfnamefont {C.-C.}\ \bibnamefont
  {Yeh}}, \bibinfo {author} {\bibfnamefont {T.-H.}\ \bibnamefont {Do}},
  \bibinfo {author} {\bibfnamefont {P.-C.}\ \bibnamefont {Liao}}, \bibinfo
  {author} {\bibfnamefont {C.-H.}\ \bibnamefont {Hsu}}, \bibinfo {author}
  {\bibfnamefont {Y.-H.}\ \bibnamefont {Tu}}, \bibinfo {author} {\bibfnamefont
  {H.}~\bibnamefont {Lin}}, \bibinfo {author} {\bibfnamefont {T.-R.}\
  \bibnamefont {Chang}}, \bibinfo {author} {\bibfnamefont {S.-C.}\ \bibnamefont
  {Wang}}, \bibinfo {author} {\bibfnamefont {Y.-Y.}\ \bibnamefont {Gao}},
  \bibinfo {author} {\bibfnamefont {Y.-H.}\ \bibnamefont {Wu}}, \bibinfo
  {author} {\bibfnamefont {C.-C.}\ \bibnamefont {Wu}}, \bibinfo {author}
  {\bibfnamefont {Y.~A.}\ \bibnamefont {Lai}}, \bibinfo {author} {\bibfnamefont
  {I.}~\bibnamefont {Martin}}, \bibinfo {author} {\bibfnamefont {S.-D.}\
  \bibnamefont {Lin}}, \bibinfo {author} {\bibfnamefont {C.}~\bibnamefont
  {Panagopoulos}}, \ and\ \bibinfo {author} {\bibfnamefont {C.-T.}\
  \bibnamefont {Liang}},\ }\href {\doibase 10.1103/PhysRevMaterials.7.114801}
  {\bibfield  {journal} {\bibinfo  {journal} {Phys. Rev. Mater.}\ }\textbf
  {\bibinfo {volume} {7}},\ \bibinfo {pages} {114801} (\bibinfo {year}
  {2023})}\BibitemShut {NoStop}%
\bibitem [{\citenamefont {Luo}\ \emph {et~al.}(2013)\citenamefont {Luo},
  \citenamefont {Zhao}, \citenamefont {Zhang}, \citenamefont {Xiong},\ and\
  \citenamefont {Quek}}]{Luo2013Aug}%
  \BibitemOpen
  \bibfield  {author} {\bibinfo {author} {\bibfnamefont {X.}~\bibnamefont
  {Luo}}, \bibinfo {author} {\bibfnamefont {Y.}~\bibnamefont {Zhao}}, \bibinfo
  {author} {\bibfnamefont {J.}~\bibnamefont {Zhang}}, \bibinfo {author}
  {\bibfnamefont {Q.}~\bibnamefont {Xiong}}, \ and\ \bibinfo {author}
  {\bibfnamefont {S.~Y.}\ \bibnamefont {Quek}},\ }\href {\doibase
  10.1103/PhysRevB.88.075320} {\bibfield  {journal} {\bibinfo  {journal} {Phys.
  Rev. B}\ }\textbf {\bibinfo {volume} {88}},\ \bibinfo {pages} {075320}
  (\bibinfo {year} {2013})}\BibitemShut {NoStop}%
\bibitem [{\citenamefont {Schreier}\ \emph {et~al.}(2008)\citenamefont
  {Schreier}, \citenamefont {Houck}, \citenamefont {Koch}, \citenamefont
  {Schuster}, \citenamefont {Johnson}, \citenamefont {Chow}, \citenamefont
  {Gambetta}, \citenamefont {Majer}, \citenamefont {Frunzio}, \citenamefont
  {Devoret}, \citenamefont {Girvin},\ and\ \citenamefont
  {Schoelkopf}}]{Schreier2008May}%
  \BibitemOpen
  \bibfield  {author} {\bibinfo {author} {\bibfnamefont {J.~A.}\ \bibnamefont
  {Schreier}}, \bibinfo {author} {\bibfnamefont {A.~A.}\ \bibnamefont {Houck}},
  \bibinfo {author} {\bibfnamefont {J.}~\bibnamefont {Koch}}, \bibinfo {author}
  {\bibfnamefont {D.~I.}\ \bibnamefont {Schuster}}, \bibinfo {author}
  {\bibfnamefont {B.~R.}\ \bibnamefont {Johnson}}, \bibinfo {author}
  {\bibfnamefont {J.~M.}\ \bibnamefont {Chow}}, \bibinfo {author}
  {\bibfnamefont {J.~M.}\ \bibnamefont {Gambetta}}, \bibinfo {author}
  {\bibfnamefont {J.}~\bibnamefont {Majer}}, \bibinfo {author} {\bibfnamefont
  {L.}~\bibnamefont {Frunzio}}, \bibinfo {author} {\bibfnamefont {M.~H.}\
  \bibnamefont {Devoret}}, \bibinfo {author} {\bibfnamefont {S.~M.}\
  \bibnamefont {Girvin}}, \ and\ \bibinfo {author} {\bibfnamefont {R.~J.}\
  \bibnamefont {Schoelkopf}},\ }\href {\doibase 10.1103/PhysRevB.77.180502}
  {\bibfield  {journal} {\bibinfo  {journal} {Phys. Rev. B}\ }\textbf {\bibinfo
  {volume} {77}},\ \bibinfo {pages} {180502} (\bibinfo {year}
  {2008})}\BibitemShut {NoStop}%
\bibitem [{\citenamefont {Queen}\ \emph {et~al.}(2013)\citenamefont {Queen},
  \citenamefont {Liu}, \citenamefont {Karel}, \citenamefont {Metcalf},\ and\
  \citenamefont {Hellman}}]{Queen2013Mar}%
  \BibitemOpen
  \bibfield  {author} {\bibinfo {author} {\bibfnamefont {D.~R.}\ \bibnamefont
  {Queen}}, \bibinfo {author} {\bibfnamefont {X.}~\bibnamefont {Liu}}, \bibinfo
  {author} {\bibfnamefont {J.}~\bibnamefont {Karel}}, \bibinfo {author}
  {\bibfnamefont {T.~H.}\ \bibnamefont {Metcalf}}, \ and\ \bibinfo {author}
  {\bibfnamefont {F.}~\bibnamefont {Hellman}},\ }\href {\doibase
  10.1103/PhysRevLett.110.135901} {\bibfield  {journal} {\bibinfo  {journal}
  {Phys. Rev. Lett.}\ }\textbf {\bibinfo {volume} {110}},\ \bibinfo {pages}
  {135901} (\bibinfo {year} {2013})}\BibitemShut {NoStop}%
\bibitem [{\citenamefont {Gunnarsson}\ \emph {et~al.}(2013)\citenamefont
  {Gunnarsson}, \citenamefont {Pirkkalainen}, \citenamefont {Li}, \citenamefont
  {Paraoanu}, \citenamefont {Hakonen}, \citenamefont
  {Sillanp{\ifmmode\ddot{a}\else\"{a}\fi}{\ifmmode\ddot{a}\else\"{a}\fi}},\
  and\ \citenamefont {Prunnila}}]{Gunnarsson2013Jul}%
  \BibitemOpen
  \bibfield  {author} {\bibinfo {author} {\bibfnamefont {D.}~\bibnamefont
  {Gunnarsson}}, \bibinfo {author} {\bibfnamefont {J.-M.}\ \bibnamefont
  {Pirkkalainen}}, \bibinfo {author} {\bibfnamefont {J.}~\bibnamefont {Li}},
  \bibinfo {author} {\bibfnamefont {G.~S.}\ \bibnamefont {Paraoanu}}, \bibinfo
  {author} {\bibfnamefont {P.}~\bibnamefont {Hakonen}}, \bibinfo {author}
  {\bibfnamefont {M.}~\bibnamefont
  {Sillanp{\ifmmode\ddot{a}\else\"{a}\fi}{\ifmmode\ddot{a}\else\"{a}\fi}}}, \
  and\ \bibinfo {author} {\bibfnamefont {M.}~\bibnamefont {Prunnila}},\ }\href
  {\doibase 10.1088/0953-2048/26/8/085010} {\bibfield  {journal} {\bibinfo
  {journal} {Supercond. Sci. Technol.}\ }\textbf {\bibinfo {volume} {26}},\
  \bibinfo {pages} {085010} (\bibinfo {year} {2013})}\BibitemShut {NoStop}%
\bibitem [{\citenamefont {B{\ifmmode\acute{e}\else\'{e}\fi}janin}\ \emph
  {et~al.}(2021)\citenamefont {B{\ifmmode\acute{e}\else\'{e}\fi}janin},
  \citenamefont {Earnest}, \citenamefont {Sharafeldin},\ and\ \citenamefont
  {Mariantoni}}]{Bejanin2021Sep}%
  \BibitemOpen
  \bibfield  {author} {\bibinfo {author} {\bibfnamefont {J.~H.}\ \bibnamefont
  {B{\ifmmode\acute{e}\else\'{e}\fi}janin}}, \bibinfo {author} {\bibfnamefont
  {C.~T.}\ \bibnamefont {Earnest}}, \bibinfo {author} {\bibfnamefont {A.~S.}\
  \bibnamefont {Sharafeldin}}, \ and\ \bibinfo {author} {\bibfnamefont
  {M.}~\bibnamefont {Mariantoni}},\ }\href {\doibase
  10.1103/PhysRevB.104.094106} {\bibfield  {journal} {\bibinfo  {journal}
  {Phys. Rev. B}\ }\textbf {\bibinfo {volume} {104}},\ \bibinfo {pages}
  {094106} (\bibinfo {year} {2021})}\BibitemShut {NoStop}%
\bibitem [{\citenamefont {Giaever}\ \emph {et~al.}(1962)\citenamefont
  {Giaever}, \citenamefont {Hart},\ and\ \citenamefont
  {Megerle}}]{Giaever1962May}%
  \BibitemOpen
  \bibfield  {author} {\bibinfo {author} {\bibfnamefont {I.}~\bibnamefont
  {Giaever}}, \bibinfo {author} {\bibfnamefont {H.~R.}\ \bibnamefont {Hart}}, \
  and\ \bibinfo {author} {\bibfnamefont {K.}~\bibnamefont {Megerle}},\ }\href
  {\doibase 10.1103/PhysRev.126.941} {\bibfield  {journal} {\bibinfo  {journal}
  {Phys. Rev.}\ }\textbf {\bibinfo {volume} {126}},\ \bibinfo {pages} {941}
  (\bibinfo {year} {1962})}\BibitemShut {NoStop}%
\bibitem [{\citenamefont {Schrieffer}\ \emph {et~al.}(1963)\citenamefont
  {Schrieffer}, \citenamefont {Scalapino},\ and\ \citenamefont
  {Wilkins}}]{Schrieffer1963Apr}%
  \BibitemOpen
  \bibfield  {author} {\bibinfo {author} {\bibfnamefont {J.~R.}\ \bibnamefont
  {Schrieffer}}, \bibinfo {author} {\bibfnamefont {D.~J.}\ \bibnamefont
  {Scalapino}}, \ and\ \bibinfo {author} {\bibfnamefont {J.~W.}\ \bibnamefont
  {Wilkins}},\ }\href {\doibase 10.1103/PhysRevLett.10.336} {\bibfield
  {journal} {\bibinfo  {journal} {Phys. Rev. Lett.}\ }\textbf {\bibinfo
  {volume} {10}},\ \bibinfo {pages} {336} (\bibinfo {year} {1963})}\BibitemShut
  {NoStop}%
\bibitem [{\citenamefont {Scalapino}\ \emph {et~al.}(1966)\citenamefont
  {Scalapino}, \citenamefont {Schrieffer},\ and\ \citenamefont
  {Wilkins}}]{Scalapino1966Aug}%
  \BibitemOpen
  \bibfield  {author} {\bibinfo {author} {\bibfnamefont {D.~J.}\ \bibnamefont
  {Scalapino}}, \bibinfo {author} {\bibfnamefont {J.~R.}\ \bibnamefont
  {Schrieffer}}, \ and\ \bibinfo {author} {\bibfnamefont {J.~W.}\ \bibnamefont
  {Wilkins}},\ }\href {\doibase 10.1103/PhysRev.148.263} {\bibfield  {journal}
  {\bibinfo  {journal} {Phys. Rev.}\ }\textbf {\bibinfo {volume} {148}},\
  \bibinfo {pages} {263} (\bibinfo {year} {1966})}\BibitemShut {NoStop}%
\bibitem [{\citenamefont {Jaklevic}\ and\ \citenamefont
  {Lambe}(1966)}]{Jaklevic1966Nov}%
  \BibitemOpen
  \bibfield  {author} {\bibinfo {author} {\bibfnamefont {R.~C.}\ \bibnamefont
  {Jaklevic}}\ and\ \bibinfo {author} {\bibfnamefont {J.}~\bibnamefont
  {Lambe}},\ }\href {\doibase 10.1103/PhysRevLett.17.1139} {\bibfield
  {journal} {\bibinfo  {journal} {Phys. Rev. Lett.}\ }\textbf {\bibinfo
  {volume} {17}},\ \bibinfo {pages} {1139} (\bibinfo {year}
  {1966})}\BibitemShut {NoStop}%
\bibitem [{\citenamefont {Schackert}(2014)}]{Schackert2014}%
  \BibitemOpen
  \bibfield  {author} {\bibinfo {author} {\bibfnamefont {M.~P.}\ \bibnamefont
  {Schackert}},\ }\emph {\bibinfo {title} {Scanning Tunneling Spectroscopy on
  Electron-Boson Interactions in Superconductors}},\ \href {\doibase
  10.5445/IR/1000041689} {\bibinfo {type} {Dissertation}},\ \bibinfo  {school}
  {Karlsruhe Institute of Technology}, \bibinfo {address} {Karlsruhe, Germany}
  (\bibinfo {year} {2014}),\ \bibinfo {note} {hochschulschrift;
  Prü­fungs­datum: 06 June 2014}\BibitemShut {NoStop}%
\bibitem [{\citenamefont {Dynes}\ \emph {et~al.}(1978)\citenamefont {Dynes},
  \citenamefont {Narayanamurti},\ and\ \citenamefont {Garno}}]{Dynes1978Nov}%
  \BibitemOpen
  \bibfield  {author} {\bibinfo {author} {\bibfnamefont {R.~C.}\ \bibnamefont
  {Dynes}}, \bibinfo {author} {\bibfnamefont {V.}~\bibnamefont
  {Narayanamurti}}, \ and\ \bibinfo {author} {\bibfnamefont {J.~P.}\
  \bibnamefont {Garno}},\ }\href {\doibase 10.1103/PhysRevLett.41.1509}
  {\bibfield  {journal} {\bibinfo  {journal} {Phys. Rev. Lett.}\ }\textbf
  {\bibinfo {volume} {41}},\ \bibinfo {pages} {1509} (\bibinfo {year}
  {1978})}\BibitemShut {NoStop}%
\bibitem [{\citenamefont {Oda}\ \emph {et~al.}(1996)\citenamefont {Oda},
  \citenamefont {Manabe},\ and\ \citenamefont {Ido}}]{Oda1996Feb}%
  \BibitemOpen
  \bibfield  {author} {\bibinfo {author} {\bibfnamefont {M.}~\bibnamefont
  {Oda}}, \bibinfo {author} {\bibfnamefont {C.}~\bibnamefont {Manabe}}, \ and\
  \bibinfo {author} {\bibfnamefont {M.}~\bibnamefont {Ido}},\ }\href {\doibase
  10.1103/PhysRevB.53.2253} {\bibfield  {journal} {\bibinfo  {journal} {Phys.
  Rev. B}\ }\textbf {\bibinfo {volume} {53}},\ \bibinfo {pages} {2253}
  (\bibinfo {year} {1996})}\BibitemShut {NoStop}%
\bibitem [{\citenamefont {Howald}\ \emph {et~al.}(2001)\citenamefont {Howald},
  \citenamefont {Fournier},\ and\ \citenamefont {Kapitulnik}}]{Howald2001Aug}%
  \BibitemOpen
  \bibfield  {author} {\bibinfo {author} {\bibfnamefont {C.}~\bibnamefont
  {Howald}}, \bibinfo {author} {\bibfnamefont {P.}~\bibnamefont {Fournier}}, \
  and\ \bibinfo {author} {\bibfnamefont {A.}~\bibnamefont {Kapitulnik}},\
  }\href {\doibase 10.1103/PhysRevB.64.100504} {\bibfield  {journal} {\bibinfo
  {journal} {Phys. Rev. B}\ }\textbf {\bibinfo {volume} {64}},\ \bibinfo
  {pages} {100504} (\bibinfo {year} {2001})}\BibitemShut {NoStop}%
\bibitem [{\citenamefont {H{\ifmmode\ddot{a}\else\"{a}\fi}hnle}\ \emph
  {et~al.}(2021)\citenamefont {H{\ifmmode\ddot{a}\else\"{a}\fi}hnle},
  \citenamefont {Kouwenhoven}, \citenamefont {Buijtendorp}, \citenamefont
  {Endo}, \citenamefont {Karatsu}, \citenamefont {Thoen}, \citenamefont
  {Murugesan},\ and\ \citenamefont {Baselmans}}]{Hahnle2021Jul}%
  \BibitemOpen
  \bibfield  {author} {\bibinfo {author} {\bibfnamefont {S.}~\bibnamefont
  {H{\ifmmode\ddot{a}\else\"{a}\fi}hnle}}, \bibinfo {author} {\bibfnamefont
  {K.}~\bibnamefont {Kouwenhoven}}, \bibinfo {author} {\bibfnamefont
  {B.}~\bibnamefont {Buijtendorp}}, \bibinfo {author} {\bibfnamefont
  {A.}~\bibnamefont {Endo}}, \bibinfo {author} {\bibfnamefont {K.}~\bibnamefont
  {Karatsu}}, \bibinfo {author} {\bibfnamefont {D.~J.}\ \bibnamefont {Thoen}},
  \bibinfo {author} {\bibfnamefont {V.}~\bibnamefont {Murugesan}}, \ and\
  \bibinfo {author} {\bibfnamefont {J.~J.~A.}\ \bibnamefont {Baselmans}},\
  }\href {\doibase 10.1103/PhysRevApplied.16.014019} {\bibfield  {journal}
  {\bibinfo  {journal} {Phys. Rev. Appl.}\ }\textbf {\bibinfo {volume} {16}},\
  \bibinfo {pages} {014019} (\bibinfo {year} {2021})}\BibitemShut {NoStop}%
\bibitem [{\citenamefont {Shan}\ and\ \citenamefont
  {Ezaki}(2024)}]{Shan2024Sep}%
  \BibitemOpen
  \bibfield  {author} {\bibinfo {author} {\bibfnamefont {W.}~\bibnamefont
  {Shan}}\ and\ \bibinfo {author} {\bibfnamefont {S.}~\bibnamefont {Ezaki}},\
  }\href {\doibase 10.1063/5.0226792} {\bibfield  {journal} {\bibinfo
  {journal} {Appl. Phys. Lett.}\ }\textbf {\bibinfo {volume} {125}},\ \bibinfo
  {pages} {112601} (\bibinfo {year} {2024})}\BibitemShut {NoStop}%
\bibitem [{\citenamefont {Buijtendorp}\ \emph {et~al.}(2025)\citenamefont
  {Buijtendorp}, \citenamefont {Endo}, \citenamefont {Jellema}, \citenamefont
  {Karatsu}, \citenamefont {Kouwenhoven}, \citenamefont {Lamers}, \citenamefont
  {van~der Linden}, \citenamefont {Rostem}, \citenamefont {Veen}, \citenamefont
  {Wollack}, \citenamefont {Baselmans},\ and\ \citenamefont
  {Vollebregt}}]{Buijtendorp2025Jan}%
  \BibitemOpen
  \bibfield  {author} {\bibinfo {author} {\bibfnamefont {B.~T.}\ \bibnamefont
  {Buijtendorp}}, \bibinfo {author} {\bibfnamefont {A.}~\bibnamefont {Endo}},
  \bibinfo {author} {\bibfnamefont {W.}~\bibnamefont {Jellema}}, \bibinfo
  {author} {\bibfnamefont {K.}~\bibnamefont {Karatsu}}, \bibinfo {author}
  {\bibfnamefont {K.}~\bibnamefont {Kouwenhoven}}, \bibinfo {author}
  {\bibfnamefont {D.}~\bibnamefont {Lamers}}, \bibinfo {author} {\bibfnamefont
  {A.~J.}\ \bibnamefont {van~der Linden}}, \bibinfo {author} {\bibfnamefont
  {K.}~\bibnamefont {Rostem}}, \bibinfo {author} {\bibfnamefont {H.~M.}\
  \bibnamefont {Veen}}, \bibinfo {author} {\bibfnamefont {E.~J.}\ \bibnamefont
  {Wollack}}, \bibinfo {author} {\bibfnamefont {J.~J.~A.}\ \bibnamefont
  {Baselmans}}, \ and\ \bibinfo {author} {\bibfnamefont {S.}~\bibnamefont
  {Vollebregt}},\ }\href {\doibase 10.1103/PhysRevApplied.23.014035} {\bibfield
   {journal} {\bibinfo  {journal} {Phys. Rev. Appl.}\ }\textbf {\bibinfo
  {volume} {23}},\ \bibinfo {pages} {014035} (\bibinfo {year}
  {2025})}\BibitemShut {NoStop}%
\bibitem [{\citenamefont {Kamal}\ \emph {et~al.}(2015)\citenamefont {Kamal},
  \citenamefont {Chakrabarti},\ and\ \citenamefont {Ezawa}}]{Kamal2015Aug}%
  \BibitemOpen
  \bibfield  {author} {\bibinfo {author} {\bibfnamefont {C.}~\bibnamefont
  {Kamal}}, \bibinfo {author} {\bibfnamefont {A.}~\bibnamefont {Chakrabarti}},
  \ and\ \bibinfo {author} {\bibfnamefont {M.}~\bibnamefont {Ezawa}},\ }\href
  {\doibase 10.1088/1367-2630/17/8/083014} {\bibfield  {journal} {\bibinfo
  {journal} {New J. Phys.}\ }\textbf {\bibinfo {volume} {17}},\ \bibinfo
  {pages} {083014} (\bibinfo {year} {2015})}\BibitemShut {NoStop}%
\bibitem [{\citenamefont {Owen}\ and\ \citenamefont
  {Scalapino}(1971)}]{Owen1971Oct}%
  \BibitemOpen
  \bibfield  {author} {\bibinfo {author} {\bibfnamefont {C.~S.}\ \bibnamefont
  {Owen}}\ and\ \bibinfo {author} {\bibfnamefont {D.~J.}\ \bibnamefont
  {Scalapino}},\ }\href {\doibase 10.1016/0031-8914(71)90320-X} {\bibfield
  {journal} {\bibinfo  {journal} {Physica}\ }\textbf {\bibinfo {volume} {55}},\
  \bibinfo {pages} {691} (\bibinfo {year} {1971})}\BibitemShut {NoStop}%
\bibitem [{\citenamefont {Kresin}(1987)}]{Kresin1987Jun}%
  \BibitemOpen
  \bibfield  {author} {\bibinfo {author} {\bibfnamefont {V.~Z.}\ \bibnamefont
  {Kresin}},\ }\href {\doibase 10.1016/0375-9601(87)90744-4} {\bibfield
  {journal} {\bibinfo  {journal} {Phys. Lett. A}\ }\textbf {\bibinfo {volume}
  {122}},\ \bibinfo {pages} {434} (\bibinfo {year} {1987})}\BibitemShut
  {NoStop}%
\bibitem [{\citenamefont {Tajik}(2018)}]{Tajik2018}%
  \BibitemOpen
  \bibfield  {author} {\bibinfo {author} {\bibfnamefont {S.}~\bibnamefont
  {Tajik}},\ }\emph {\bibinfo {title} {Effect of Rattling Phonons on
  Superconductivity of KOs$_2$O$_6$}},\ \href
  {https://brocku.scholaris.ca/items/9dadd2fc-a49d-4d0c-a5b1-365f91636fb4/full}
  {\bibinfo {type} {{M.Sc. Thesis}}},\ \bibinfo  {school} {Brock University},
  \bibinfo {address} {St. Catharines, Ontario, Canada} (\bibinfo {year}
  {2018})\BibitemShut {NoStop}%
\bibitem [{\citenamefont {Anderson}(1959)}]{Anderson1959Sep}%
  \BibitemOpen
  \bibfield  {author} {\bibinfo {author} {\bibfnamefont {P.~W.}\ \bibnamefont
  {Anderson}},\ }\href {\doibase 10.1016/0022-3697(59)90036-8} {\bibfield
  {journal} {\bibinfo  {journal} {J. Phys. Chem. Solids}\ }\textbf {\bibinfo
  {volume} {11}},\ \bibinfo {pages} {26} (\bibinfo {year} {1959})}\BibitemShut
  {NoStop}%
\bibitem [{\citenamefont {Carbotte}(1987)}]{Carbotte1987}%
  \BibitemOpen
  \bibfield  {author} {\bibinfo {author} {\bibfnamefont {J.~P.}\ \bibnamefont
  {Carbotte}},\ }in\ \href {\doibase 10.1007/978-1-4613-1937-5_9} {\emph
  {\bibinfo {booktitle} {{Novel Superconductivity}}}}\ (\bibinfo  {publisher}
  {Springer, Boston, MA},\ \bibinfo {address} {Boston, MA, USA},\ \bibinfo
  {year} {1987})\ pp.\ \bibinfo {pages} {73--81}\BibitemShut {NoStop}%
\bibitem [{\citenamefont {Carbotte}\ \emph {et~al.}(1986)\citenamefont
  {Carbotte}, \citenamefont {Marsiglio},\ and\ \citenamefont
  {Mitrovi{\ifmmode\acute{c}\else\'{c}\fi}}}]{Carbotte1986May}%
  \BibitemOpen
  \bibfield  {author} {\bibinfo {author} {\bibfnamefont {J.~P.}\ \bibnamefont
  {Carbotte}}, \bibinfo {author} {\bibfnamefont {F.}~\bibnamefont {Marsiglio}},
  \ and\ \bibinfo {author} {\bibfnamefont {B.}~\bibnamefont
  {Mitrovi{\ifmmode\acute{c}\else\'{c}\fi}}},\ }\href {\doibase
  10.1103/PhysRevB.33.6135} {\bibfield  {journal} {\bibinfo  {journal} {Phys.
  Rev. B}\ }\textbf {\bibinfo {volume} {33}},\ \bibinfo {pages} {6135}
  (\bibinfo {year} {1986})}\BibitemShut {NoStop}%
\bibitem [{\citenamefont {Marsiglio}\ \emph {et~al.}(1988)\citenamefont
  {Marsiglio}, \citenamefont {Schossmann},\ and\ \citenamefont
  {Carbotte}}]{Marsiglio1988Apr}%
  \BibitemOpen
  \bibfield  {author} {\bibinfo {author} {\bibfnamefont {F.}~\bibnamefont
  {Marsiglio}}, \bibinfo {author} {\bibfnamefont {M.}~\bibnamefont
  {Schossmann}}, \ and\ \bibinfo {author} {\bibfnamefont {J.~P.}\ \bibnamefont
  {Carbotte}},\ }\href {\doibase 10.1103/PhysRevB.37.4965} {\bibfield
  {journal} {\bibinfo  {journal} {Phys. Rev. B}\ }\textbf {\bibinfo {volume}
  {37}},\ \bibinfo {pages} {4965} (\bibinfo {year} {1988})}\BibitemShut
  {NoStop}%
\bibitem [{\citenamefont {Bardeen}\ \emph
  {et~al.}(1957{\natexlab{b}})\citenamefont {Bardeen}, \citenamefont {Cooper},\
  and\ \citenamefont {Schrieffer}}]{Bardeen1957Dec}%
  \BibitemOpen
  \bibfield  {author} {\bibinfo {author} {\bibfnamefont {J.}~\bibnamefont
  {Bardeen}}, \bibinfo {author} {\bibfnamefont {L.~N.}\ \bibnamefont {Cooper}},
  \ and\ \bibinfo {author} {\bibfnamefont {J.~R.}\ \bibnamefont {Schrieffer}},\
  }\href {\doibase 10.1103/PhysRev.108.1175} {\bibfield  {journal} {\bibinfo
  {journal} {Phys. Rev.}\ }\textbf {\bibinfo {volume} {108}},\ \bibinfo {pages}
  {1175} (\bibinfo {year} {1957}{\natexlab{b}})}\BibitemShut {NoStop}%
\end{thebibliography}%

\end{document}


\title{
Supplemental Material:  Tailoring Superconductivity with Two-Level Systems
}

\author{Joshuah T. Heath}
\email{joshuah.t.heath@su.se}
\author{Alexander C. Tyner}

\affiliation{Nordita, Stockholm University and KTH Royal Institute of Technology, Hannes Alfvéns väg 12, SE-106 91 Stockholm, Sweden}
\affiliation{Department of Physics, University of Connecticut, Storrs, Connecticut 06269, USA}


\author{\\ S. Pamir Alpay}
\affiliation{Institute of Material Science, University of Connecticut, Storrs, Connecticut 06269, USA}
\affiliation{Department of Materials Science and Engineering, University of Connecticut, Storrs, Connecticut 06269, USA}

\author{Peter Krogstrup}

\affiliation{NNF Quantum Computing Programme, Niels Bohr Institute, University of Copenhagen, Denmark}


\author{Alexander V. Balatsky}
\email{balatsky@kth.se}
\affiliation{Nordita, Stockholm University and KTH Royal Institute of Technology, Hannes Alfvéns väg 12, SE-106 91 Stockholm, Sweden}
\affiliation{Department of Physics, University of Connecticut, Storrs, Connecticut 06269, USA}

\date{\today}

\begin{abstract}
\noindent 
\end{abstract}

\pacs{1}

\maketitle

\tableofcontents




\section{Analytic calculation of the critical temperature $T_c$}

\subsection{Derivation of the Leavens-Carbotte Formula in the bulk superconductor}

As explained in the main text, we identify two-level systems (TLSs) as a localized soft-frequency mode (vibron). For this reason, the full effect of TLSs on the superconductor will not be captured by BCS theory, which is agnostic towards the details of the underlying phonons. If we wish to go beyond BCS theory and incorporate the details of the underlying phonon spectrum in an analytical calculation of $T_c$, we must turn to Eliashberg theory~\cite{Eliashberg1960, Eliashberg1961, RickayzenBook, RevModPhys.62.1027, Marsiglio2020Jun}. In this respect, it is quite typical to take the Allen-Dynes formula~\cite{Allen1975Aug}, which is a highly accurate formula for $T_c$ in bulk superconductors. However, this formula is dependent on specific parameters found by numerically fitting to bulk, isotropic data, and therefore may not be appropriate in the thin-film limit. From this reasoning, we instead turn to the result of Leavens and Carbotte~\cite{Leavens2011Feb}, which is nearly identical to the expression for $T_c$ given in the Allen-Dynes equation, without any fitting parameters. Our negligence of any fine-tuned fitting parameters is justified by looking at the ratios of physical quantities of interest (i.e., $\Delta_1$ and $T_c$) with their "clean" counterparts (i.e., $\Delta_1^{(0)}$ and $T_c^{(0)}$, respectively). Looking at $\Delta_1/\Delta_1^{(0)}$ and $T_c/T_c^{(0)}$ quantifies the change in these quantities induced by a finite TLS population without the need for including fine-tuned fitting parameters. Note that, in this way, we assume that such fitting parameters do not depend on a large TLS density, and thus such dimensionless ratios should capture the salient effects of introducing dynamical defects. 

Before turning to the TLS-modified equation for $T_c$, we will re-derive the Leavens-Carbotte equation. We begin by recalling the real frequency axis Eliashberg equations, as derived by Scalapino, Schrieffer, and Wilkins (SSW)~\cite{Schrieffer1963Apr,Parks1}:

\begin{align}
\Delta(\omega)Z(\omega)=\int_0^{\omega_c}d\omega' \Re\left[\dfrac{\Delta(\omega')}{\sqrt{\omega'^2-\Delta^2(\omega')}}\right]\int_0^\infty d\nu\,\alpha^2 F(\nu)\bigg\{
&[n(\nu)+f(-\omega')]\bigg(
\dfrac{1}{\omega+\omega'+\nu+i\delta}-\dfrac{1}{\omega-\omega'-\nu+i\delta}
\bigg)\notag\\
-&[n(\nu)+f(\omega')]\left(
\dfrac{1}{\omega+\nu-\omega'+i\delta}-\dfrac{1}{\omega-\nu+\omega'+i\delta}
\right)
\bigg\}\notag\\
-\mu\int_0^{\omega_c}d\omega' \Re\bigg[\dfrac{\Delta(\omega')}{\sqrt{\omega'^2-\Delta^2(\omega')}}\bigg](1-2f(\omega')).\label{SSW1}
\end{align}

\begin{align}
Z(\omega)&=1-\dfrac{1}{\omega}\int_0^\infty d\omega' \Re \left[\dfrac{\omega'}{\sqrt{\omega'^2-\Delta^2(\omega')}}\right]
\int_0^\infty d\nu \alpha^2 F(\nu) \bigg\{
[n(\nu)+f(-\omega')]\bigg(\dfrac{1}{\omega+\omega'+\nu+i\delta}+\dfrac{1}{\omega-\omega'-\nu+i\delta}\bigg)\notag\\
&\phantom{=1-\dfrac{1}{\omega}\int_0^\infty d\omega' \Re \left[\dfrac{\omega'}{\sqrt{\omega'^2\Delta^2(\omega')}}\right]
\int_0^\infty d\nu \alpha^2 F(\nu) \bigg\{l}
+[n(\nu)+f(\omega')]\bigg(
\dfrac{1}{\omega-\omega'+\nu+i\delta}+\dfrac{1}{\omega-\nu+\omega'+i\delta}
\bigg)
\bigg\}.\label{SSW2}
\end{align}

\noindent Note that while the above equations will prove useful in our present analytical calculation, the expression for the real axis gap is quite tedious and difficult to compute numerically~\cite{Scalapino1965Jan,Marsiglio1988Apr}. For the real-axis numerics to be considered later, we will turn to the exact analytical continuation of the imaginary-axis solutions to the real frequency axis derived by Marsiglio, Schossmann, and Carbotte (MSC)~\cite{Marsiglio1988Apr}.

Let us focus on the equation for the gap function. In the limit of $T=T_c$ and ignoring the boson factors (i.e., neglecting real boson scattering processes), we obtain the linearised SSW real-axis equation for $\Delta(\omega)$:
\begin{align}
\Delta(\omega)Z(\omega)=\int_0^{\omega_c}d\omega' \dfrac{\Delta(\omega')}{|\omega'|}\int_0^\infty d\nu\,\alpha^2 F(\nu)\bigg\{
&
\dfrac{f(-\omega')}{\omega+\omega'+\nu+i\delta}-\dfrac{f(-\omega')}{\omega-\omega'-\nu+i\delta}
\notag\\
-&
\dfrac{f(\omega')}{\omega+\nu-\omega'+i\delta}+\dfrac{f(\omega')}{\omega-\nu+\omega'+i\delta}
\bigg\}
-\mu\int_0^{\omega_c}d\omega' \dfrac{\Delta(\omega')}{|\omega'|}(1-2f(\omega')).
\end{align}

\noindent We shall now take the two-step approach of Leavens and Carbotte:

\begin{align}
    \Delta(\omega)=\begin{cases}
&\Delta_0,\quad 0<\omega<\omega_0\notag\\
&\Delta_\infty,\quad \phantom{0<}\omega>\omega_0,
    \end{cases}
\end{align}

\begin{align}
    Z(\omega)=\begin{cases}1+\lambda,\quad  1<\omega<\omega_0\notag\\
    1,\phantom{+\lambda\,\,\,}\quad \phantom{1<\,}\omega>\omega_0.
    \end{cases}
\end{align}

\noindent We assume that for the frequency regime $0<\omega<\omega_0$ for some $\omega_0<\omega_c$ that the gap is equal to its $\omega=0$ value. Above $\omega_0$, the gap is equal to its value at the highest phonon frequency, which we take to be $\omega_c$. 

From the above assumption, we can read off each form of $\Delta_0$ and $\Delta_\infty$ as follows:

\begin{align}
    \Delta_\infty=-\mu \bigg[
    \Delta_0 \int_0^{\omega_0}\dfrac{d\omega'}{\omega'}(1-2f(\omega'))
    +\Delta_\infty \int_{\omega_0}^{\omega_c}\dfrac{d\omega'}{\omega'}(1-2f(\omega'))
    \bigg],
\end{align}

\begin{align}
(1+\lambda)\Delta_0&=\int_0^{\omega_0} d\omega' \dfrac{\Delta_0}{\omega'}\int_0^\infty d\nu \alpha^2 F(\nu)\bigg[
\dfrac{2f(-\omega')}{\omega'+\nu}+\dfrac{2f(\omega')}{\omega'-\nu}
\bigg]+\int_{\omega_0}^{\omega_c}d\omega' \dfrac{\Delta_\infty}{\omega'}\int_0^\infty d\nu \alpha^2 F(\nu)\dfrac{2}{\omega'+\nu}\notag\\
&-\mu \bigg[
\Delta_0 \int_0^{\omega_0}\dfrac{d\omega'}{\omega'}(1-2f(\omega'))+\Delta_\infty \int_{\omega_0}^{\omega_c}\dfrac{d\omega'}{\omega'}(1-2f(\omega'))
\bigg]\notag\\
&\notag\\
&=\int_0^{\omega_0} d\omega' \dfrac{\Delta_0}{\omega'}\int_0^\infty d\nu \alpha^2 F(\nu)\bigg[
\dfrac{2f(-\omega')}{\omega'+\nu}+\dfrac{2f(\omega')}{\omega'-\nu}
\bigg]+\int_{\omega_0}^{\omega_c}d\omega' \dfrac{\Delta_\infty}{\omega'}\int_0^\infty d\nu \alpha^2 F(\nu)\dfrac{2}{\omega'+\nu}+ \Delta_\infty.
\end{align}

\noindent In the equation for $\Delta_\infty$, only terms related to $\mu$ remain as we take $\omega \rightarrow \infty$. Likewise, in the equation for $\Delta_0$, the regime of the integral over the region $\omega_0$ to $\omega_c$ is at a high enough frequency we assume $f(-\omega')=1$ and $f(\omega')=0$. Note that the equation for $\Delta_\infty$ may be rearranged to yield

\begin{align}
    \Delta_\infty&=-\Delta_0 \int_0^{\omega_0}\dfrac{d\omega'}{\omega'}\tanh\left({\beta_c\omega'}/{2}\right)\dfrac{\mu}{1+\mu \int_{\omega_0}^{\omega_c}\dfrac{d\omega'}{\omega'}\tanh(\beta_c \omega'/2)}\notag\\
    &\approx-\Delta_0\log(1.13\omega_0\beta_c)\dfrac{\mu}{
1+\mu \log(\omega_c/\omega_0)
    }\notag\\
    &\equiv -\Delta_0\mu^*\log(1.13\omega_0\beta_c)
\end{align}

\noindent where we have taken $1-2f(\omega')=\tanh(\omega'\beta_c/2)$ where $\beta_c=1/(k_BT_c)$.  The second term in the equation for $\Delta_0$ can likewise be solved as follows:

\begin{align}
\int_{\omega_0}^{\omega_c}d\omega'\dfrac{\Delta_\infty}{\omega'}\int_0^\infty d\nu\,\alpha^2 F(\nu)\dfrac{2}{\omega'+\nu}&\approx \Delta_\infty \int_0^{\infty}d\nu \alpha^2 F(\nu)\dfrac{2}{\nu} \log(1+\nu/\omega_0)\notag\\
&\approx \Delta_\infty \dfrac{2}{\omega_0}\int_0^\infty d\nu \alpha^2F(\nu)\notag\\
&\equiv \dfrac{2\Delta_\infty A}{\omega_0},
\end{align}

\noindent where in the second line we have assumed $\omega_c\gg\omega_0$ and the most important regime of $\alpha^2 F(\nu)$ is dependent on frequencies $\nu\ll \omega_c$, and in the last line we have assumed $\nu\ll \omega_0$. We have also taken $A\equiv \int_0^\infty d\nu\,\alpha^2 F(\nu)$ to be the area under the spectral density. This all reduces the equation for $\Delta_0$ to the following:

\begin{align}
1+\lambda&=2\int_0^{\omega_0}  \dfrac{d\omega'}{\omega'}\int_0^\infty d\nu \alpha^2 F(\nu)\bigg[
\dfrac{f(-\omega')}{\omega'+\nu}+\dfrac{f(\omega')}{\omega'-\nu}
\bigg]
-\left(\dfrac{2 A}{\omega_0}
+1\right)\mu^*\log(1.13\omega_0\beta_c).
\end{align}
The first insight of Leavens and Carbotte comes into play when we consider the typical form of the real component of the real-frequency axis gap function, $\Delta_1(\omega)$, found by numerically solving Eqns.~\eqref{SSW1} and \eqref{SSW2} with $\alpha^2 F(\nu)$ derived from tunnelling experiment data. Upon increasing the frequency, $\Delta_1(\omega)$ increases until it reaches the maximum phonon frequency of $\alpha^2 F(\nu)$, after which $\Delta_1(\omega)$ drops from it's maximum and becomes negative before gradually increasing towards some small negative value at large frequency. As such, approximating the gap function with $\Delta_0$ up to the maximum phonon frequency $\omega_0$ {\it underestimates} the contribution of this integral~\cite{Leavens1974Jan}, and thus to remedy this we drop the first term proportional to $\Delta_\infty$ in Eqn.~\eqref{SSW1}. Thus, the previous equation becomes 

\begin{align}
1+\lambda&=2\int_0^{\omega_0}  \dfrac{d\omega'}{\omega'}\int_0^\infty d\nu \alpha^2 F(\nu)\bigg[
\dfrac{f(-\omega')}{\omega'+\nu}+\dfrac{f(\omega')}{\omega'-\nu}
\bigg]
-\mu^*\log(1.13\omega_0\beta_c).
\end{align}

The second insight of Leavens and Carbotte comes into play by noting that the function $(1/\omega')f(\omega')/(\nu-\omega')$ is peaked around $\omega'\approx 0$, and thus we can replace this factor by $(1/\omega')f(\omega')/(\nu+\omega')$ if the imporant frequency in $\alpha^2 F(\nu)$ are large compared to $k_B T_c$. Thus the above is further simplified to 

\begin{align}
    1+\lambda =2\int_0^{\omega_0}\dfrac{d\omega'}{\omega'}\int_0^\infty d\nu \alpha^2 F(\nu)\dfrac{\tanh(\omega'\beta_c/2)}{\omega'+\nu}-\mu^* \log(1.13\omega_0\beta_c),
\end{align}
where we have used the fact that $f(-\omega')-f(\omega')=\tanh(\beta_c\omega'/2)$. We will now simplify the first integral:

\begin{align}
&\phantom{=}2\int_0^\infty d\nu \alpha^2F(\nu)\int_0^{\omega_0}\dfrac{d\omega'}{\omega'}\dfrac{\tanh(\omega'\beta_c/2}{\omega'+\nu}\notag\\
&=2\int_0^\infty d\nu \alpha^2F(\nu)\int_0^{\omega_0}\tanh(\omega'\beta_c/2)\left\{
\dfrac{1}{\nu\omega'}-\dfrac{1}{\nu(\omega'+\nu)}
\right\}\notag\\
&\approx2\int_0^\infty d\nu \alpha^2F(\nu)\int_0^{\omega_0}\left\{
\dfrac{\tanh(\omega'\beta_c/2)}{\nu\omega'}-\dfrac{\tanh((\omega'+\nu)\beta_c/2)}{\nu(\omega'+\nu)}
\right\}\notag\\
&=2\int_0^\infty d\nu \dfrac{\alpha^2F(\nu)}{\nu}\int_0^{\omega_0}
\dfrac{\tanh(\omega'\beta_c/2)}{\omega'}-
2\int_0^\infty d\nu \dfrac{\alpha^2F(\nu)}{\nu}\int_\nu^{\nu+\omega_0}\dfrac{\tanh(x\beta_c/2)}{x}\notag\\
&=\lambda \log(1.13\omega_0\beta_c)-\bar{\lambda},
\end{align}
where we have used once again the fact that $1/(\omega'(\nu+\omega'))$ is peaked about $\omega'$, we have invoked the form of $\lambda$ given in the main text, and we have additionally defined a new moment of $\alpha^2 F(\nu)$:
\begin{align}
\bar{\lambda}=2\int_0^{\infty}d\nu \dfrac{\alpha^2F(\nu)}{\nu}\log(1+\omega_0/\nu).
\end{align}
The above is the third major insight of Leavens and Carbotte. The equation for $T_c$ can then easily be solved to yield

\begin{align}
k_B T_c =1.13 \omega_0 \exp \left(-\dfrac{1+\lambda +\bar{\lambda}}{\lambda-\mu^*}\right).
\end{align}

We can simplify further to eliminate the parameter $\omega_0$. Assuming again that $\nu\ll \omega_0$, then $\bar{\lambda}$ can be rewritten in terms of the log average phonon frequency:
\begin{align}
\bar{\lambda}\approx 2\int_0^{\infty}d\nu \dfrac{\alpha^2F(\nu)}{\nu}\log(\omega_0/\nu)=\lambda \log(\omega_0/\omega_{\ln}).
\end{align}
As such,
\begin{align}
    T_c & \approx 1.13\dfrac{\omega_0}{k_B} \exp\left(-\dfrac{1+\lambda +\lambda \log(\omega_0/\omega_{\ln})}{\lambda-\mu^*}\right)\notag\\
    & =1.13\dfrac{\omega_0}{k_B}\exp\bigg(\lambda \dfrac{\log(\omega_{\ln}/\omega_0)}{\lambda-\mu^*}\bigg) \exp\left(-\dfrac{1+\lambda }{\lambda-\mu^*}\right)\notag\\
    &\approx 1.13\dfrac{\omega_{\ln}}{k_B}\exp\left(-\dfrac{1+\lambda }{\lambda-\mu^*}\right),
\end{align}
where we have assumed that $\lambda/(\lambda-\mu^*)\approx 1$. In this way, we have explicitly derived the dependence of $T_c$ on $\omega_{\ln}$ without resorting to numerically-fitted parameters or empirically-motivated ansatzes.

\subsection{The Leavens-Carbotte formula in the thin-film superconductor}

In the above, we have derived the Leavens-Carbotte formula assuming a bulk superconductor. However, as discussed in our article, we mostly concern our study with the thin-film superconductor down to 1 monolayer of aluminium, and thus we want to ensure that the Leavens-Carbotte formula remains appropriate as we approach the monolayer limit. To do this, we'll build off recent work on quantum confinement in thin film superconductors~\cite{Travaglino2023Jan,Ummarino2024Nov}. As we progress towards the monolayer limit, the Fermi surface becomes severely distorted below a critical length scale $L_c=(2\pi/n_0)^{1/3}$ (where $n_0$ is the carrier density), thus resulting in an electronic density of states that can no longer be approximated as a constant. As such, if we include an energy-dependent density of states and a finite bandwidth, the real-axis SSW equation become

\begin{align}
&\Delta(\omega)Z(\omega)
=\notag\\
&\int_0^{\omega_c}d\omega' \mathcal{B}(\omega')\Re\left[\dfrac{\Delta(\omega')}{\sqrt{\omega'^2-\Delta^2(\omega')}}\right] \int_0^\infty d\nu\,\alpha^2 F(\nu)\bigg\{
[n(\nu)+f(-\omega')]\bigg(
\dfrac{1}{\omega+\omega'+\nu+i\delta}-\dfrac{1}{\omega-\omega'-\nu+i\delta}
\bigg)\notag\\
&\phantom{\int_0^{\omega_c}d\omega' \mathcal{B}(\omega')\Re\left[\dfrac{\Delta(\omega')}{\sqrt{\omega'^2-\Delta^2(\omega')}}\right] \int_0^\infty d\nu\,\alpha^2 F(\nu)}-[n(\nu)+f(\omega')]\left(
\dfrac{1}{\omega+\nu-\omega'+i\delta}-\dfrac{1}{\omega-\nu+\omega'+i\delta}
\right)
\bigg\}\notag\\
&-\mu\int_0^{\omega_c}d\omega' \mathcal{B}(\omega')\Re\bigg[\dfrac{\Delta(\omega')}{\sqrt{\omega'^2-\Delta^2(\omega')}}\bigg](1-2f(\omega')),
\end{align}

\noindent where
\begin{align}
    \mathcal{B}(\omega')=\Re\left[N(\epsilon(\omega))\dfrac{2}{\pi}\arctan\left(
    \dfrac{W}{2\epsilon(\omega)}
    \right)\right]
\end{align}
with  $\epsilon(\omega)=Z(\omega)\sqrt{\omega^2-\Delta^2(\omega)} $ and we have assumed a simple symmetrical electronic DoS. If we assume that $L<L_c$ (as is the case for $1-6$ layers of Al) and we take the reference energy the Fermi energy (which is taken as the zero of the energy in the Eliashberg equations), then the SSW equations are modified by the term
\begin{align}
    N(x) \propto 1-C |x|
\end{align}
Here, $C\sim \epsilon_F^{-1}$ is a constant and we have ensured that $\lambda$ is defined as in the case of a constant density of states, and thus the main modification is an energy-dependent shift of the electronic DoS.
As $L<L_c$ for the $1-6$ layer system under consideration, we will always have $\epsilon^*>\epsilon_F$, and thus the above is appropriate. 

At $T=T_c$, the factor $\mathcal{B}(\omega')$ is simplified as follows:

\begin{align}
    \mathcal{B}(\omega')
    &=\Re\left[N(|\omega|Z(\omega))\dfrac{2}{\pi}\arctan\left(
    \dfrac{W}{2|\omega|Z(\omega)}
    \right)\right]\notag\\
    &\approx 1-C\omega' Z(\omega')\left\{
\dfrac{8}{\pi }+1
    \right\}\notag\\
    &\equiv 1 -\omega'Z(\omega')C.
\end{align}
where we have assumed that the bandwidth $W$ is much larger than typical energies considered. At $T=T_c$, the SSW real-axis equation for the gap becomes

\begin{align}
\Delta(\omega)Z(\omega)=\int_0^{\omega_c}d\omega' \dfrac{\Delta(\omega')}{|\omega'|}\int_0^\infty d\nu\,\alpha^2 F(\nu)\bigg\{
&
\dfrac{f(-\omega')}{\omega+\omega'+\nu+i\delta}-\dfrac{f(-\omega')}{\omega-\omega'-\nu+i\delta}
\notag\\
-&
\dfrac{f(\omega')}{\omega+\nu-\omega'+i\delta}+\dfrac{f(\omega')}{\omega-\nu+\omega'+i\delta}
\bigg\}
-\mu \int_0^{\omega_c}d\omega' \dfrac{\Delta(\omega')}{|\omega'|}(1-2f(\omega'))\notag\\
-C\int_0^{\omega_c}d\omega' Z(\omega')\Delta(\omega')\int_0^\infty d\nu\,\alpha^2 F(\nu)\bigg\{
&
\dfrac{f(-\omega')}{\omega+\omega'+\nu+i\delta}-\dfrac{f(-\omega')}{\omega-\omega'-\nu+i\delta}
\notag\\
-&
\dfrac{f(\omega')}{\omega+\nu-\omega'+i\delta}+\dfrac{f(\omega')}{\omega-\nu+\omega'+i\delta}
\bigg\}
+C\mu \int_0^{\omega_c}d\omega' Z(\omega')\Delta(\omega')(1-2f(\omega')).
\end{align}

We can now take the same two-step form of the Leavens-Carbotte for the above, in which case we find new equations for $\Delta_0$ and $\Delta_\infty$

\begin{align}
    \Delta_\infty&=-\mu \bigg[
    \Delta_0 \int_0^{\omega_0}\dfrac{d\omega'}{\omega'}(1-2f(\omega'))
    +\Delta_\infty \int_{\omega_0}^{\omega_c}\dfrac{d\omega'}{\omega'}(1-2f(\omega'))
    \bigg]\notag\\
    &\phantom{=}+C\mu \bigg[\Delta_0(1+\lambda)\int_0^{\omega_0}d\omega'(1-2f(\omega'))+\Delta_\infty\int_{\omega_0}^{\omega_c}d\omega'(1-2f(\omega'))  \bigg].
\end{align}

\begin{align}
(1+\lambda)\Delta_0&=\int_0^{\omega_0} d\omega' \dfrac{\Delta_0}{\omega'}\int_0^\infty d\nu \alpha^2 F(\nu)\bigg[
\dfrac{2f(-\omega')}{\omega'+\nu}+\dfrac{2f(\omega')}{\omega'-\nu}
\bigg]+\int_{\omega_0}^{\omega_c}d\omega' \dfrac{\Delta_\infty}{\omega'}\int_0^\infty d\nu \alpha^2 F(\nu)\dfrac{2}{\omega'+\nu}\notag\\
&-C(1+\lambda)\Delta_0\int_0^{\omega_0} d\omega' \int_0^\infty d\nu \alpha^2 F(\nu)\bigg[
\dfrac{2f(-\omega')}{\omega'+\nu}+\dfrac{2f(\omega')}{\omega'-\nu}
\bigg]-C\Delta_\infty\int_{\omega_0}^{\omega_c}d\omega' \int_0^\infty d\nu \alpha^2 F(\nu)\dfrac{2}{\omega'+\nu}+\Delta_\infty.
\end{align}

\noindent Starting with the equation for $\Delta_\infty$, we can rearrange and solve the integrals to find

\begin{align}
    \Delta_\infty &=-\Delta_0 \mu 
    \dfrac{
    \log(1.13\omega_0\beta_c)-C \dfrac{2}{\beta}\log[\cosh(\omega_0\beta/2)]
    }{
1+ \mu \log(\omega_c/\omega_0)-C \mu \dfrac{2}{\beta}\log\bigg[\dfrac{\cosh(\omega_c\beta_c/2)}{\cosh(\omega_0\beta_c/2)}\bigg]
    }\notag\\
&\equiv -\Delta_0 \bar{\mu}^*\log(1.13\omega_0\beta_c).
\end{align}
We thus see that, in the thin film limit, we obtain a new equation for $\Delta_\infty$ dependence on $\bar{\mu}^*$ as defined above. Turning now to the equation for $\Delta_0$, note that we can write

\begin{align}
    \Delta_\infty+\Delta_\infty\int_{\omega_0}^{\omega_c}d\omega'\int_0^\infty d\nu \alpha^2 F(\nu)\dfrac{2}{\omega'+\nu}&\approx \Delta_\infty+\Delta_\infty \int_0^\infty d\nu \alpha^2 F(\nu)\dfrac{2}{\nu} \log(1+\nu/\omega_0)\notag\\
    &\approx-\Delta_0\bar{\mu}^*\log(1.13\omega_0\beta_c)\left(
1+\dfrac{2A}{\omega_0}
    \right),
\end{align}

\begin{align}
 -C \Delta_\infty \int_{\omega_0}^{\omega_c}d\omega'\int_0^{\infty}d\nu\alpha^2 F(\nu)\dfrac{2}{\omega'+\nu} &\approx-C \Delta_\infty\int_0^\infty d\nu \alpha^2 F(\nu) 2\log\left(\dfrac{\omega_c}{\omega_0+\nu}\right)\notag\\
 &\approx2\Delta_0C \bar{\mu}^* \log(1.13\omega_0\beta_c) \left(
A \log(\omega_c/\omega_0)-\dfrac{1}{\omega_0}\int_0^\infty d\nu \nu \alpha^2 F(\nu)
 \right),
\end{align}

\begin{align}
1+\lambda&=2\int_0^{\omega_0}  \dfrac{d\omega'}{\omega'}\int_0^\infty d\nu \alpha^2 F(\nu)\bigg[
\dfrac{f(-\omega')}{\omega'+\nu}+\dfrac{f(\omega')}{\omega'-\nu}
\bigg]
-\left(\dfrac{2 A}{\omega_0}
+1\right)\bar{\mu}^*\log(1.13\omega_0\beta_c)\notag\\
&-2C(1+\lambda)\int_0^{\omega_0}d\omega'\int_0^\infty d\nu \alpha^2 F(\nu)\Bigg[
\dfrac{f(-\omega')}{\omega'+\nu}+\dfrac{f(\omega')}{\omega'-\nu}
\bigg]\notag\\
&+2C \bar{\mu}^* \log(1.13\omega_0\beta_c) \left(
A \log(\omega_c/\omega_0)-\dfrac{1}{\omega_0}\int_0^\infty d\nu \nu \alpha^2 F(\nu)
 \right).
\end{align}
Following the approximation of Leavens and Carbotte, we'll drop the term proportional to $A$ in the $\Delta_\infty$ term, and thus we arrive at

\begin{align}
1+\lambda&=2\int_0^{\omega_0}  \dfrac{d\omega'}{\omega'}\int_0^\infty d\nu \alpha^2 F(\nu)\bigg[
\dfrac{f(-\omega')}{\omega'+\nu}+\dfrac{f(\omega')}{\omega'-\nu}
\bigg]
-\bar{\mu}^*\log(1.13\omega_0\beta_c)\notag\\
&-2C(1+\lambda)\int_0^{\omega_0}d\omega'\int_0^\infty d\nu \alpha^2 F(\nu)\Bigg[
\dfrac{f(-\omega')}{\omega'+\nu}+\dfrac{f(\omega')}{\omega'-\nu}
\bigg]\notag\\
&-2C \bar{\mu}^* \log(1.13\omega_0\beta_c) 
\dfrac{1}{\omega_0}\int_0^\infty d\nu\, \nu \alpha^2 F(\nu).
\end{align}
We already have seen that the first integral yields $\lambda \log(1.13\omega_0 \beta_c)-\bar{\lambda}$. The second integral is evaluated to be the following:
\begin{align}
\int_0^{\omega_0}d\omega ' \bigg[\dfrac{f(-\omega')}{\omega'+\nu}+\dfrac{f(\omega')}{\omega'-\nu}\bigg]&=\int_0^{\omega_0}d\omega ' \dfrac{f(-\omega')}{\omega'+\nu}+\int_0^{-\omega_0}d\omega'\dfrac{f(-\omega')}{\omega'+\nu}\notag\\
&=\dfrac{1}{2}\int_0^{\omega_0}d\omega'\dfrac{1+\tanh(\beta_c \omega/2)}{\omega'+\nu}
+\dfrac{1}{2}\int_0^{-\omega_0}d\omega'\dfrac{1+\tanh(\beta_c \omega/2)}{\omega'+\nu}\notag\\
&\approx \dfrac{2}{\beta\nu}\log\bigg[\cosh(\beta_c\nu/2)\bigg],
\end{align}

\begin{align}
1+\lambda&=\left(\lambda\log(1.13\omega_0\beta_c)-\bar{\lambda}\right)
-\bar{\mu}^*\log(1.13\omega_0\beta_c)\notag\\
&-2\dfrac{C}{\beta_c}(1+\lambda)\int_0^\infty d\nu \alpha^2 F(\nu)\bigg[
\dfrac{2}{\nu}\log(\cosh(\beta_c\nu/2))
\bigg]
\notag\\
&-2\dfrac{C}{\omega_0} \bar{\mu}^* \log(1.13\omega_0\beta_c) 
\int_0^\infty d\nu \,\nu \alpha^2 F(\nu).
\end{align}

\noindent Turning our focus to the integral proportional to $\log(\cosh(...))$, we note that this term will have the most weight about $\nu=0$, where $\cosh(...)$ will be around unity. Thus, we ignore this term. Looking at the other term gained in the thin film limit, we find


\begin{align}
-\lambda  \dfrac{C}{\omega_0}\bar{\mu}^* \log(1.13\omega_0\beta_c)\dfrac{2}{\lambda}\int_0^\infty d\nu \nu \alpha^2 F(\nu)&=-\lambda  \dfrac{C}{\omega_0}\bar{\mu}^*\log(1.13\omega_0\beta_c)\langle \omega^2\rangle.
\end{align}
Therefore, the equation for $T_c$ now becomes

\begin{align}
1+\lambda
&=\log(1.13\omega_0\beta_c)\left(\lambda-\bar{\mu}^*\left(1+\dfrac{\lambda C}{\omega_0}\langle \omega^2\rangle\right)\right)-\bar{\lambda}.
\end{align}

\noindent Now, we can redefine

\begin{align}
    \widetilde{\mu}^*=\mu\dfrac{\left(1+\dfrac{\lambda C}{\omega_0}\langle \omega^2\rangle\right)
\left(
1-\dfrac{2C}{\beta_c}\dfrac{\log[\cosh(\omega_0\beta_c/2)]}{\log(1.13\omega_0\beta_c)} \right)
    }{
1+ \mu \left(\log(\omega_c/\omega_0)- \dfrac{2C}{\beta_c}\log\bigg[\dfrac{\cosh(\omega_c\beta_c/2)}{\cosh(\omega_0\beta_c/2)}\bigg]  \right)
    }.
\end{align}
As $C$ scales as the inverse Fermi energy, we find that $\mu^*$ is, at most, moderately changed by the effects of quantum confinement. Since ab initio calculations reveal $\mu^*\approx 0.1$ in bulk Al and $\mu^*$ is predicted to be bounded above by around $0.2$, this appears to be a reasonable approximation~\cite{Simonato2023Aug,Zaccone2024Apr}.

Turning back to our equation for $T_c$, we now find that 

\begin{align}
    1+\lambda\approx \log(1.13\omega_0\beta_c)\left(\lambda -\mu^*\right)-\bar{\lambda}.
\end{align}
This is the same form we had before, as the above shows the energy-dependent portion of the density of states drops out. As such, up to reasonable accuracy the addition of the energy-dependence on the electronic DoS will still result in
\begin{align}
T_c&\approx 1.13\dfrac{\omega_{\ln}}{k_B}\exp\left(-\dfrac{1+\lambda }{\lambda-\mu^*}\right).
\end{align}
\noindent We therefore predict that the salient features of $T_c$ in the thin film limit should appear by virtue of the more complicated structure of $\alpha^2 F(\omega)$ (which will enter $\omega_{\ln}$ and $\lambda$), as opposed to the energy-dependence of the electronic density of states near the Fermi surface. As the thin film modification of the electronic DoS goes as an energy shift which scales as $\epsilon_F^{-1}$, and $\epsilon_F$ tend to either remain the same or slightly increase in the monolayer limit,~\cite{Feibelman1984Jun,Nguyen2019May,Allen2024Jun}, we will ignore the effects of electronic quantum confinement in both the Leavens-Carbotte analytic result and the numerical solution to the Eliashberg equations.




\subsection{The TLS contribution to $T_c$ in the Einstein approximation}

With the analytical form of $T_c$ established, we will now turn to the TLS contribution to the critical temperature. In order to incorporate the effects of a finite TLS density into quantities of physical interest, we will write the total electron-phonon spectral function follows:


\begin{align}
    \alpha^2 F(\nu)=\alpha^2F^{(0)}(\nu)+n_{TLS}\alpha^2 F_{TLS}(\nu).
\end{align}

\noindent Here, we define $n_{TLS}$ as the total fraction of TLSs that make up the bosonic spectral weight in the material; i.e., the ratio of the TLS density to the density of bulk phonons in the optimal frequency regime. Note that $n_{TLS}$ as we have defined it is not in units of GHz$^{-1}$ $\mu m^{-3}$ as it is typically reported, but rather is a unitless number.

We can get an estimate of this number by looking at reported values for the TLS density per GHz $\langle g_{TLS}\rangle$ in Al and related materials. Whereas typical densities of TLS range from $10^2-10^4$ GHz$^{-1}$ $\mu m^{-3}$~\cite{Muller2019Oct,Martinis2005Nov,Schreier2008May,Gunnarsson2013Jul}, high-end estimates lie within the range of $\langle g_{TLS}\rangle\sim 8.5\times 10^4-2.5\times 10^5$ GHz$^{-1}$ $\mu m^{-3}$~\cite{Queen2013Mar,Wang2025Feb}, with some rare reports of very high densities around $\langle g_{TLS}\rangle\sim 6\times 10^5$ GHz$^{-1}$ $\mu m^{-3}$. The average density of phonons may be estimated within the Debye model for a certain frequency interval $[\nu_2,\nu_1]$ as 

\begin{align}
    \langle g_{ph}\rangle =\dfrac{3n}{\nu_D^3(\nu_2-\nu_1)}(\nu_2^3-\nu_1^3)
\end{align}
where $n\approx 6.02\times 10^{23}$ atoms $\mu m^{-3}$ and the above is derived by taking the integral of the phonon density of states $(9n/\nu_D^3)\nu^2$ in the Debye approximation and dividing by the frequency interval. In the bulk system, the majority of spectral weight is upwards of $5$ THz, however in the monolayer limit a large portion of the spectral weight is shifted downward below 5 THz and eventually, for a single monolayer, to below 1 THz. This yields $\langle g_{ph}\rangle\sim  8.5\times 10^6$ GHz$^{-1}$ $\mu m^{-3}$ and $3\times 10^5$ GHz$^{-1}$ $\mu m^{-3}$, respectively. If we take $\langle g_{TLS}\rangle\sim 10^5$ GHz$^{-1}$ $\mu m^{-3}$ based upon the typical values quoted above, then we obtain $n_{TLS}\sim 0.01-0.3$ as we approach the monolayer limit (with the larger value applicable for single monolayer). Note that this upper limit of $n_{TLS}\sim 0.3$ assumes that the physical density of TLSs is around $10^5$ GHz$^{-1}$ $\mu m^{-3}$, which is quite high but not unreasonable. For more typical TLS densities of $10^2$ GHz$^{-1}$ $\mu m^{-3}$, then $n_{TLS}\approx 0.0001$.

We can similarly estimate $n_{TLS}$ by considering the density of TLSs per unit cell, assuming a "dangling oxygen bond" origin for TLSs. Assuming we have (111) Al, then there are $4\times 3=12$ phonons per unit cell. As we approach the monolayer limit, there will be one Al atom on the surface per unit cell. As the monolayer oxidizes, at most we would then expect one dangling oxygen bond per unit cell (one covalent bond forms with Al, leaving the other one dangling). In the case of a surface alumina layer on bulk Al, we'd expect 3 O atoms for every 2 Al atoms, and thus a maximum of $1-1.5$ dangling oxygen bonds per unit cell. As such, we would expect at most $1-1.5$ TLSs per unit cell, and thus at most $3-4.5$ average vibrons per unit cell. In this simple estimate, we can approximate the maximum $n_{TLS}$ as the ratio of the maximum number of TLSs per unit cell to the number of total modes per unit cell, which is $\sim 0.3$, which is in agreement with our estimate given above based on experimental values for the TLS density. As such, we will assume $n_{TLS}=0.3$ as a maximum value for the unitless TLS density. From the above argument, this roughly corresponds to about one TLS (or equivalently one vibron) per unit cell.

The change in the electron-phonon spectral function modifies both the electron-phonon coupling and the log average frequency. Recalling the form of the Leavens-Carbotte formula for $T_c$, we will rewrite it as
\begin{align}
    T_c k_B &=1.13 \omega_{\ln}\exp\left(-\dfrac{1+\lambda}{\lambda-\mu^*}\right)\notag\\
    &=1.13 \omega_{\ln}\exp\left(-\dfrac{1+\lambda}{\lambda}\dfrac{1}{1-\mu^*/\lambda}\right).
\end{align}
When we include the contribution from TLSs, then both $\mu^*$ will increase from $0.091$ to a maximum of $0.188$, with $\lambda$ increasing from $\sim 0.4$ to a maximum $\sim 5.5$. As such, the contribution proportional to $1/(1-\mu^*/\lambda)$ is at most around $1.3$, with the addition of TLSs resulting in this term becoming closer to unity. As a consequence, ignoring this contribution will give a good approximation to the $T_c$ in the presence of TLSs. 

In deriving the TLS contribution to the critical temperature $T_c$, we'll write the electron-phonon coupling strength as follows, and subsequently take the Einstein phonon model to gain insight:

\begin{align}
    \lambda &=\lambda^{(0)}+n_{TLS}\lambda_{TLS}\notag\\
    &=\lambda^{(0)}+n_{TLS}\lambda^{(0)}\dfrac{\lambda^{TLS}}{\lambda^{(0)}}\notag\\
    &=\lambda^{(0)}\bigg(1+n_{TLS}\dfrac{\omega_E}{\omega_{TLS}}\bigg).
\end{align}

\noindent There is a subtlety in regards to the limit of vanishing TLSs. As $n_{TLS}\rightarrow 0$, we see that $\lambda \rightarrow \lambda^{(0)}$, as expected. However, as $\omega_{TLS}\rightarrow 0$, $\lambda$ diverges for any finite $n_{TLS}$. To remedy this, we note that there is a minimum value for $\omega_{TLS}$, which is directly proportional to the speed of sound in the material and inversely proportional to the sample size. Given that experiments on thin film (111) Al exhibit "kinks" in the I-V curve for frequencies down to $\sim 0.15$ meV, we'd expect that the relevant vibron frequency is bounded below by around $0.01-0.1$ THz. As a consequence of this and the locations of the peaks of $\alpha^2 F(\omega)$ (approximations to the Einstein frequency) to be $\propto 10$ meV, it's reasonable to expect (in the Einstein model) a maximum increase of the electron-phonon coupling from TLSs to be around an order of magnitude. If we take the lowest possible TLS frequency and instead take $n_{TLS}\approx 0.01$, then $\lambda/\lambda^{(0)}\sim 1.67$. If we consider typical TLS frequencies of around $10^2$ GHz$^{-1}$ $\mu m^{-3}$, then $n_{TLS}\approx 0.0001$ as stated before, and $\lambda/\lambda^{(0)}\approx 1.007$.

Ignoring the contribution $\mu^*$ and including the contribution from TLSs in $\lambda$ only, we obtain 

\begin{align}
    k_B T_c&=1.13 \omega_{\ln}\exp\left(-\dfrac{1+\lambda}{\lambda}\right)\notag\\
    &=1.13\omega_{\ln} \exp\left[-\dfrac{1+\lambda^{(0)}+n_{TLS}\lambda_{TLS}}{\lambda^{(0)}+n_{TLS}\lambda_{TLS} }\right]\notag\\
    %
&=k_BT_c^{(0)}\dfrac{\omega_{\ln}}{\omega_{\ln}^{(0)}}\exp\left[\dfrac{1}{\lambda^{(0)}\left(\dfrac{\lambda^{(0)}}{\lambda_{TLS}}n_{TLS}^{-1}+1\right)}\right].
\end{align}

\noindent Note that the above is a general expression that does not dependent on a specific form for the phonon spectrum.

Now, we can solve for the TLS contribution to the log average phonon frequency. To find the vibron contribution to the log average frequency, recall that the log average may be defined as the average of $\omega^n$ to the power of $1/n$ as $n\rightarrow 0$. As such, for a one-boson system,

\begin{align}
\omega_{\ln}&=\lim_{n\rightarrow0}\langle\omega^{n}\rangle^{1/n}\notag \\ 
&=\omega_{0}\lim_{n\rightarrow0}\left(\dfrac{2}{\lambda}\int\dfrac{1}{\nu}d\nu\alpha^{2}F(\nu)(\nu/\omega_{0})^{n}\right)^{1/n}\notag\\
&=\dfrac{2}{\lambda}\omega_{0}\lim_{n\rightarrow0}\left(\int\dfrac{1}{\nu}d\nu\alpha^{2}F(\nu)e^{n\log(\nu/\omega_{0})}\right)^{1/n}\notag \\
&\approx\dfrac{2}{\lambda}\omega_{0}\lim_{n\rightarrow0}\left(\int\dfrac{1}{\nu}d\nu\alpha^{2}F(\nu)\bigg[1+n\log(\nu/\omega_{0})\bigg]\right)^{1/n}\notag \\
&=\omega_{0}\lim_{n\rightarrow0}\left(1+\dfrac{2}{\lambda}n\int\dfrac{1}{\nu}d\nu\alpha^{2}F(\nu)\log(\nu/\omega_{0})\bigg]\right)^{1/n}\notag \\&=\omega_{0}\exp\bigg[\dfrac{2}{\lambda}\int d\nu\log(\nu/\omega_{0})\dfrac{1}{\nu}\alpha^{2}F(\nu)\bigg].
\end{align}

\noindent Using this definition of the log average frequency being the nth root of $\langle \omega^n\rangle^{1/n}$, we find the following for the system in the presence of TLSs:

\begin{align}
   \omega_{\ln}
    &=\lim_{n\rightarrow0}\left(\dfrac{2}{\lambda^{(0)}+\lambda_{TLS}n_{TLS}}\int d\nu\dfrac{1}{\nu}\bigg[\alpha^{2}F_{0}(\nu)+n_{TLS}\alpha^{2}F_{TLS}(\nu)\bigg]\nu^{n}\right)^{1/n}\notag\\
&=\omega_{0}\lim_{n\rightarrow0}\left(\dfrac{2}{\lambda^{(0)}+\lambda_{TLS}n_{TLS}}\int d\nu\dfrac{1}{\nu}\alpha^{2}F_{0}(\nu)e^{n\log(\nu/\omega_{0})}+\dfrac{2n_{TLS}}{\lambda^{(0)}+\lambda_{TLS}n_{TLS}}\int d\nu\dfrac{1}{\nu}\alpha^{2}F_{TLS}(\nu)e^{n\log(\nu/\omega_{0})}\bigg]\right)^{1/n}.
\end{align}

\noindent Looking at each term separately, we can expand each to find the following:

\begin{align}
    \dfrac{2}{\lambda^{(0)}+\lambda_{TLS}n_{TLS}}\int d\nu\dfrac{1}{\nu}\alpha^{2}F_{0}(\nu)e^{n\log(\nu/\omega_{0})}\approx\dfrac{\lambda^{(0)}}{\lambda^{(0)}+\lambda_{TLS}n_{TLS}}+\dfrac{2}{\lambda^{(0)}+\lambda_{TLS}n_{TLS}}n\int d\nu\dfrac{1}{\nu}\alpha^{2}F_{0}(\nu)\log(\nu/\omega_{0}),
\end{align}

\begin{align}
    \dfrac{2n_{TLS}}{\lambda^{(0)}+\lambda_{TLS}n_{TLS}}\int d\nu\dfrac{1}{\nu}\alpha^{2}F_{TLS}(\nu)e^{n\log(\nu/\omega_{0})}\approx\dfrac{\lambda_{TLS}n_{TLS}}{\lambda^{(0)}+\lambda_{TLS}n_{TLS}}+\dfrac{2n_{TLS}}{\lambda^{(0)}+\lambda_{TLS}n_{TLS}}n\int d\nu\dfrac{1}{\nu}\alpha^{2}F_{TLS}(\nu)\log(\nu/\omega_{0}).
\end{align}
The sum of the two simplifies to 

\begin{align}
&\dfrac{\lambda_{0}}{\lambda^{(0)}+\lambda_{TLS}n_{TLS}}+\dfrac{2}{\lambda^{(0)}+\lambda_{TLS}n_{TLS}}n\int d\nu\dfrac{1}{\nu}\alpha^{2}F_{0}(\nu)\log(\nu/\omega_{0})\notag \\&+\dfrac{\lambda_{TLS}n_{TLS}}{\lambda^{(0)}+\lambda_{TLS}n_{TLS}}+\dfrac{2n_{TLS}}{\lambda^{(0)}+\lambda_{TLS}n_{TLS}}n\int d\nu\dfrac{1}{\nu}\alpha^{2}F_{TLS}(\nu)\log(\nu/\omega_{0})\notag \\ &\notag\\
&=1+\dfrac{2}{\lambda^{(0)}+\lambda_{TLS}n_{TLS}}n\int d\nu\dfrac{1}{\nu}\alpha^{2}F_{0}(\nu)\log(\nu/\omega_{0})+\dfrac{2n_{TLS}}{\lambda^{(0)}+\lambda_{TLS}n_{TLS}}n\int d\nu\dfrac{1}{\nu}\alpha^{2}F_{TLS}(\nu)\log(\nu/\omega_{0})\notag\\ &\notag\\ 
&=1+n\bigg[\dfrac{2}{\lambda^{(0)}+\lambda_{TLS}n_{TLS}}\bigg(\int d\nu\dfrac{1}{\nu}\log(\nu/\omega_{0})\bigg[\alpha^{2}F_{0}(\nu)+n_{TLS}\alpha^{2}F_{TLS}(\nu)\bigg]\bigg)\bigg].
\end{align}


\noindent As a consequence, plugging into our previous equation and taking the limit of $n\rightarrow 0$, we find

\begin{align}
\omega_{\ln}
&=\omega_{0}\lim_{n\rightarrow0}\bigg(1+n\bigg[\dfrac{2}{\lambda^{(0)}+\lambda_{TLS}n_{TLS}}\bigg(\int d\nu\dfrac{1}{\nu}\log(\nu/\omega_{0})\bigg[\alpha^{2}F_{0}(\nu)+n_{TLS}\alpha^{2}F_{TLS}(\nu)\bigg]\bigg)\bigg]\bigg)^{1/n}\notag \\& 
\notag \\
&=\omega_{0}\exp\bigg[\dfrac{2}{\lambda^{(0)}+\lambda_{TLS}n_{TLS}}\bigg(\int d\nu\dfrac{1}{\nu}\log(\nu/\omega_{0})\alpha^{2}F_{0}(\nu)\bigg)\bigg]\exp\bigg[\dfrac{2n_{TLS}}{\lambda^{(0)}+\lambda_{TLS}n_{TLS}}\bigg(\int d\nu\dfrac{1}{\nu}\log(\nu/\omega_{0})\alpha^{2}F_{TLS}(\nu)\bigg)\bigg].
\end{align}


\noindent Using the fact that

\begin{align}
    \dfrac{2}{\lambda^{(0)}+\lambda_{TLS}n_{TLS}}=-\dfrac{2\lambda_{TLS}n_{TLS}}{\lambda^{(0)}(\lambda^{(0)}+\lambda_{TLS}n_{TLS})}+\dfrac{2}{\lambda^{(0)}},
\end{align}

\noindent we simplify our final expression to the following:

\begin{align}
    \omega_{\ln}&=\omega_{0}\exp\bigg[\dfrac{2}{\lambda^{(0)}+\lambda_{TLS}n_{TLS}}\bigg(\int d\nu\dfrac{1}{\nu}\log(\nu/\omega_{0})\alpha^{2}F_{0}(\nu)\bigg)\bigg]\exp\bigg[\dfrac{2n_{TLS}}{\lambda^{(0)}+\lambda_{TLS}n_{TLS}}\bigg(\int d\nu\dfrac{1}{\nu}\log(\nu/\omega_{0})\alpha^{2}F_{TLS}(\nu)\bigg)\bigg]\notag\\
    &\notag\\
    &=\omega_{\ln}^{(0)}\exp\bigg[\dfrac{2n_{TLS}}{\lambda^{(0)}+\lambda_{TLS}n_{TLS}}\int d\nu\dfrac{1}{\nu}\log(\nu/\omega_{0})\bigg(\alpha^{2}F_{TLS}(\nu)-\dfrac{\lambda_{TLS}}{\lambda^{(0)}}\alpha^{2}F_{0}(\nu)\bigg)\bigg].
\end{align}

\noindent The above is general for some $\alpha^2 F_{TLS}(\nu)$ and $\alpha^2 F_0(\nu)$. As with $\lambda$, we can write the log average frequency assuming an Einstein model for both the phonon and vibron:

\begin{align}
\omega_{\ln}&=\omega_{\ln}^{(0)}\exp\bigg[\dfrac{2n_{TLS}}{\lambda^{(0)}+\lambda_{TLS}n_{TLS}}\int d\nu\dfrac{1}{\nu}\log(\nu/\omega_{0})\bigg(\alpha^{2}F_{TLS}(\nu)-\dfrac{\lambda_{TLS}}{\lambda^{(0)}}\alpha^{2}F_{0}(\nu)\bigg)\bigg]\notag \\&
\notag \\
&=\omega_E\exp\bigg[\dfrac{2n_{TLS}}{\lambda^{(0)}+\lambda_{TLS}n_{TLS}}\int d\nu\dfrac{1}{\nu}\log(\nu/\omega_{E})\bigg(N(0)g^{2}\delta(\omega-\omega_{TLS})-\dfrac{\lambda_{TLS}}{\lambda^{(0)}}\delta(\omega-\omega_E)\bigg)\bigg]\notag \\&\notag \\
&=\omega_E\exp\bigg[\dfrac{n_{TLS}\lambda_{TLS}}{\lambda^{(0)}+\lambda_{TLS}n_{TLS}}\log(\omega_{TLS}/\omega_{E})\bigg]\notag \\
&=\omega_{E}\left(\dfrac{\omega_{TLS}}{\omega_{E}}\right)^{\dfrac{n_{TLS}}{\omega_{TLS}/\omega_{E}+n_{TLS}}}.
\end{align}

In the limit of $n_{TLS}\rightarrow 0$, we see that the log average frequency approaches $\omega_E$, as expected for an Einstein phonon model. However, as $\omega_{TLS}\rightarrow 0$ for finite $n_{TLS}$, the log average goes to zero as a consequence of the low-frequency vibrons causing this quantity to be severely suppressed. As with the $\lambda$ calulation, we can get some estimates by estimating the lower bound of $\omega_{TLS}\approx 0.15$ meV. For $n_{TLS}=0.3$, we'd therefore expect $\omega_{\ln}/\omega_{\ln}^{(0)}\approx 0.02$. For $n_{TLS}\approx 0.01$, then $\omega_{\ln}/\omega_{\ln}^{(0)}\approx 0.2$, while for common densities of TLSs $n_{TLS}\approx 0.0001$, then $\omega_{\ln}/\omega_{\ln}^{(0)}\approx 0.97$. Note that small values of $\omega_{\ln}$ is an artifact of the Leavens-Carbotte equation and the fact that it assumes all frequency regimes of $\alpha^2 F(\omega)$ equally influence $T_c$, which is not correct~\cite{Bergmann1973Aug}. As we'll see later, numerics suggest $T_c/T_c^{(0)}\rightarrow 1$ in the limit of very small TLS frequencies, as required by Anderson's theorem~\cite{Anderson1959Sep}.



From the above form of $\omega_{\ln}$, $T_c$ becomes the following:

\begin{align}
    T_{c}&=T_{c}^{(0)}\dfrac{\omega_{\ln}}{\omega_{\ln}^{(0)}}\exp\bigg[\dfrac{1}{\lambda^{(0)}\left(\dfrac{\lambda^{(0)}}{\lambda_{TLS}}n_{TLS}^{-1}+1\right)}\bigg]
    \notag\\
    &\notag \\
    &=T_{c}^{(0)}\exp\bigg[\dfrac{n_{TLS}\lambda_{TLS}}{\lambda^{(0)}(\lambda^{(0)}+\lambda_{TLS}n_{TLS})}\bigg\{1+\dfrac{\lambda^{(0)}}{\lambda_{TLS}}\int d\nu\dfrac{2}{\nu}\log(\nu/\omega_{0})\bigg(\alpha^{2}F_{TLS}(\nu)-\dfrac{\lambda_{TLS}}{\lambda^{(0)}}\alpha^{2}F_{0}(\nu)\bigg)\bigg\}\bigg].
\end{align}

\noindent Trivially, we obtain $T_c=T_c^{(0)}$ for $n_{TLS}=0$. For an Einstein phonon model, the above simplifies further:

\begin{align}
    \dfrac{T_c}{T_c^{(0)}}
    &=\exp\bigg[\dfrac{n_{TLS}\lambda_{TLS}}{\lambda^{(0)}(\lambda^{(0)}+\lambda_{TLS}n_{TLS})}\bigg\{1+\dfrac{\lambda^{(0)}}{\lambda_{TLS}}\int d\nu\dfrac{2}{\nu}\log(\nu/\omega_{E})\bigg(\alpha^{2}F_{TLS}(\nu)-\dfrac{\lambda_{TLS}}{\lambda^{(0)}}\alpha^{2}F_{0}(\nu)\bigg)\bigg\}\bigg]\notag\\
    &=\left({\omega_{TLS}}/{\omega_E}
    \right)^{\dfrac{1}{1+(\omega_{TLS}/\omega_E)n_{TLS}^{-1}}}\exp\left(
\dfrac{1}{\lambda^{(0)}}\dfrac{1}{1+(\omega_{TLS}/\omega_E)n_{TLS}^{-1}}
    \right).
\end{align}

\begin{figure}[t!]
    \centering
        \centering
\includegraphics[width=0.75\linewidth]{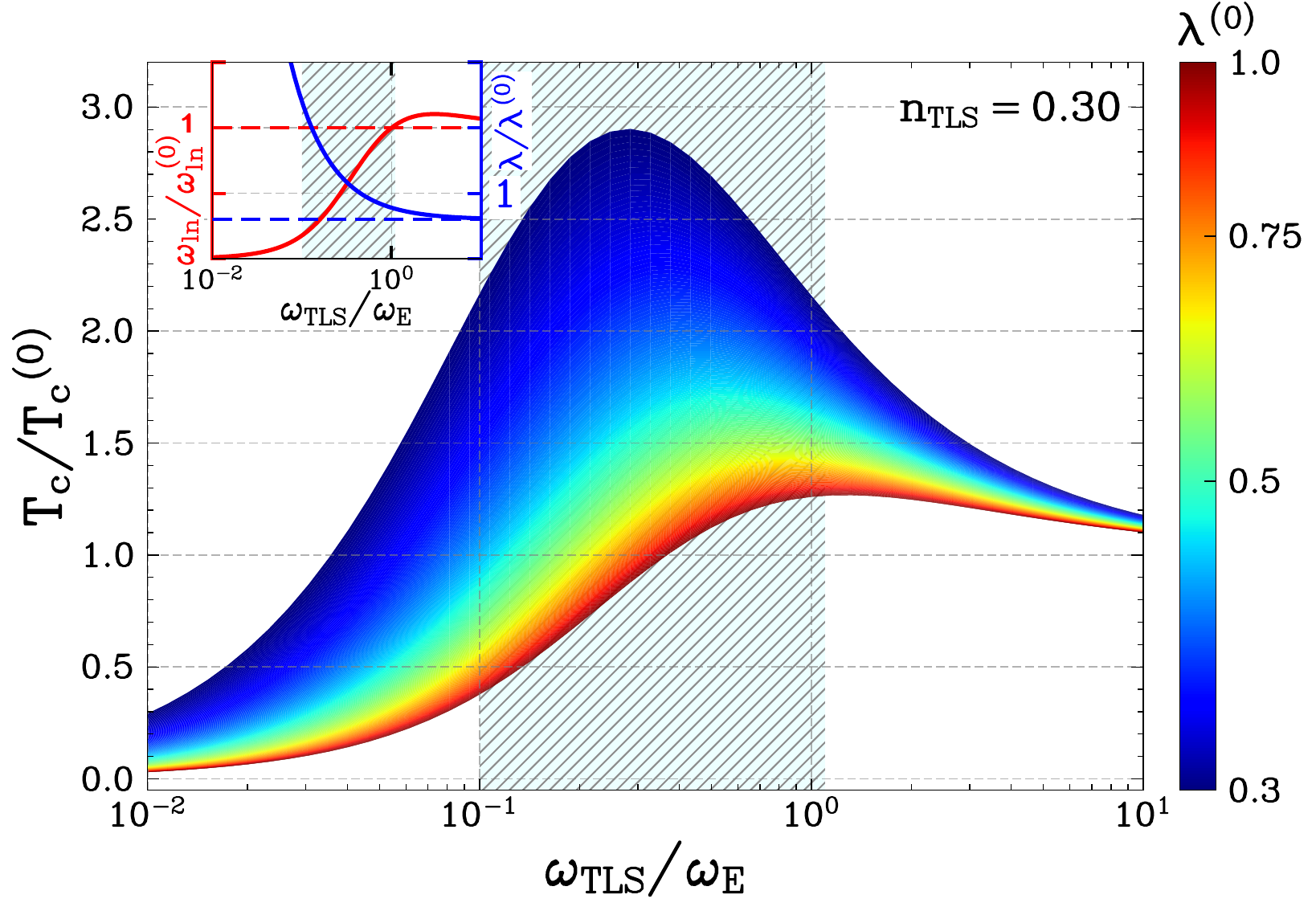}
        \label{fig:S1aHeath}
    \caption{\small 
    Change in the critical temperature $T_c/T_c^{(0)}$ versus TLS (vibron) frequency $\omega_{TLS}$ normalized by Einstein frequency $\omega_E$. An Einstein boson model is used for both the bulk phonon and vibron, and $T_c$ is found from the Leavens-Carbotte formula. The hatched cyan region corresponds to the frequency regime of amplified $T_c$, which becomes more pronounced for weaker coupling. Inset: Comparison of $\omega_{\ln}/\omega_{\ln}^{(0)}$ (left axis, red) and $\lambda/\lambda^{(0)}$ (right axis, blue) as $\omega_{TLS}$ changes.}
    \label{fig:S1aHeath}
\end{figure}

\noindent Once more, we can consider the case of certain values of physical parameters. If the minimum vibron frequency is $\omega_{TLS}\approx 0.15$ meV (or around $36$ GHz) and we approximate $\omega_E\sim 10$ meV, then $T_c/T_c^{(0)}\sim 0.25$ for $n_{TLS}\sim 0.3$. For $n_{TLS}\sim 0.01$, $T_c/T_c^{(0)}\sim 0.56$ and for $n_{TLS}\sim 0.0001$, $T_c/T_c^{(0)}\sim 0.99$. As such, for more common TLS densities, the reduction of $T_c$ is negligible. TLS densities would have to be synthetically increased above their naturally occurring quantity in order to see modification of the critical temperature. Also note that the above simplifies to the following, simpler form:

\begin{align}
    \dfrac{T_c}{T_c^{(0)}}=\left[\dfrac{\omega_{\ln}}{\omega_E}\exp\left(
    \dfrac{1}{\lambda^{(0)}}
    \right)\right]^{1-\lambda^{(0)}/\lambda}.
\end{align}

\noindent From the above equation for $T_c/T_c^{(0)}$ and assuming $n_{TLS}\not=0$, we can take the derivative with respect to $\omega_{TLS}/\omega_E$ and find the maximum, which yields the average TLS frequency $\omega_{TLS}^*$ that maximizes the critical temperature:

\begin{align}
\omega_{TLS}^*=\omega_E \dfrac{n_{TLS}}{W_0\left[n_{TLS}\exp\left(-1+\dfrac{1}{\lambda^{(0)}}\right)\right]},
\end{align}
where $W_0(...)$ is the Lambert W function. Identifying the righthand side as some number $\alpha$ and $\omega_E$ as $\omega_{\textrm{peak}}$ for some more complicated $\alpha^2F(\nu)$, we obtain the expression for $\omega_{TLS}^*$ given in the main text.

\begin{figure}[h!]
    \centering
    \includegraphics[width=0.5\linewidth]{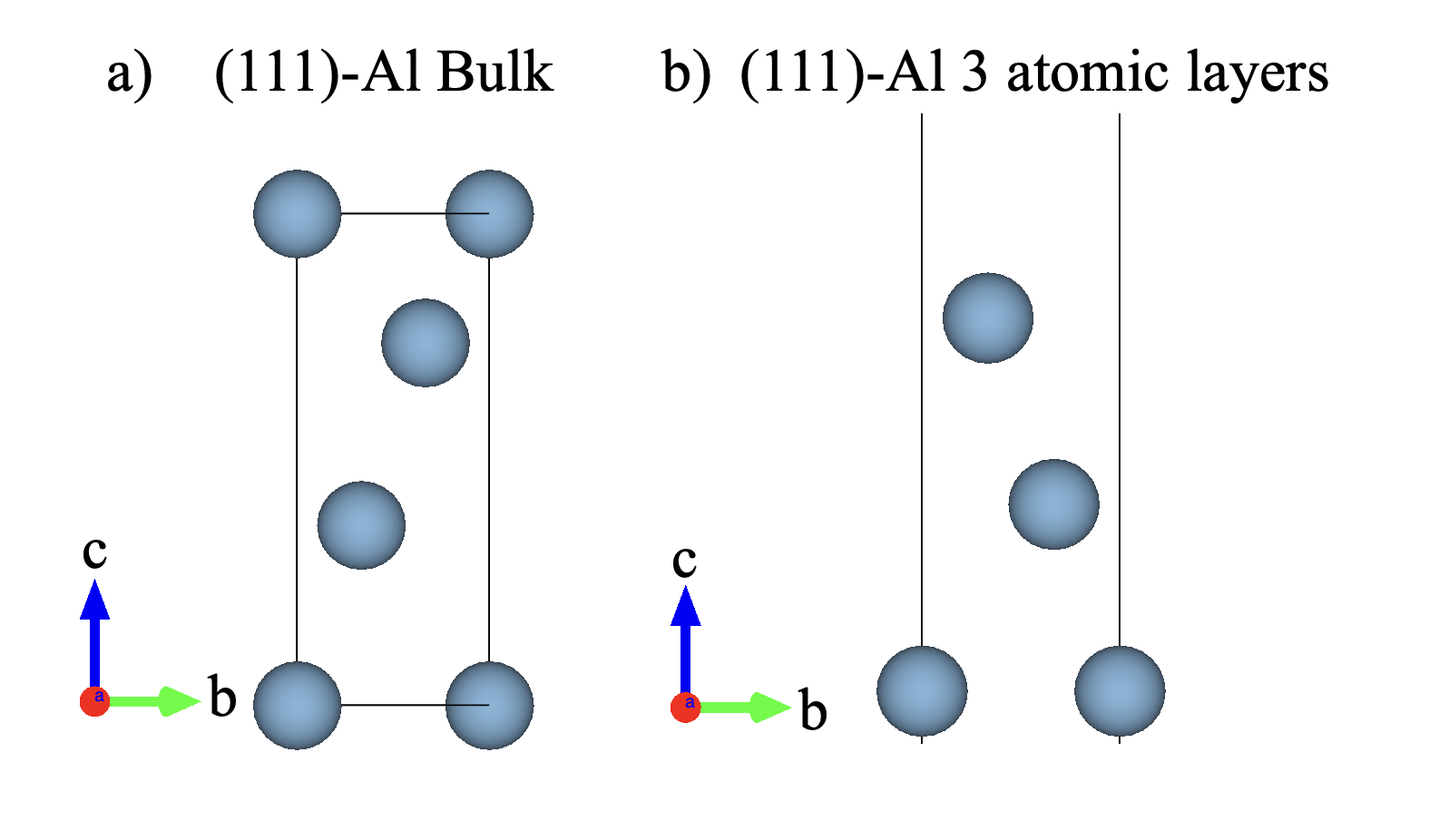}
    \caption{(a) Bulk atomic structure of (111) oriented aluminium. (b) A (111)-Al slab of three atomic layers. Along the $c$-axis $20 \AA$ of vacuum is included but not shown in its entirety for clarity.  }
    \label{fig:Al_Structures}
\end{figure}

In Fig.~\ref{fig:S1aHeath}, we plot $T_c/T_c^{(0)}$ in the Einstein approximation versus $\omega_{TLS}/\omega_E$ for $n_{TLS}=0.3$. As the TLS frequency becomes much smaller than $\omega_E$, the TLSs will reduce $T_c$ to zero. Note that such severe suppression of $T_c$ is an artifact of the Leavens-Carbotte formulation of $T_c$. This is because the the LC solution does not take into account the varying sensitivity of $T_c$ on the fine-structure of $\alpha^2 F(\omega)$. As shown by Bergmann and Rainer~\cite{Bergmann1973Aug}, $\delta T_c/\delta \alpha^2 F(\omega)\sim \omega$ at very low frequencies, and thus the functional derivative goes to zero in this frequency regime. As a consequence of this, there is a reduced emphasis of vibrons with $\omega_{TLS}\ll 2\pi T_c$ on the determination of $T_c$. Thus, as $\omega_{TLS}\rightarrow 0$, we should obtain $T_c/T_c^{(0)}\rightarrow 1$ in the numerical solution, as the TLS then becomes a static non-magnetic impurity which should not affect the $T_c$~\cite{Anderson1959Sep}.

As we will see later in this Supplemental Material and as mentioned in the main text, we find that i) in the absence of TLSs, Leavens-Carbotte formula is in good agreement with the numerical solution of $T_c/T_c^{(0)}$ for all Al monolayers excluding the $3$ monolayer system, ii) in the scenario of a high density of low-frequency TLSs, the numerical estimation of $T_c/T_c^{(0)}$ goes to unity, iii) for higher frequency TLSs, the Leavens-Carbotte solution matches the numerical solution for $T_c$ to very good accuracy, and iv) upon increasing $\omega_{TLS}$, maximising $T_c/T_c^{(0)}$ roughly corresponds to TLS frequencies around $0.2-0.4\times$ the largest salient peaks in $\alpha^2 F(\omega)$, in agreement with our Einstein phonon estimation.

\section{First-principles computation of $\alpha^{2}F(\omega)$}

Density functional theory computations were performed to extract $\alpha^{2}F(\omega)$ for bulk and $(111)$ oriented slabs of layered Al using the Quantum Espresso software package\cite{QE-2020}. Exchange-correlation potentials use the Perdew-Burke-Ernzerhof (PBE) parametrization of the generalized gradient approximation (GGA) \cite{kresse1999VASP,Perdew1996}. For all computations we employ fully relativistic norm-conserving pseudopotentials from the PseudoDojo\cite{van2018pseudodojo} database with a plane wave cutoff of 80 Ry. For the (111) orientated slabs a 12 x 12 x 1 Monkhorst-Pack grid\cite{PhysRevB.13.5188} of $\mathbf{k}$-points is used to fully relax the structures. The bulk structure as well as a three atomic layer structure are shown in Fig.~\eqref{fig:Al_Structures}. When considering the layered structure 20 $\AA$ of vaccum are inserted along the $c$-axis.
\par 
After relaxation the electron-phonon computations are performed via density functional perturbation theory as implemented in Quantum Espresso. For bulk Al, we perform convergence testing of $\alpha^{2}F(\omega)$, considering a $3\times 3 \times 3$ and $6\times 6 \times 6$ $\mathbf{q}$-mesh. We then utilize the EPW software package\cite{Noffsinger2010Dec,lee2023electron,Pizzi2020} and recompute $\alpha^{2}F(\omega)$. The EPW software package is capable of interpolating the $\mathbf{q}$-mesh to achieve an extremely fine grid. For the bulk computations we use a $20 \times 20 \times 20$ grid within EPW.

\begin{figure}[p]
    \centering
    \includegraphics[width=0.85\linewidth]{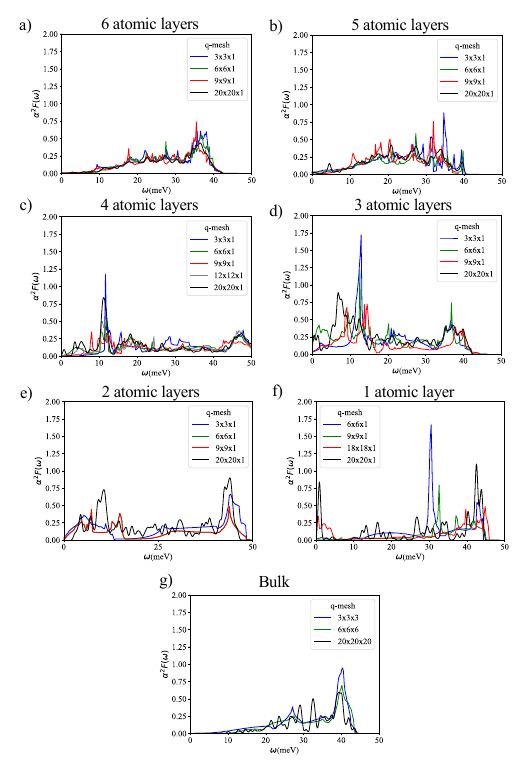}
    \caption{(a)-(f) $\alpha^2F(\omega)$ computed for $N=1$ to $N=6$ layer $(111)$-Al slabs respectively as a function of varying the $\mathbf{q}$-mesh invoked for density functional perturbation theory computations. (g) Result of bulk computation of $\alpha^2F(\omega)$ as a function of varying the $\mathbf{q}$-mesh.   }
    \label{fig:DFT_Result_Summary}
\end{figure}

\begin{figure}[p]
    \centering
    
    \begin{subfigure}{0.45\textwidth}
        \centering
        \includegraphics[width=\linewidth]{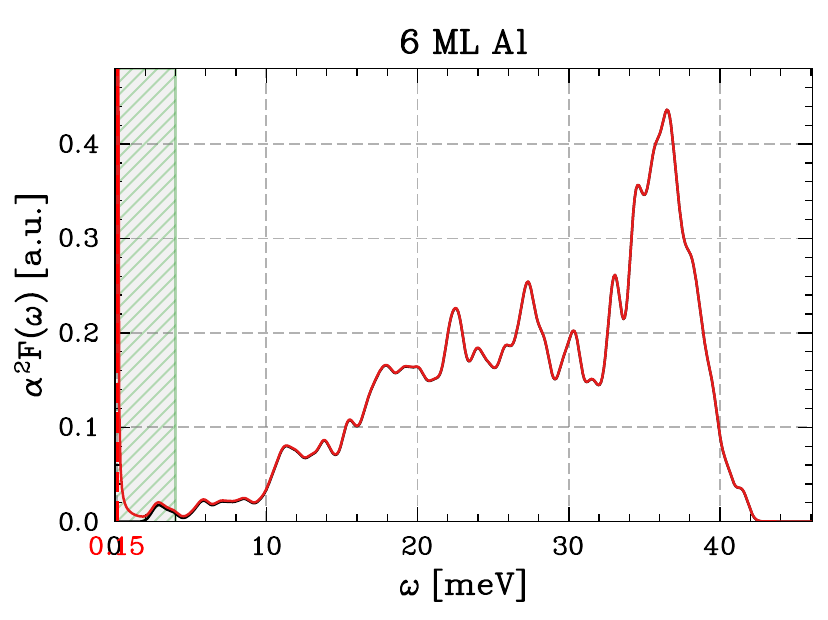}
        \caption{$\alpha^2 F(\omega)$ versus frequency for 6 ML}
        \label{fig:sub6}
    \end{subfigure}
    \hspace{0.02\textwidth}
    \begin{subfigure}{0.45\textwidth}
        \centering
        \includegraphics[width=\linewidth]{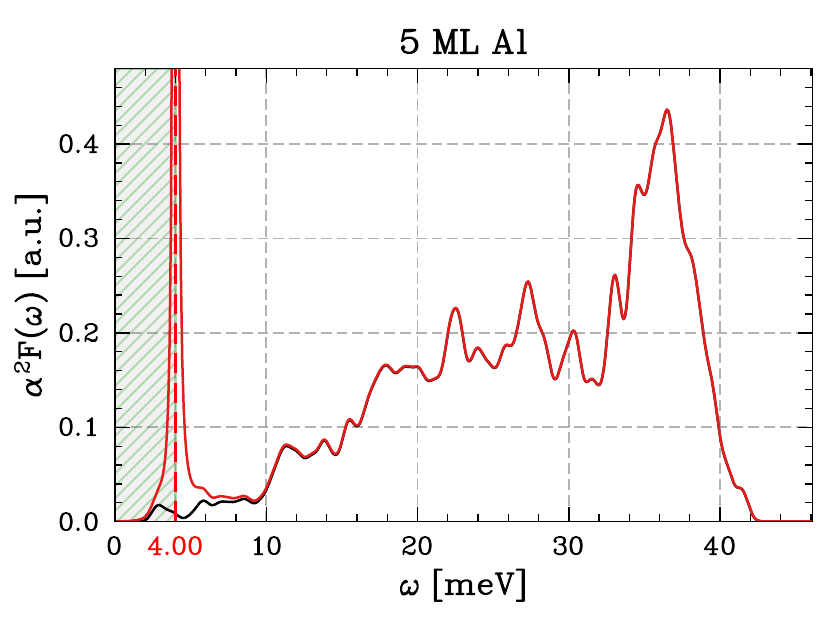}
        \caption{$\alpha^2 F(\omega)$ versus frequency for 5 ML}
        \label{fig:sub5}
    \end{subfigure}
    
    \vskip\baselineskip
    \begin{subfigure}{0.45\textwidth}
        \centering
        \includegraphics[width=\linewidth]{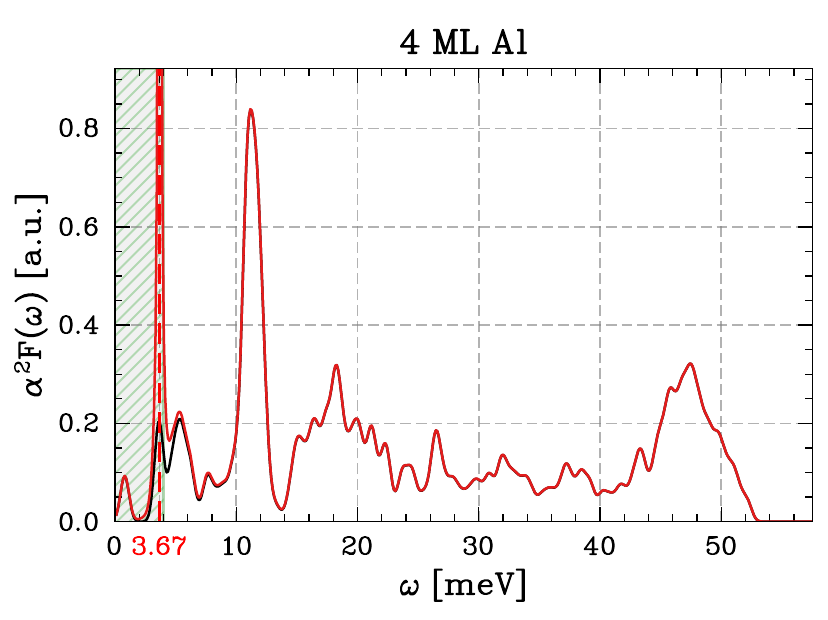}
        \caption{$\alpha^2 F(\omega)$ versus frequency for 4 ML}
        \label{fig:sub4}
    \end{subfigure}
    \hspace{0.02\textwidth}
    \begin{subfigure}{0.45\textwidth}
        \centering
        \includegraphics[width=\linewidth]{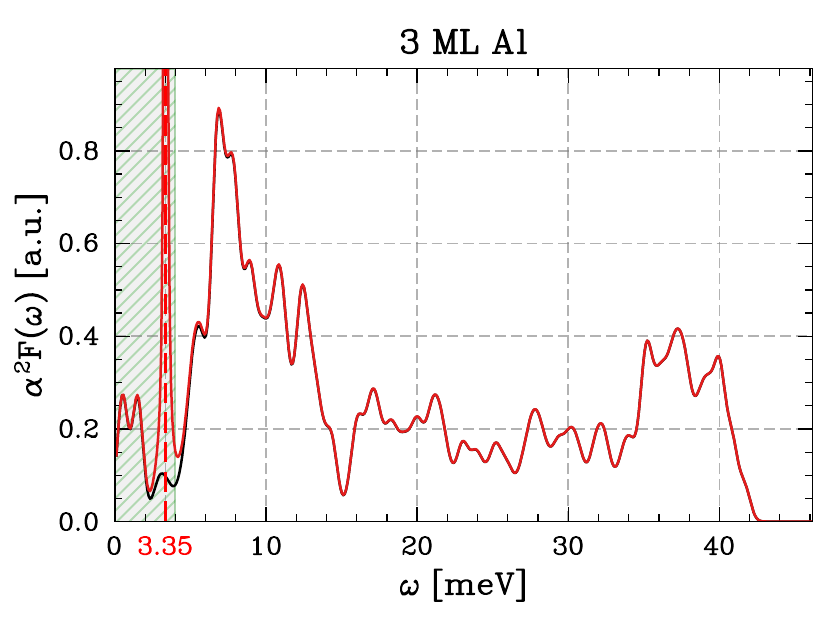}
        \caption{$\alpha^2 F(\omega)$ versus frequency for 3 ML}
        \label{fig:sub3}
    \end{subfigure}
    
    \vskip\baselineskip
    \begin{subfigure}{0.45\textwidth}
        \centering
        \includegraphics[width=\linewidth]{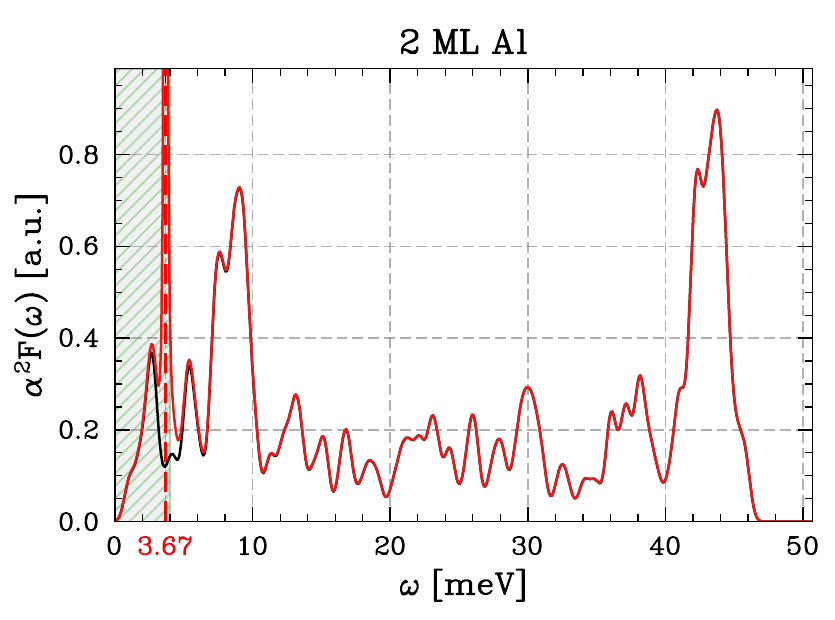}
        \caption{$\alpha^2 F(\omega)$ versus frequency for 2 ML}
        \label{fig:sub2}
    \end{subfigure}
    \hspace{0.02\textwidth}
    \begin{subfigure}{0.45\textwidth}
        \centering
        \includegraphics[width=\linewidth]{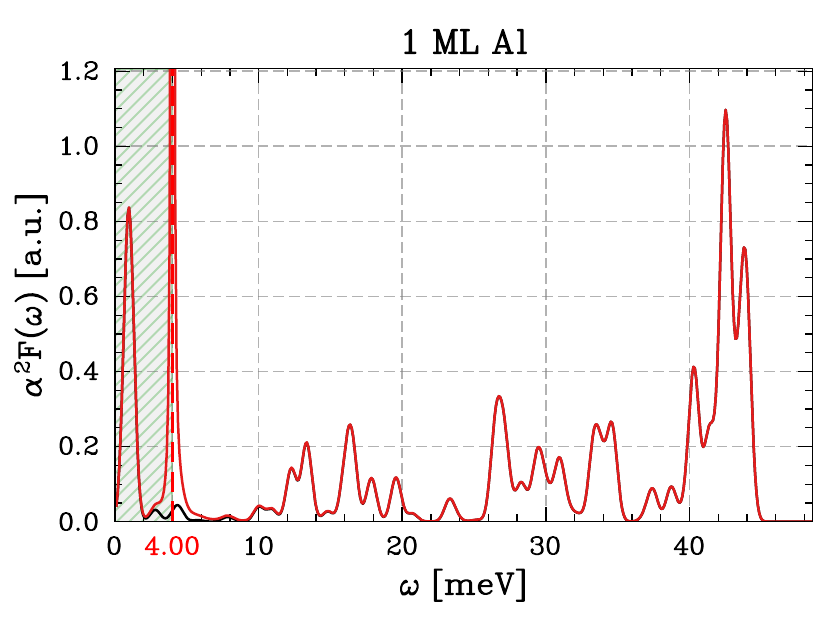}
        \caption{$\alpha^2 F(\omega)$ versus frequency for 1 ML}
        \label{fig:sub1}
    \end{subfigure}

    \caption{Spectral function $\alpha^2F(\omega)$ for 6–1 Al monolayers (TLS), overlaid with the contribution from the TLSs. For all cases, the TLS parameters are chosen such that $T_c/T_c^{(0)}$ is maximised.
    }
    \label{fig:Heath_TLS1}
\end{figure}

When considering $N$-atomic layer thick where $N \in [1,6]$, (111)-oriented slabs of Al, we repeat this process. In each case we vary the $\mathbf{q}$-mesh between $3\times 3 \times 1$ and $18 \times 18 \times 1$ within Quantum Espresso until the Frechet distance between the derived $\alpha^2F(\omega)$ curves is less than two. We then utilize the EPW software package to access an interpolated $20 \times 20 \times 1$ $\mathbf{q}$-mesh and recompute $\alpha^2F(\omega)$. Within the EPW software package we utilize $s$ and $p$ orbitals in construction of the Wannier basis. The results are shown in Fig.~\eqref{fig:DFT_Result_Summary}.

In our investigation we find that convergence testing process as a function of the $\mathbf{q}$-mesh size is critical, particularly for the $N$-atomic layer slabs to capture the low-frequency peaks in $\alpha^2F(\omega)$. This is best perhaps best exemplified in the data gathered for the monolayer limit. In this data we note that only upon increasing the $\mathbf{q}$-mesh to $18 \times 18 \times 1$ within the DFPT computations carried out in Quantum Espresso, we see the development of a peak near zero frequency. It is important to note that a similar peak is observed if an EPW computation is completed using a smaller $\mathbf{q}$-mesh of $6 \times 6 \times 1$ computed via DFPT as input and subsequently interpolating to a mesh of size $20 \times 20 \times 1$ within EPW for computation of $\alpha^{2}F(\omega)$. However, in this case the Frechet distance between the EPW result and result obtained directly from Quantum Espresso will be increase significantly. An increase in the Frechet distance between the EPW result and that computed directly from Quantum Espresso indicates inconsistency among the levels of theory and need for further convergence testing.

Indeed the results from EPW and DFPT represent two levels of theory as the EPW computation does not represent a stand-alone first-principles approach. Rather, it uses the Wannier90 software package\cite{Pizzi2020} to project the plane-wave basis constructed by Quantum Espresso into a localized Wannier basis from which a tight binding model can be constructed to interpolate the electronic and phonon spectra over an arbitrary $\mathbf{k}$ and $\mathbf{q}$ mesh. The capability of extrapolating to dense $\mathbf{k}$ and $\mathbf{q}$ meshes without incurring significant computational expense is a primary advantage of EPW. However, if the resulting data demonstrates significant deviation from that produced by a direct, first-principles, DFPT approach within Quantum Espresso limited confidence can be placed in the result. This is why we pursue the computationally expensive route of convergence testing. 
\par 
In the resulting data it is important to emphasize that low-frequency peaks appear in increasing prominence as the number of layers is reduced towards the monolayer limit. This suggests softening of phonon modes at the crystalline surfaces. Softening of surface phonon modes has been studied previously~\cite{Strongin1968Oct,Allen1974May,Leavens1981Nov,Tamura1993Mar,Bose2009Apr,bozyigit2016soft,Lozano2019Feb,Houben2020Mar} and can be understood in part by the reduced coordination number of atoms on the surface which in turn reduces the restoring force for displacement. As the number of layers is increased towards the bulk limit we observe a shift with high-frequency peaks dominating the $\alpha^{2}F(\omega)$ distribution. 

In Fig.~\ref{fig:Heath_TLS1}, we plot each individual $\alpha^2 F(\omega)$ for 6 to 1 AL monolayers alongside the addition of the TLS contribution as an Einstein phonon-like peak located at the TLS frequency that maximizes $T_c$ and assuming $n_{TLS}=0.3$. As mentioned before, the TLS frequency which raises $T_c$ does not necessarily correspond to the low-frequency peaks of $\alpha^2 F(\omega)$.

\section{Numerical calculation of the gap edge $\Delta_1$ and critical temperature $T_c$}

\subsection{Numerically finding the gap edge in the bulk and monolayer limits}

In the previous sections of this appendix, we have considered the real axis Eliashberg equations of Scalapino, Schrieffer, and Wilkins. This formulation of the Eliashberg equations is highly useful in regards to analytical calculations, as we have seen in the case of the Leavens-Carbotte estimation of $T_c$. However, numerically solving these equations is non-trivial~\cite{Scalapino1965Jan}. 

To remedy the issues in directly solving the SSW equations on the real axis, we will instead consider the Eliashberg equations on the imaginary axis, and then analytically continue to the real axis exactly via the method of Marsiglio, Schossmann, and Carbotte (MSC)~\cite{Marsiglio1988Apr}. On the Matsubara frequency axis, the Eliashberg equations which determine the anomalous part of the self energy $\phi(i\omega_n)$ and the normalised frequency $\widetilde{\omega}_n$ are given by ~\cite{Eliashberg1960,Eliashberg1961,RickayzenBook}

\begin{subequations}
\begin{equation}
\widetilde{\omega}_n=\omega_n+\pi k_BT_c\sum_{m=-M}^{M-1}\lambda(i\omega_n-i\omega_{m})\dfrac{\widetilde{\omega}_m}{\sqrt{
\widetilde{\omega}_{m}^2+\phi^2(\omega_{m})
}},
\end{equation}
\begin{align}
    \phi(i\omega_n)=\pi k_BT_c\sum_{m=-M}^{M-1}\lambda(i\omega_n-i\omega_m)\dfrac{\phi(i\omega_m)}{\sqrt{\widetilde{\omega}_{m}^2+\phi^2(\omega_{m})}}.
\end{align}
\end{subequations}
where the Matsubara frequencies are given by $\omega_n =(2n+1)\pi k_BT_c$, $\widetilde{\omega}_n=\omega_n Z(i\omega_n)$, $\phi(i\omega_n)=\Delta(i\omega_n)Z(i\omega_n)$, and
\begin{align}
    \lambda(i\omega_m-i\omega_{m'})\equiv \int_0^\infty d\nu\dfrac{2\nu \alpha^2 F(\nu)}{\nu^2+(\omega_m-\omega_{m'})^2},
\end{align}


\noindent where $\Delta(i\omega_n)$ is the gap function and $Z(i\omega_n)$ is the renormalization function. The maximum index in the sum $M$ is taken to be large enough to ensure no appreciable change in the gap function occurs upon changing $M$. Similar criterion is taken to ensure $T$ is small enough in order to interpret the solution to the above as the $T=0$ solution to the Eliashberg equations. 

The imaginary axis Eliashberg equations can be analytically continued to the real axis to yield the following~\cite{Marsiglio1988Apr}:

\begin{align}
\phi(z)&=\pi k_B T\sum_{m=-\infty}^\infty \bigg[\lambda(z-i\omega_m)-\mu^*\Theta(\omega_c-|\omega_m|)\bigg]\dfrac{\phi(i\omega_m)}{\sqrt{\widetilde{\omega}_m^2+\phi^2(i\omega_m)}}\notag\\
&+i\pi \int_0^\infty d\nu\, \alpha^2 F(\nu)\bigg\{
[n_b(\nu)+n_f(\nu-z)]\dfrac{\phi(z-\nu)}{\sqrt{(\widetilde{z-\nu})^2-\phi^2(z-\nu)}}+[n_b(\nu)+n_f(\nu+z)]\dfrac{\phi(z+\nu)}{\sqrt{(\widetilde{z+\nu})^2-\phi^2(z+\nu)}}
\bigg\},
\end{align}

\begin{align}
Z(z)&=1+\dfrac{i\pi k_B T}{z}\sum_{m=-\infty}^\infty \lambda(z-i\omega_m)\dfrac{\omega_m}{\sqrt{\widetilde{\omega}_m^2+\phi^2(i\omega_m)}}\notag\\
&+i\pi \int_0^\infty d\nu\, \alpha^2 F(\nu)\bigg\{
[n_b(\nu)+n_f(\nu-z)]\dfrac{\widetilde{z-\nu}}{\sqrt{(\widetilde{z-\nu})^2-\phi^2(z-\nu)}}+[n_b(\nu)+n_f(\nu+z)]\dfrac{\widetilde{z+\nu}}{\sqrt{(\widetilde{z+\nu})^2-\phi^2(z+\nu)}}
\bigg\}
\end{align}
\noindent where $z\equiv \omega+i\delta$, the tilde means that the relevant quantity is multiplied by $Z(i\omega_n)$ and $n_b(\nu)$ ($n_f(\nu)$) is the Bose-Einstein (Fermi-Dirac) distribution function. In both expressions, as the integral term is zero for $0<z<\omega_E$ in the Einstein phonon model, it is typically called the inhomogeneous contribution, while the integral over the purely real-frequency terms is known as the homogeneous contribution. As with the case for solving $\Delta(i\omega_n)$ and $Z(i\omega_n)$, $\Delta(z)$ and $Z(z)$ are found via numerical iteration until convergence. The Eliashberg equivalent of the superconducting gap is then identified with the gap edge~\cite{Fibich1965Apr}, which is defined as $\Delta_1\equiv \Re(\Delta(\omega=\Delta_1))$.

\subsection{Numerically finding $T_c$ with the imaginary axis Eliashberg equations}

To estimate the superconducting critical temperature in Eliashberg theory. we linearize the Eliashberg equations near $T=T_c$:

\begin{subequations}
    \begin{equation}
        \widetilde{\omega}_n=\omega_n+\pi k_B T_c \sum_{m=-\infty}^\infty\lambda(i\omega_n-i\omega_m)\textrm{sgn}(\widetilde{\omega}_m),
    \end{equation}
    \begin{equation}
        \phi(i\omega_n)=\pi k_B T_c\sum_{m=-\infty}^\infty \lambda(i\omega_n-i\omega_m)\dfrac{\phi(i\omega_m)}{|\widetilde{\omega}_m|}
    \end{equation}
\end{subequations}

\noindent Note that contributions for $n=m$ (i.e., $\lambda(0)$) will not contribute to the determination of the gap function and, as a consequence, $T_c$~\cite{Marsiglio1991Mar,Chubukov2020Jun}. In the above and for what follows below, we take $\mu^*=0$, but the algorithm we provide for finding $T_c$ is quite general and works identically for $\mu^*\not=0$. Representing our equations in terms of the gap and renormalization functions and restricting the sum to only positive Matsubara indices, then we obtain

\begin{align}
    Z(i\omega_n)&=1+\dfrac{\pi k_B T_c}{\omega_n}\sum_{m=0}^\infty \bigg\{
    \lambda(i\omega_n-i\omega_m)-\lambda(i\omega_n+i\omega_m)
    \bigg\}\notag\\
    &=1+\dfrac{\pi k_B T_c}{\omega_n}\sum_{m=0}^{2n}\lambda(i\omega_n-i\omega_m),
\end{align}
\begin{align}
    Z(i\omega_n)\Delta(i\omega_n)=\pi k_B T_c\sum_{m=0}^\infty \bigg\{
\lambda(i\omega_n-i\omega_m)+\lambda(i\omega_n+i\omega_m)
    \bigg\}\dfrac{\Delta(i\omega_m)}{\omega_m}
\end{align}

\begin{figure}[t!]
    \centering
        \centering
        \includegraphics[width=0.75\linewidth]{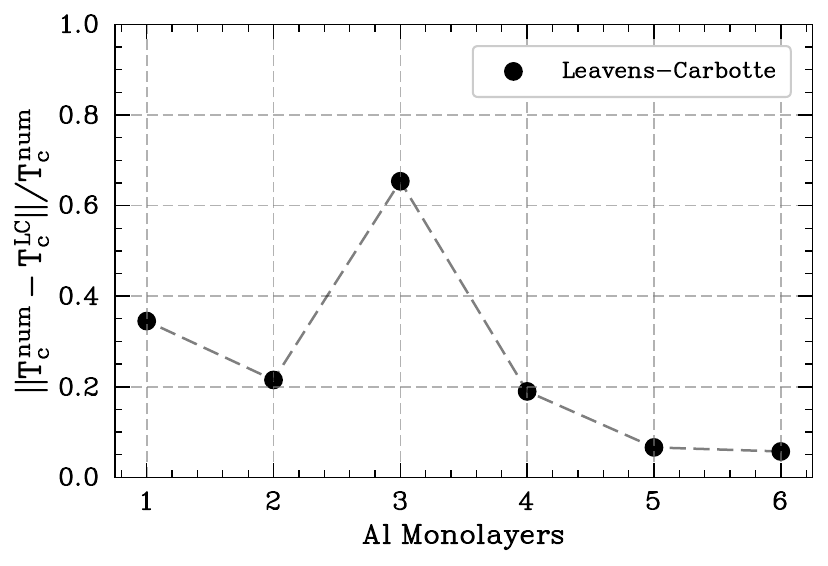}
        
    \caption{\small 
    Difference of the Leavens-Carbotte (LC) and numerical solution for $T_c$ in the absence of TLSs, divided by the numerical solution. The only system where the analytical and numerical solutions differ significantly is the $3$ monolayer system, which may be attributed to the structure of $\alpha^2 F(\omega)$.}
    \label{LCComparison}
\end{figure}

Focusing on the equation for the gap, we can rewrite this in the form

\begin{align}
\Delta(i\omega_n)-\pi k_B T_c \sum_{m=0}^\infty\dfrac{
1
    }{\omega_m}\bigg\{
\lambda(i\omega_n-i\omega_m)+\lambda(i\omega_n+i\omega_m)
-\delta_{nm}\sum_{\ell=0}^{2m}\lambda(i\omega_m-i\omega_\ell)\bigg\}\Delta(i\omega_m)
=0
\end{align}
This corresponds to an eigenvalue equation~\cite{Owen1971Oct,Kresin1987Jun,Marsiglio2020Jun}, with a vector composed of the gap function at different Matsubara frequencies, an eigenvalue of unity, and a matrix given by

\begin{align}
    M_{nm}=\dfrac{
\pi k_B T_c
    }{\omega_m}\bigg\{
\lambda(i\omega_n-i\omega_m)+\lambda(i\omega_n+i\omega_m)
-\delta_{nm}\sum_{\ell=0}^{2m}\lambda(i\omega_m-i\omega_\ell)\bigg\}
\end{align}

\noindent Thus, the critical temperature $T_c$ corresponds to the temperature where the largest eigenvalue of the matrix $M_{nm}$ is unity. 

Instead of directly solving for the eigenvalues of the equation above, we will use the power iteration method (or Von Mises iteration) on the matrix $M$. This derivation follows from Marsiglio's review of Eliashberg theory~\cite{Marsiglio2020Jun}, who attributes this method to Ewald Schachinger. We will start by consider some general equation for the whole eigenspectrum of the matrix $M$:
\begin{align}
    M|i\rangle =r_i|i\rangle
\end{align}
\noindent where $r_i$ is the eigenvalue of eigenvector $|i\rangle$. As $M$ is real and symmetric, it's eigenvectors form an orthonormal basis. As a consequence, writing coefficients of the eigenbasis as $c_i$, then

\begin{figure}[p]
    \centering
    
    \begin{subfigure}{0.45\textwidth}
        \centering
        \includegraphics[width=\linewidth]{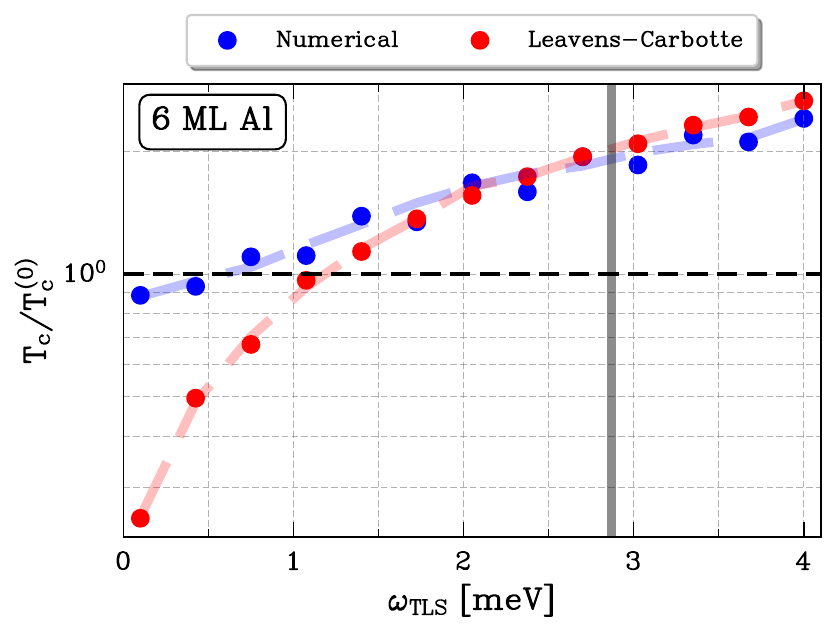}
        \caption{\small 6 Al MLs: Max $T_c/T_c^{(0)}\approx 2.4$ at $\omega_{TLS}=4$ meV}
        \label{fig:sub6}
    \end{subfigure}
    \hspace{0.02\textwidth}
    \begin{subfigure}{0.45\textwidth}
        \centering
        \includegraphics[width=\linewidth]{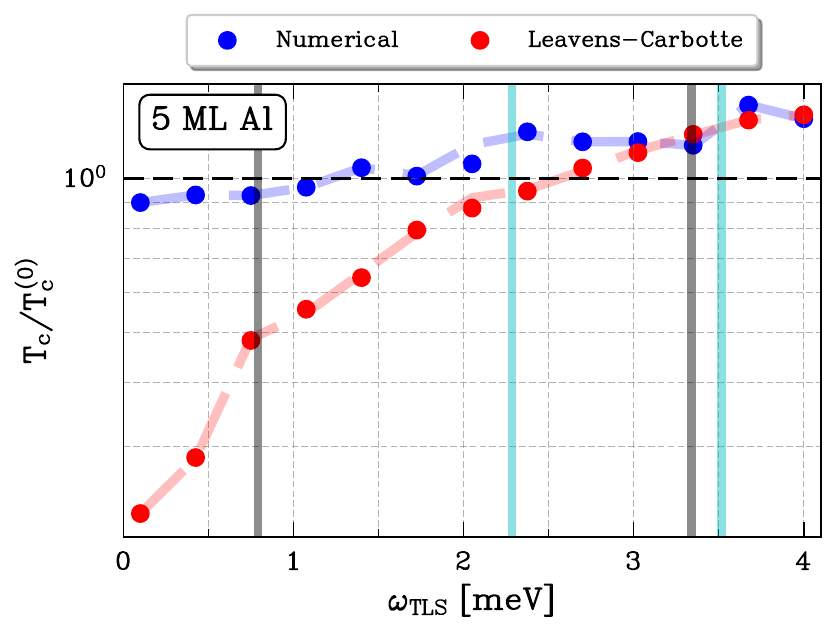}
        \caption{\small 5 Al MLs: Max $T_c/T_c^{(0)}\approx 1.4$ at $\omega_{TLS}=3.7$}
        \label{fig:sub5}
    \end{subfigure}
    
    \vskip\baselineskip
    \begin{subfigure}{0.45\textwidth}
        \centering
        \includegraphics[width=\linewidth]{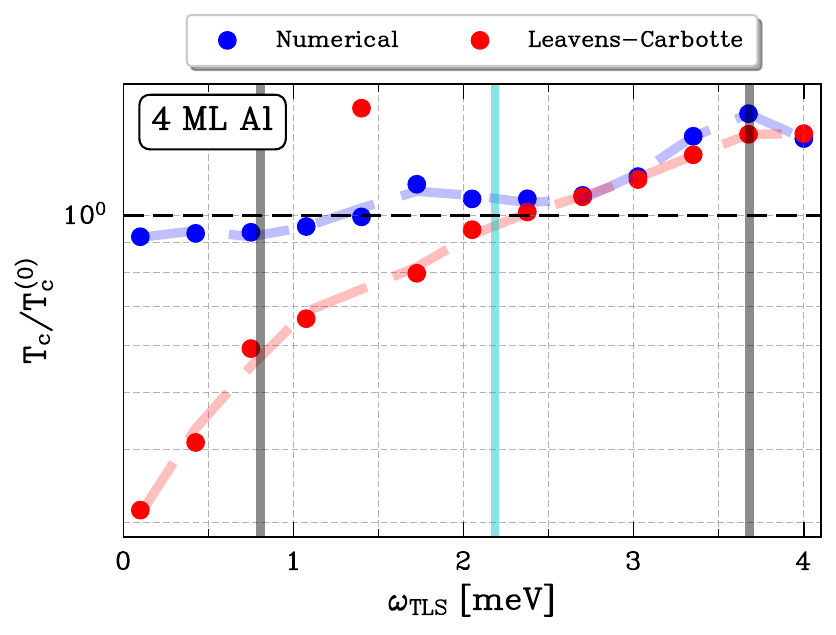}
        \caption{\small 4 Al MLs: Max $T_c/T_c^{(0)}\approx 1.5$ at $\omega_{TLS}\approx 3.7$ meV}
        \label{fig:sub4}
    \end{subfigure}
    \hspace{0.02\textwidth}
    \begin{subfigure}{0.45\textwidth}
        \centering
        \includegraphics[width=\linewidth]{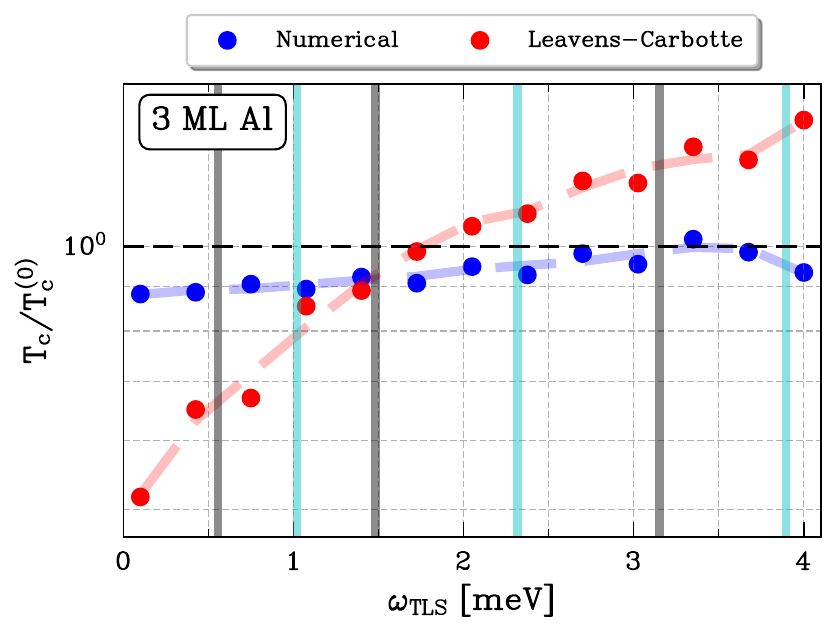}
        \caption{\small 3 Al MLs: Max $T_c/T_c^{(0)}\approx 1.02$ at $\omega_{TLS}=3.35$ meV}
        \label{fig:sub3}
    \end{subfigure}
    
    \vskip\baselineskip
    \begin{subfigure}{0.45\textwidth}
        \centering
        \includegraphics[width=\linewidth]{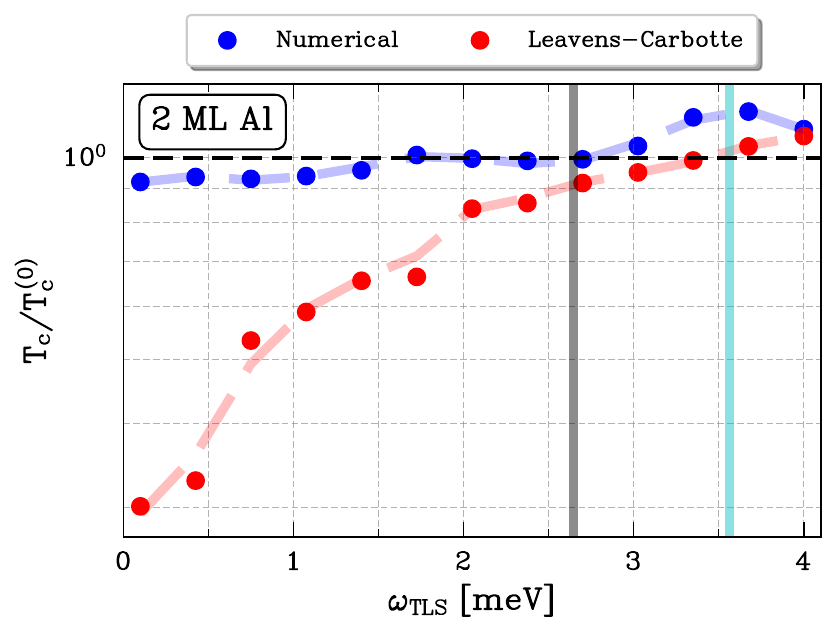}
        \caption{\small 2 Al MLs: Max $T_c/T_c^{(0)}\approx 1.2$ at $\omega_{TLS}\approx 3.7$ meV}
        \label{fig:sub2}
    \end{subfigure}
    \hspace{0.02\textwidth}
    \begin{subfigure}{0.45\textwidth}
        \centering
        \includegraphics[width=\linewidth]{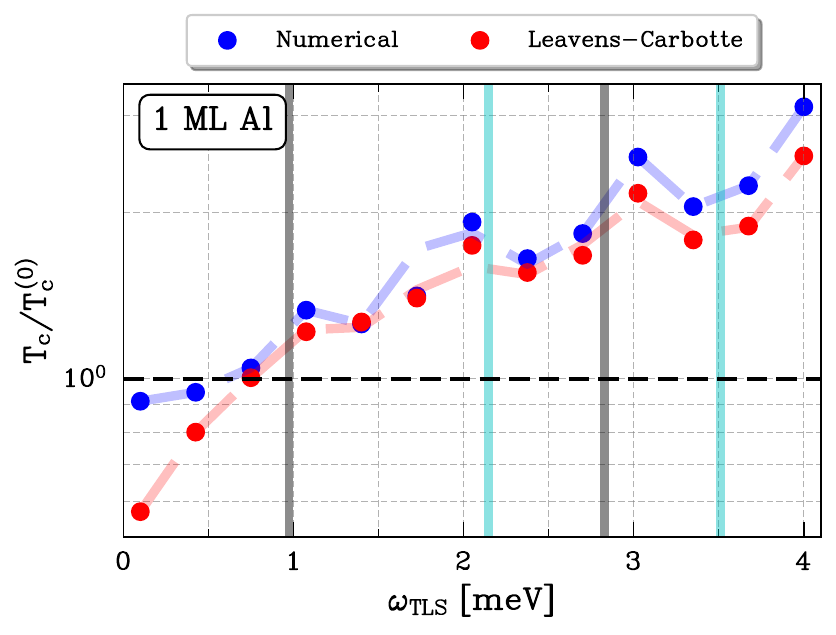}
        \caption{\small 1 Al MLs: Max $T_c/T_c^{(0)}\approx 3.1$ at $\omega_{TLS}=4$ meV}
        \label{fig:sub1}
    \end{subfigure}

    \caption{\small Comparison of the numerical result for $T_c/T_c^{(0)}$ found from the linearised Eliashberg equations (blue dots) and the Leavens-Carbotte analytical form of $T_c/T_c^{(0)}$ (red dots). Both the numerical and analytical form of $T_c/T_c^{(0)}$ use the form of $\alpha^2F(\omega)$ found via ab initio simulation. The blue and red dashed lines are used as guides for the eyed. The vertical grey (cyan) line corresponds to a local maximum (minimum) in the corresponding $\alpha^2F(\omega)$ structure. As the TLS frequency grows, agreement between numerics and analytics is nearly perfect. The only exception is the 3 monolayer case, which may be attributed to the form of $\alpha^2F(\omega)$ in the 3 monolayer system. 
    }
    \label{fig:LCCompare_with_TLS}
\end{figure}

\begin{align}
    M|i\rangle =c_1r_1|1\rangle+c_2r_2|2\rangle+c_3r_3|3\rangle+...
\end{align}
Applying $M$ many, many times will lead to the largest eigenvalue dominating over the other smaller eigenvalues, and thus for some large $n$,
\begin{align}
    M^n|i\rangle=r_1^n c_1|1\rangle\equiv |v_n\rangle
\end{align}
If we want to ensure that this vector doesn't blow up, then it must be normalized after each application of $M$. Noting that $|v_{n+1}\rangle=M|v_n\rangle$, then this normalization factor should tend towards $\lambda_1$ as $n\rightarrow \infty$. As $|1\rangle$ corresponds to the largest eigenvalue, we expect this normalization factor to reasonably be approximated by $\Delta(i\omega_0)$. As a consequence, we may identify the largest eigenvalue of the matrix $M$ as this normalization factor upon increasingly applying the matrix $M$, which is equivalent to the converged value of $\Delta(i\omega_0)$ upon repeated iteration in the Eliashberg equations. 

The closeness of the converged $\Delta(i\omega_0)$ to unity will then be used to find $T_c$. We now want to get an estimate for $T_c$ via linear interpolation~\cite{Tajik2018}. We begin by taking two temperatures $T_1$ and $T_2\gtrapprox T_1$ which we expect to be reasonably close to some temperature $T_c$. We iteratively solve the Eliashberg equations to obtain the corresponding values of gap functions at zero Matsubara index for $T_1$ and $T_2$, which we'll denote as $\Delta_0(T_1)$ and $\Delta_0(T_2)$, respectively. As $T_1$ and $T_2$ is close to $T_c$, we'll assume that $\Delta_0(T_c)\approx 1$ can be approximated as a linear interpolation between $T_1$ and $T_2$:
\begin{align}
    1\approx \Delta_0(T_2)+\dfrac{\Delta_0(T_2)-\Delta_0(T_1)}{T_2-T_1}(T_c-T_2)
\end{align}

\noindent Note that the above linear interpolation method is violated for $T\ll T_c$, as then the gap will have more non-linear dependence on the temperature. Identifying the first term on the right-hand side as the y-intercept $b$ and the second term as proportional to slope $m$ when plotting $\Delta_0(T)$ vs. $T$, we can then solve for $T_c$:
\begin{align}
1\approx b+m(T_c-T_2)\rightarrow T_c=T_2+\dfrac{1-b}{m}
\end{align}
As a consequence, from $T_1$, $T_2$, $\Delta_0(T_1)$, and $\Delta_0(T_2)$, we can get an estimate of $T_c$. An iteration process can then be used to update $T_1$ and $T_2$ with the previous estimate of $T_c$ so as to eventually reach the convergence condition $\Delta(i\omega_0)\rightarrow 1$, which corresponds to the true $T_c$ as determined by the Eliashberg equations on the Matsubara frequency axis.

In Fig.~\ref{LCComparison}, we see how in the absence of TLSs the Leavens-Carbotte solution is generally a very good approximation to the numerical result found by solving the linearized Eliashberg equations. The only notable exception is the 3 monolayer case, where the agreement between numerics and analytics is quite bad. This can be seen as a consequence of the form of $\alpha^2 F(\omega)$ in the 3 layer case, which is characterised by certain features not seen in other layers (the absence of a large peak in $\alpha^2F(\omega)$ at high frequency, more equal spread of $\alpha^2 F(\omega)$ over the whole frequency range, etc.). In Fig.~\ref{fig:LCCompare_with_TLS}, we see a similar comparison of the Leavens-Carbotte solution for $T_c/T_c^{(0)}$ and the numerical solution, but now as a function of the TLS frequency for $6-1$ MLs of aluminium. $n_{TLS}=0.3$ is taken throughout. As mentioned before in this Supplement, there is poor agreement between analytics and numerics when there is a large concentration of TLSs at low frequency due to the fact the LC solution does not take into account the sensitivity of $T_c$ on different frequency regimes of $\alpha^2F(\omega)$~\cite{Bergmann1973Aug}. For higher TLS frequencies, we see excellent agreement between numerics and analytics, excluding the case for 3 monolayers. In addition, we see that maximal $T_c/T_c^{(0)}$ occurs around $0.2-0.4\times$ the frequencies corresponding to the most salient low-frequency peaks in $\alpha^2 F(\omega)$, and therefore agrees with our prediction from the Einstein phonon model. Those TLS frequencies that correspond to a maximised value of $T_c/T_c^{(0)}$ are then re-used for the determination of the maximum gap edge induced from TLSs.

\bibliography{main}{}